\documentclass[12pt]{article}
\usepackage{latexsym}
\def\beq{\begin{equation}}
\def\eeq{\end{equation}}
\def\beqs{\begin{equation*}}
\def\eeqs{\end{equation*}}
\def\bea{\begin{eqnarray}}
\def\eea{\end{eqnarray}}
\def\ba{\begin{array}}
\def\ea{\end{array}}
\def\beas{\begin{eqnarray*}}
\def\eeas{\end{eqnarray*}}
\def\bq{\begin{quote}}
\def\eq{\end{quote}}
\def\p{\partial}

\def\da{{\dot a}}

\def\db{{\dot b}}
\def\c{\gamma}
\def\dc{{\dot c}}
\def\d{\delta}
\def\dd{\dot{ d}}
\def\e{\varepsilon}
\def\ep{\epsilon}
\def\l{\lambda}
\def\bl{\bar\lambda}
\def\th{\theta}
\def\bth{\bar\theta}

\def\s{\sigma}
\def\bs{{\bar {\s}}}

\def\dh{\dot h}
\def\x{\xi}
\def\bx{\bar {\xi}}
\def\bz{\bar {z}}
\def\nn{\nonumber}
\def\o{\overline}
\def\dgr{\dagger}
\def\tr{{\rm Tr~}}
\def\hc{{\rm h.c.}}
\def\t{\tilde}
\def\g{\grave}
\def\Tr{{\rm Tr}}
\def\half{\frac{1}{2}}
\def\thalf{\textstyle{ {1 \over 2} } }

\def\third{\frac{1}{3}}
\def\fourth{\frac{1}{4}}
\def\ihalf{{\frac{i}{2}}}
\def\sqr2{\sqrt{2}}
\def\rhalf{\frac{1}{\sqrt{2}}}
\def\irhalf{\frac{i}{\sqrt{2}}}
\def\dcx{d^3\!x}

\def\dcp{d^3\!p}
\def\dck{d^3\!k}
\def\vp{\varphi}
\def\cA{\mathcal A}
\def\cD{\mathcal D}
\def\cL{\mathcal L}

\def\G{Grassmann }
\def\qm{quantum mechanics }
\def\qmp{quantum mechanics}
\def\susy{supersymmetry }

\def\susyc{supersymmetric }
\def\susyp{supersymmetry}
\def\sf{superfield }
\def\sfs{superfields }

\def\and{\quad {\hbox \mathrm{and}} \quad}
\def\for{\quad {\hbox \mathrm{for}} \quad}
\def\as{\quad {\hbox \mathrm{as}} \quad}
\def\whence{\quad {\hbox \mathrm{whence}} \quad}
\def\gappeq{\mathrel{\rlap {\raise.5ex\hbox{$>$}}
{\lower.5ex\hbox{$\sim$}}}}
\def\lappeq{\mathrel{\rlap{\raise.5ex\hbox{$<$}}
{\lower.5ex\hbox{$\sim$}}}}

\begin{document}
\pagestyle{empty}
\begin{flushright}
NMCPP/97-11\\
hep-ph/9907295\\
\end{flushright}
\vspace*{1cm}
\begin{center}
{\bf\Large Elements of Supersymmetry\footnote{Research 
partially supported by the U.S. Department of Energy}}\\
\vspace*{1cm}
{\bf Kevin Cahill\footnote{kevin@kevin.phys.unm.edu
\quad http://kevin.phys.unm.edu/~kevin/}}\\
\vspace{.5cm}
New Mexico Center for Particle Physics\\
Department of Physics and Astronomy\\
University of New Mexico\\
Albuquerque, New Mexico 87131-1156\\
\vspace*{1cm}
{\bf Abstract}
\end{center}
\begin{quote}
These notes are intended to provide an introduction
to supersymmetry.
They begin with \susyc 
quantum mechanics and the basic properties
of spinor fields. 
The \susy of simple theories
of spin-zero and spin-one-half fields
is discussed with emphasis upon the charges
that generate the transformations of \susyp.
Abelian and non-abelian supersymmetric gauge theories 
are characterized in their simpler and more general forms.
Superfields are defined, and the concise notation
they make possible is described.
The minimal \susyc standard model 
is discussed with particular attention to the
Higgs fields and the electro-weak superpotential.
\end{quote}
\vfill
\begin{flushleft}
\today\\
\end{flushleft}
\vfill\eject

\setcounter{equation}{0}
\pagestyle{plain}

\section{Introduction}
Most articles and books on \susy
are written for experts.
These notes on \susy are
explicitly intended for 
graduate students in experimental high-energy physics
and were written as part of a seminar
which Michael Gold and I taught
at the University of New Mexico
during the spring of 1997.
\par
Because papers about \susy
are often written in \sf notation
and because \sfs are defined in terms
of \G variables, most students
get the impression that \susy is harder
than quantum field theory.
In fact \susyc theories are nothing more than 
particular quantum field theories 
in which the fields and parameters 
satisfy special relations that make the
theories \emph{simpler, more symmetric,
and easier to understand}.
\par
The relative simplicity of \susy is illustrated 
by the case of \susyc quantum mechanics,
which is discussed in section 2.
The exact description of the
ground state, which typically in ordinary \qm is impossible,
becomes trivial in \susyc \qmp.  
\par
Section 3 is a laconic list of the notation
that will be employed in this work.
Section 4 contains a careful discussion
of spin-one-half fields. 
Special attention is given to left-handed
and right-handed two-component spinors
and to Majorana and Dirac fields.
\par
A discussion of the simplest \susyc
field theories is given in section 5
with emphasis upon the fermionic charges
that generate the transformations of \susyp.
The most general \susyc theories of
scalar and spinor fields are described
in section 6.
\par
Section 7 is an explicit description
of the simpler supercharges and of the
magic anti-commutation relations 
that they satisfy.
The transformations they generate are 
characterized in section 9.
Some formal properties of the
supercharges of free \susyc theories
are discussed in section 10.
\par
The simpler supersymmetric gauge theories 
and the Fayet-Iliopoulos $ D $ term are 
defined in section 11.
More general \susyc field theories
are described in section 12.
The breaking of gauge symmetry
and supersymmetry in super \emph{QED\/} 
is worked out.
Superfield notation is introduced
in section 13.
\par
The minimal \susyc standard model
is discussed in section 14.
The three reasons why two Higgs doublets
are needed rather than one as in the
standard model are explained.
The electro-weak superpotential
is studied in some detail.

\section{Supersymmetric Quantum Mechanics}

Supersymmetric quantum mechanics
was invented Nicolai~\cite{Nicolai}
and independently by Witten~\cite{Witten}\null.
A quantum-mechanical system is supersymmetric 
if it has $ N $ charges $ Q_i $ that 
commute with the hamiltonian
\beq
[ H , Q_i ] = 0 \for i = 1, \dots, N
\eeq
and satisfy the algebra
\beq
\{ Q_i , Q_j \} = \d_{ij} H .
\eeq
\par
The simplest example is for $ N=2 $
and describes a spin-one-half particle
moving on a line.
The wave function $ \psi(x) $ is
\beq
\langle x | \psi \rangle =
\psi(x) = \left( \ba{c} \phi_1(x) \\
                        \phi_2(x) \ea \right) .
\eeq
The charges $ Q_i $ are
\bea
Q_1 & = & \half \left( \s_1 p + \s_2 W(x) \right) \nn\\
Q_2 & = & \half \left( \s_2 p - \s_1 W(x) \right) ,
\eea
where the $ \s $'s are the Pauli matrices (\ref {sigmas})
and 
\beq
\langle x | p | \psi \rangle = { \hbar \over i }
{ d \over dx } \psi(x) .
\eeq
They are hermitian
\beq
Q_1^\dgr = Q_1 \and Q_2^\dgr = Q_2 .
\eeq
The superpotential $ W(x) $ may be any function of $ x $
that grows sufficiently at large $ | x | $ 
\beq
| W(x) | \to \infty \as |x| \to \infty 
\eeq
to ensure that the spectrum of $ H $ is discrete.
\par
Explicit computation shows that
\beq
\{ Q_1 , Q_1 \} = 2 Q_1^2 = H 
= \half p^2 + \half W^2(x) 
+ { \hbar \over 2 } \s_3 { d W(x) \over dx } ,
\eeq
that
\beq
\{ Q_2 , Q_2 \} = 2 Q_2^2 = H 
= \half p^2 + \half W^2(x) 
+ { \hbar \over 2 } \s_3 { d W(x) \over dx } ,
\eeq
and that
\beq
\{ Q_1 , Q_2 \} = 0 .
\eeq
\par
Supersymmetry is unbroken if either charge $ Q_1 $ or $ Q_2 $
annihilates a normalizable state $ \psi_0(x) $
\beq
Q_1 \psi_0(x) = 0 .
\eeq
In this case the state $ \psi_0(x) $ has zero energy
\beq
H \psi_0(x) = 2 Q_1^2 \psi_0(x) = 0 ,
\eeq
and the other charge $ Q_2 $ must also annihilate
the state $ \psi_0(x) $
\beq
\left( Q_2 \psi_0 , Q_2 \psi_0 \right) = 
\left( \psi_0 , Q_2^\dgr Q_2 \psi_0 \right) =
\left( \psi_0 , Q_2^2 \psi_0 \right) =
\left( \psi_0 , \half H \psi_0 \right) = 0 .
\eeq
Since both charges annihilate the state $ \psi_0(x) $,
that state is left invariant under the unitary
transformation
\beq
U( \th ) = e^{ -i \th_i Q_i } ,
\eeq
and so supersymmetry is unbroken.
\par
Note that all the energies $ E_n $ must be
positive or zero since
\beq
H = 2 Q_1^2 = 2 Q_1 ^\dgr Q_1 .
\eeq
\par
The state $ \psi_0 $ is easy to find if it exists:
\bea
Q_1 \psi_0 (x) & = & 0 \nn\\
\half \left( \s_1 p + \s_2 W(x) \right) \psi_0 (x) & = & 0 \nn\\
\s_1 p \psi_0 (x) & = & - \s_2 W(x) \psi_0 (x) \nn\\
\hbar { d \psi_0 (x) \over dx } & = &  
\s_3  W(x) \psi_0 (x) \nn\\
\psi_0 (x) & = &  
\exp \left( \int_0^x dy \hbar^{-1} \, W(y) \s_3 \right) \psi_0 (0) .
\eea
Now if the superpotential $ W (x) $ satisfies
\beq
W(x) \to \infty \as x \to \infty
\eeq
and
\beq
W(x) \to - \infty \as x \to - \infty ,
\eeq
then the normalizable state $ \psi_0 $ is
\beq
\psi_0 (x) = \exp \left( \int_0^x dy \hbar^{-1} \, W(y) \right)
\left( \ba{c} 0 \\
              1 \ea \right) .
\eeq 
If $ W(x) $ is continuous, then $ W (x_0 ) = 0 $
for some $ x_0 $.
\par
Equivalently if the superpotential $ W (x) $ satisfies
\beq
W(x) \to - \infty \as x \to \infty
\eeq
and
\beq
W(x) \to \infty \as x \to - \infty ,
\eeq
then the normalizable state $ \psi_0 $ is
\beq
\psi_0 (x) = \exp \left( \int_0^x dy \hbar^{-1} \, W(y) \right)
\left( \ba{c} 1 \\
              0 \ea \right) .
\eeq
Again if $ W(x) $ is continuous, then $ W (x_0 ) = 0 $
for some $ x_0 $.
\par
But if the superpotential $ W (x) $ satisfies
\beq
W(x) \to \infty \as |x| \to \infty 
\eeq
or
\beq
W(x) \to - \infty \as |x| \to \infty ,
\eeq
then there is no normalizable state $ \psi_0 $
that is annihilated by $ Q_1 $ or by $ Q_2 $.
In this case, supersymmetry is dynamically broken.

\section{Notation}
We shall use an Anglicized version
of the notation
of Bagger and Wess,
which they describe in the
appendices of their book~\cite{BW}\null.
Space-time indices are labeled
by letters like $ l, m, n $\null.
Spatial indices are labeled
by letters like $ i, j, k $\null.
The flat metric $ \eta $ is 
\begin{equation}
\left( \eta^{mn} \right) = \left( \matrix { -1&0&0&0\cr
                              0&1&0&0\cr 
                              0&0&1&0\cr 
                              0&0&0&1\cr } \right). 
\eeq
\par
Dotted and undotted spinor indices
run from 1 to 2 and 
are denoted by the early letters
of the English alphabet. 
Spinor indices are raised and lowered
by the $ \e $ tensors
\begin{equation}
\left( \e^{a b} \right) = \left( \matrix { 0&1\cr
                                           -1&0\cr } \right)
\label {epsup}
\eeq
and
\begin{equation}
\left( \e_{a b} \right) = \left( \matrix { 0&-1\cr
                                            1&0\cr } \right).
\label {epsdown}
\eeq
For example, $ \psi^a = \e^{a b} \psi_b $
and $ \chi_a = \e_{a b} \chi^b $\null.
If the spinors $ \psi $ and $ \chi $ anti-commute,
then the product $ \psi \chi $ is
\begin{equation}
\psi \chi = \psi^a \chi_a = - \psi_a \chi^a
= \chi^a \psi_a = \chi \psi.
\eeq
The identity 
\beq
\psi^a \psi^b = - \half \, \psi \psi \, \e^{ab}
\label {thth}
\eeq
is occasionally useful.
\par
The hermitian conjugates of the spinors
$ \psi $ and $ \chi $ are $ \bar \psi $ and $ \bar \chi $:
\begin{eqnarray}
\bar \psi^\da &=& \left( \psi^a \right)^\dgr \nn \\
\bar \chi_\da &=& \left( \chi_a \right)^\dgr.
\eea
Their dotted indices are raised and lowered
by the tensors $ \e^{\da\db} $ and $ \e_{\da\db} $
which are equal to their undotted counterparts
$ \e^{ab} $ and $ \e_{ab} $\null.
The product $ \bar \psi \bar \chi $ is 
\begin{equation}
\bar \psi \bar \chi = \bar \psi_\da \bar \chi^\da
= - \bar \psi^\da \bar \chi_\da  = \bar \chi_\da \bar \psi^\da
= \bar \chi \bar \psi.
\eeq
These definitions allow us to write
\begin{equation}
\left( \chi \psi \right)^\dgr = \left( \chi^a \psi_a \right)^\dgr
= \bar \psi_\da \bar \chi^\da = \bar \psi \bar \chi = \bar \chi \bar \psi
\eeq
and
\beq
\bar \psi^\da \bar \psi^\db = \half \, \bar \psi \bar \psi \, \e^{\da\db}.
\label {thbarthbar}
\eeq
\par
The Pauli matrices $ \s^m_{a\db} $ are
\begin{equation}
\matrix{ \s^0 = \left( \matrix { -1&0\cr
                                  0&-1\cr } \right)
       & \s^1 = \left( \matrix { 0&1\cr
                                 1&0\cr } \right)\cr
         \s^2 = \left( \matrix {  0&-i\cr
                                  i&0\cr } \right)
       & \s^3 = \left( \matrix { 1&0\cr
                                 0&-1\cr } \right)\cr }
\label {sigmas}
\eeq
in which the rule $ \s^0 = - I $ 
is occasioned by the choice of metric $ \eta = ( -1, 1, 1, 1) $\null.
The barred Pauli matrices $ \bar \s^{m\da b} $ 
are defined as
\beq
\bar \s^{m\da b} = \e^{ \da \dc } \e^{ b d } \s^m_{ d \dc } \, ,
\whence \s^n_{ a \db } = \e_{ a c } \, \e_{ \db \dd } \, \bs^{n \dd c } \, ,
\eeq
and are related to the unbarred ones by
\begin{eqnarray}
\bar \s^0 &=& \s^0 = - I \nn\\
\vec { \bar \s } &=& \mbox{} - \vec \s.
\eea
\par
The generators of the Lorentz group in the spinor
representation are
\bea
\s^{nm\;\; c}_{\quad a} & = & \fourth
\left( \s_{a\db}^n \bs^{m\db c}
- \s_{a\db}^m \bs^{n\db c} \right) \\
\bs^{nm\da}_{\quad\;\;\;\dc} & = & \fourth
\left( \bs^{n\da b} \s^m_{b \dc}
- \bs^{m \da b} \s^n_{b \dc} \right).
\label {snm}
\eea
\par
The Pauli matrices satisfy the relations
\begin{equation}
\left( \s^m \bar \s^n + \s^n \bar \s^m \right)_a^{~b} 
= -2 \eta^{mn} \d_a^{~b},
\label {Pauli}
\eeq
\begin{equation}
\left( \bar \s^m \s^n + \bar \s^n \s^m \right)^\da_{~\db}
= -2 \eta^{mn} \d^\da_{~\db},
\eeq
the trace identities
\begin{equation}
\tr \s^m \bar \s^n = - 2 \eta^{mn}
\quad {\hbox \mathrm{and}} \quad
\tr \s^m \s^n = 2 \d^{mn} ,
\label {traces}
\eeq
and the completeness relation
\begin{equation}
\s_{a\dc}^{~m} \bar \s_m^{~\dd b} = -2 \d_a^{~b} \d_\dc^{~\dd},
\label {comp}
\eeq
which implies that
\beq
\left( \psi \phi \right) \bar \chi_\db = 
- \half \phi \s^n \bar \chi \left( \psi \s_n \right)_\db
= - \half \phi \s^n \bar \chi \, \psi^a \s_{ n a \db } .
\label {fcomp}
\eeq
The adjoint of this relation is
\beq
\chi_b \left( \bar \phi \bar \psi \right) =
- \half \left( \s_n \bar \psi \right)_b \, \chi \s^n \phi .
\label {afcomp}
\eeq
\par
Other useful identities are
\beq
\left( \bx \, \bs^m \right) _a = \e_{ ab } \left( \bx \, \bs^m \right) ^b
= - \s^m_{ a \dc } \, \bx ^ \dc ,
\label {bxbs}
\eeq
\beq
\left( \bs^m \, \x \right) ^\da = - \left( \x \, \s^m \right) ^\da
= - \e^{ \da \dc } \, \x^d \, \s^m _{ d \dc } ,
\label {bsx}
\eeq
\beq
\chi \s^n \bar \psi = - \bar \psi \bar \s^n \chi ,
\label {trans}
\eeq
\beq
\chi \s^m \bar \s^n \psi = \psi \s^n \bar \s^m \chi ,
\label {sstrans}
\eeq
\begin{equation}
\left( \chi \s^m \bar \psi \right)^\dgr
= \psi \s^m \bar \chi ,
\label {dagid}
\eeq
and
\beq
\left( \chi \s^m \bar \s^n \psi \right)^\dgr
= \bar \psi \bar \s^n \s^m \bar \chi .
\label {ssdagid}
\eeq

\section{Spinor Fields}
All known matter fields are of spin one-half.
Yet spinor fields are confusing, and the authors
of most books choose to assume that the topic
has been worked out somewhere else. 
\subsection{The Right-Handed Massless Field}
\par
The Lagrange density
\begin{equation}
{ \cal L}_K = \frac{ i }{ 2 } \partial_n \bar \psi \bar \s^n \psi
- \frac{ i }{ 2 } \bar \psi \bar \s^n \partial_n \psi
\label {1/2}
\eeq
describes a massless two-component spinor field $ \psi $
of spin one-half which satisfies the equation
\beq
\bar \s^n \partial_n \psi = 0.
\label {1/2 eq}
\eeq
This field transforms according to the $ ( \thalf , 0 ) $
representation of the Lorentz group
and so carries undotted indices.
We may expand $ \psi $ in terms of 
the operators $ b(p,s) $
that annihilate the particles of the field $ \psi $
and the operators $ b_c^\dgr(p,s) $ that create
the particles that are the anti-particles  
of the particles of the field $ \psi $:
\beq
\psi_a( x ) =
\sum_s \int \!
\frac{ d^3p }{ \sqrt{ ( 2 \pi )^3 } } \,
\left[
e^{ ip \cdot x } u_a(p,s) b(p,s) 
+ e^{ - ip \cdot x } v_a(p,s) b_c^\dgr(p,s)
\right],
\label {1/2 exp}
\eeq
in which $ p^0 = | \vec p |$.
We shall see that
the states $ b^ \dgr ( p , s ) | 0 \rangle $ 
and $ b_c ^ \dgr ( p , s ) | 0 \rangle $
are different, because they have opposite helicity
whether or not the field $ \psi $ carries a charge.
But circumstances and conventions
will determine whether we call these states 
particles and anti-particles 
or just different helicity states of the same particle.
\par
Since the free field $ \psi $
satisfies the free-field equation (\ref{1/2 eq}),
the spinors $ u(p,s) $ and $ v(p,s) $
must be eigenvectors of $ \vec \s \cdot \vec p $
\beq
\vec \s \cdot \vec p \, u(p,s) =  p^0 \, u(p,s)
\quad {\hbox \mathrm{and}} \quad
\vec \s \cdot \vec p \, v(p,s) = p^0 \, v(p,s)
\label {1/2 ev eqs}
\eeq
with positive eigenvalue $ p^0 $. 
Now $ \vec \s \cdot \vec p $ has two eigenvalues $ \pm | \vec p |$,
and there is only one eigenvector of $ \vec \s \cdot \vec p $
with positive eigenvalue $ p^0 = | \vec p | $.
Normalized to unity,
this eigenvector $ u(p) $ is
\beq
u(p) = \frac{ 1 }{ \sqrt{ 2 p^0 ( p^0 + p^3 ) } }
\left( \ba{c} p^0 + p^3 \\
              p^1 + i p^2 
\ea \right)
\label {u(p)}
\eeq
apart from a phase factor.
\par
Thus in the expansion (\ref{1/2 exp})
only one helicity state survives,
and the expansion of the massless field $ \psi $
takes the form
\beq
\psi_a( x ) =
\int \!
\frac{ d^3p }{ \sqrt{ ( 2 \pi )^3 } } \,
\left[
e^{ ip \cdot x } u_a(p,\thalf) b(p,\thalf)
+ e^{ - ip \cdot x } v_a(p,-\thalf) b_c^\dgr(p,-\thalf)
\right]
\label {actual 1/2 exp}
\eeq
in which \( u_a(p,\thalf) = v_a(p,-\thalf) = u_a ( p ) \).
Equivalently, we may use the briefer notation
\beq
\psi_a( x ) =
\int \!
\frac{ d^3p }{ \sqrt{ ( 2 \pi )^3 } } \,
u_a(p)
\left[
e^{ ip \cdot x } b(p)
+ e^{ - ip \cdot x } b_c^\dgr(p)
\right].
\label {short 1/2 exp}
\eeq
We impose the anti-commutation relations
\beq
\{ b( p , s ) , b^\dgr( k , s' ) \} = \d_{s s'} \d( \vec p - \vec k )
\and
\{ b_c( p , s ) , b_c^\dgr( k , s' ) \} = \d_{s s'} \d( \vec p - \vec k )
\label {bbdag}
\eeq
or equivalently
\beq
\{ b( p ) , b^\dgr( k ) \} = \d( \vec p - \vec k )
\and
\{ b_c( p ) , b_c^\dgr( k ) \} = \d( \vec p - \vec k )
\label {short bbdag}
\eeq
and require all other anti-commutators
of $ b $ and $ b_c $ operators to vanish.
It follows then from these relations and from
the explicit form (\ref{u(p)}) of the spinor $ u( p ) $
that the field $ \psi $ satisfies the 
equal-time anti-commutation relations
\beq
\{ \psi_a( t, \vec x ) , \psi_b^\dgr ( t, \vec y ) \}
= \d_{ab} \, \d( \vec x - \vec y ),
\label {1/2 1/2 dag}
\eeq
and 
\beq
\{ \psi_a( t, \vec x ) , \psi_b( t, \vec y ) \} = 0.
\label {1/2 1/2}
\eeq
\par
The spin part of the angular-momentum operator is
\beq
\vec \Sigma = \int \! d^3\!x \,
: \psi^\dgr (x) \, \half \vec \s \, \psi (x) :
\label {Sigma}
\eeq
in which the colons indicate (fermionic) normal ordering.
By using the anti-commutation relations 
(\ref {1/2 1/2 dag}) and (\ref {1/2 1/2}),
one may show that the spin operator obeys 
the angular-momentum commutation relations
\beq
\left[ \Sigma_i , \Sigma_j \right] 
= i \ep_{ i j k } \Sigma_k
\label {spin cr}
\eeq
in which the Levi-Civita tensor $\ep_{ i j k }$ 
is totally anti-symmetric with $\ep_{123} = 1$.
If we substitute the briefer expansion (\ref {short 1/2 exp})
of the field $ \psi $ into the spin operator $\Sigma$,
then we may write it as
\beq
\vec \Sigma = 
\int \! \dck 
\left[ b ^\dgr ( k ) b ( k ) - b_c ^\dgr ( k ) b_c ( k ) \right]
u ^\dgr ( k ) \, \half \vec \s \, u ( k ) .
\label {Sigmabb}
\eeq
By using the anti-commutation relations (\ref {short bbdag})
and the relation 
$ \vec p \cdot \vec \s \, u(p) = | \vec p | u(p) $,
we may show that the state \( b ^\dgr ( p ) | 0 \rangle \)
is an eigen-state of 
the helicity operator $ \vec P \cdot \vec \Sigma $
with eigen-value $ | \vec p |/2 $:
\bea
\lefteqn{
\vec P \cdot \vec \Sigma 
\, b ^\dgr ( p ) \, | 0 \rangle } \hspace {.5in} \nn \\
& = & \mbox{} \int \! \dck 
\left[ b ^\dgr ( k ) b ( k ) - b_c ^\dgr ( k ) b_c ( k ) \right]
u ^\dgr ( k ) \, \half \vec P \cdot \vec \s \, u ( k ) 
b ^\dgr ( p ) | 0 \rangle \nn\\
& = & \mbox{} \int \! \dck
b ^\dgr ( k ) b ( k )
u ^\dgr ( k ) \, \half \vec p \cdot \vec \s \, u ( k )
b ^\dgr ( p ) | 0 \rangle  \nn\\
& = & \mbox{} \int \! \dck
b ^\dgr ( k ) 
\d ( \vec k - \vec p )
u ^\dgr ( k ) \, \half \vec p \cdot \vec \s \, u ( k )
| 0 \rangle  \nn\\
& = & \mbox{} 
\left( u ^\dgr ( p ) \, \half \vec p \cdot \vec \s \, u ( p ) \right)
\, b ^\dgr ( p ) | 0 \rangle 
= \left( \frac{ | \vec p | }{ 2 } \right) 
\, b ^\dgr ( p ) | 0 \rangle .
\label {mv pS b}
\eea
We say that the $ \psi $-particle and the 
state $ | b, \vec p \rangle $ are right handed,
or have positive helicity,
because their momentum and spin are parallel.
A similar computation for the state 
\( b_c ^\dgr ( p ) | 0 \rangle \)
gives
\beq
\vec P \cdot \vec \Sigma 
\, b_c ^\dgr ( p ) \, | 0 \rangle 
= - \left( \frac{ | \vec p | }{ 2 } \right) 
\, b_c ^\dgr ( p ) \, | 0 \rangle.
\label {mv pS b_c}
\eeq
The anti-$ \psi $-particle 
and the state $ | b_c, \vec p \rangle $ 
are said to be left handed,
or to have negative helicity,
because their momentum and spin are anti-parallel.
\par
The field $ \psi $ is called right handed
because it annihilates particles that are right handed
and creates particles that are left handed.
Since the state $ | b, \vec p \rangle $
and the state $ | b_c, \vec p \rangle $
have opposite helicities,
they can not be the same.
Yet it is conventional to distinguish the 
two states by the words ``particle'' and ``anti-particle''
only if the field $ \psi $ carries a charge.
For instance, we speak of neutrino and anti-neutrino,
but not of photon and anti-photon.
\subsection{The Left-Handed Massless Field}
\par
If the spinor $ \psi_a $ is right handed,
then the conjugate spinor $ \bar \psi^\da $
is left handed. 
We might emphasize this change in handedness
by using the letter $ \chi $ instead of $ \psi $
and writing explicitly
\bea
\bar \chi^\da & = & \e^{ \da \db } \, \bar \psi_\db 
= \e^{ \da \db } \, \left( \psi_b \right)^\dag 
\label {lhspinorbar} \\
\chi^a & = & \e^{ a b } \, \psi_b .
\label {lhspinor}
\eea
By making these substitutions
in the action density (\ref {1/2}),
using the identity (\ref {trans}), and 
ignoring any c-number terms arising from 
the use of that identity,
we may arrive at the Lagrange density
for a massless, left-handed,
two-component spinor field $ \bar \chi $
\begin{equation}
{ \cal L}_K = \frac{ i }{ 2 } \partial_n \chi \s^n \bar \chi
- \frac{ i }{ 2 } \chi \s^n \partial_n \bar \chi .
\label {L1/2}
\eeq
This field of spin one-half satisfies the equation
\beq
\s^n \p_n \bar \chi = 0.
\label {L1/2 eq}
\eeq
The bar indicates that the field \( \bar \chi \)
transforms according to the $ ( 0 , \thalf ) $
representation of the Lorentz group and so
carries dotted indices.  (Whether this dotty notation
is worth the effort is unclear.)
Now in the briefer expansion 
\beq
\bar \chi_a( x ) =
\int \!
\frac{ d^3p }{ \sqrt{ ( 2 \pi )^3 } } \,
v_a(p)
\left[
e^{ ip \cdot x } c(p)
+ e^{ - ip \cdot x } c_c^\dgr(p)
\right],
\label {L1/2 exp}
\eeq
the spinor $ v( p ) $
must be an eigenvector of $ \vec \s \cdot \vec p $
with eigenvalue $ - p^0 = - | \vec p | $
\beq
\vec \s \cdot \vec p \, v( p ) = - | \vec p | \, v ( p ) .
\label {lspeq}
\eeq
Normalized to unity,
this eigenvector $ v(p) $ is
\beq
v(p) = \frac{ 1 }{ \sqrt{ 2 p^0 ( p^0 + p^3 ) } }
\left( \ba{c} 
              p^1 - i p^2 \\
             - p^0 - p^3 
\ea \right)
\label {Lv(p)}
\eeq
up to an arbitrary phase factor
here chosen so that
\beq
v(p) ^ \da = \e ^{ \da \db } \left( u_b \right)^* 
\label {v=eustar}
\eeq
in conformity with (\ref {lhspinorbar}).
\par
We impose the anti-commutation relations
\beq
\{ c( p , s ) , c^\dgr( k , s' ) \} = \d_{s s'} \d( \vec p - \vec k )
\and
\{ c_c( p , s ) , c_c^\dgr( k , s' ) \} = \d_{s s'} \d( \vec p - \vec k )
\label {ccdag}
\eeq
or equivalently
\beq
\{ c( p ) , c^\dgr( k ) \} = \d( \vec p - \vec k )
\and
\{ c_c( p ) , c_c^\dgr( k ) \} = \d( \vec p - \vec k )
\label {short ccdag}
\eeq
which together with the explicit form 
(\ref{Lv(p)}) of the spinor $ v( p ) $ imply that
that the field $ \chi $ satisfies the
equal-time anti-commutation relations
\beq
\{ \chi_a( t, \vec x ) , \chi_b^\dgr ( t, \vec y ) \}
= \d_{ab} \, \d( \vec x - \vec y ),
\label {L1/2 1/2 dag}
\eeq
and
\beq
\{ \chi_a( t, \vec x ) , \chi_b( t, \vec y ) \} = 0.
\label {L1/2 1/2}
\eeq
\par
The spin part of the angular momentum
operator is
\beq
\vec \Sigma = \int \! d^3\!x \,
: \chi (x) \, \half \vec \s \, \bar \chi (x) :
\label {SigmaChi}
\eeq
in which the colons indicate (fermionic) normal ordering.
As in the case of the right-handed field,
one may derive the eigen-value equations
\bea
\vec P \cdot \vec \Sigma \, c ^\dgr ( p ) \, | 0 \rangle
& = & \mbox{} - \left( \frac{ | \vec p | }{ 2 } \right)
\, c ^\dgr ( p ) | 0 \rangle \label {spinc} \\
\vec P \cdot \vec \Sigma \, c_c ^\dgr ( p ) \, | 0 \rangle
& = & \mbox{} \left( \frac{ | \vec p | }{ 2 } \right)
\, c_c ^\dgr ( p ) | 0 \rangle \label {spincbar}
\eea
which show that the particles 
annihilated by the field $ \chi $ 
are left handed and that their
anti-particles are right handed.
That is, the state $ c^\dgr( p ) | 0 \rangle $
is left handed, or of negative helicity,
in the sense that its momentum and its spin are 
in opposite directions, while the
state $ c_c^\dgr( p ) | 0 \rangle $
is right handed, or of positive helicity,
because its momentum is parallel to its spin.

\subsection{Majorana Mass}
Let us now return to the massless right-handed action density (\ref{1/2}) 
and add to it a Majorana-mass term and its hermitian conjugate 
\begin{equation}
{ \cal L}_K = \frac{ i }{ 2 } \partial_n \bar \psi \bar \s^n \psi
- \frac{ i }{ 2 } \bar \psi \bar \s^n \partial_n \psi
- \half m \psi \psi - \half m \bar \psi \bar \psi,
\label {Mm1/2}
\eeq
in which we have absorbed a possible phase
in the Majorana mass $ m $ into the 
definition of the field $ \psi $.
If the mass $ m $ is really a constant,
and not a field, then the field $ \psi $
can not carry a charge, at least not one
whose conservation is formally implied 
by a symmetry of the action.
It is possible that the mass $ m $ is actually
a field $ \hat m $
that assumes a mean value $ m $ in the vacuum.
And in this case, the field $ \psi $ may carry
a charge $ q $ if the mass field $ \hat m $ carries
charge $ - 2 q $.  But if the mass field $ \hat m $
acquires the non-zero mean value $ m $ in the vacuum,
then the symmetry of the action
that formally implies the conservation
of the charge $ q $ would be spontaneously broken.
So the field $ \psi $ must either be neutral
or the carrier of a charge of a broken symmetry.
\par
We shall consider here the case in which $ m $
is really a constant.
The field $ \psi $ then satisfies the equation
\beq
-i \bar \s^{n \da b} \partial_n \psi_b = m \bar \psi^\da
= m \e^{ \da \db } \bar \psi_\db .
\label {M1/2 eq}
\eeq
In a simpler notation, with $ \e^{ \da \db } = i \s^2_{ a \db } $,
this equation is
\beq
i \left( \partial_0 + \vec \s \cdot \nabla \right) \psi
= i m \s^2 \bar \psi.
\label {sM1/2 eq}
\eeq
\par
If we try to solve this equation
by expanding the field $ \psi $
in the form
\beq
\psi_a( x ) =
\sum_s \int \!
\frac{ d^3p }{ \sqrt{ ( 2 \pi )^3 } } \,
\left[
e^{ ip \cdot x } u_a(p,s) b(p,s)
- i \s^2 e^{ - ip \cdot x } v_a^* (p,s) b_c^\dgr(p,s)
\right],
\label {1/2 exp again}
\eeq
then we find 
\bea
\left( p^0 - \vec \s \cdot \vec p \right) u( p , s ) b( p , s )
& = & m v( p , s ) b_c ( p , s ) \\
\left( p^0 - \vec \s \cdot \vec p \right) 
\left( -i \s^2 v^* ( p , s ) \right) b_c^\dgr ( p , s ) 
& = & - i m \s^2 u^* ( p , s ) b^\dgr ( p , s ).
\label {p-spacesoln}
\eea
Evidently the particles described by this field, $ \psi $,
must be the same as their anti-particles
\beq
b( p , s ) = b_c( p , s ) .
\eeq
By using the familiar relation 
$ \s^i \s^j = \d^{ij} + i \e_{ijk} \s^k $,
we may rewrite the spinor equations as
\bea
\left( p^0 - \vec \s \cdot \vec p \right) u ( p , s ) 
& = & m v( p , s ) \label {ueq} \\
\left( p^0 + \vec \s \cdot \vec p \right) v ( p , s )
& = &  m u( p , s ).
\label {veq}
\eea
By combining these equations, we find
\bea
\left( ( p^0 )^2 - ( \vec p )^2 \right) u( p , s ) & = &
\left( p^0 + \vec \s \cdot \vec p \right)
\left( p^0 - \vec \s \cdot \vec p \right) u( p , s ) \nn\\
& = & m \left( p^0 + \vec \s \cdot \vec p \right) v ( p , s )
= m^2 u( p , s ),
\label {psq=msq}
\eea
which shows that $ ( p^0 )^2 = ( \vec p )^2 + m^2 $.
The spinor equations (\ref{ueq}) and (\ref{veq})
together form the ``positive-mass" Dirac equation
\beq
\left( i p^n \c_n + m \right) U( p , s ) = 0,
\label {D+}
\eeq
and the spinors $ u ( p , s ) $ and $ v ( p , s ) $
may be identified with the lower and upper
components of its solution $ U( p , s ) $
\beq
U( p , s ) = \left( \ba{c}
       v ( p , s ) \\
       u ( p , s ) \ea \right).
\label {MD}
\eeq
\par
The addition of Majorana mass terms 
has converted the action density of a massless
right-handed field to that of a massive field
whose particles and anti-particles are the same
or differ by a charge of a broken symmetry.
\par
We may also add Majorana mass terms to
the action density (\ref {L1/2})
of the left-handed field \( \chi \)
\beq
{ \cal L}_K = \frac{ i }{ 2 } \partial_n \chi \s^n \bar \chi
- \frac{ i }{ 2 } \chi \s^n \partial_n \bar \chi
- \half m \chi \chi - \half m \bar \chi \bar \chi ,
\label {MmL1/2}
\eeq
in which we have absorbed a possible phase
in the Majorana mass $ m $ into the
definition of the field $ \chi $.
The equation of motion is now
\beq
- i \s ^n _{ a \db } \p _n \bar \chi^\db = m \chi_a 
= m \e _{ a b } \chi ^b .
\label {MmL1/2 eq}
\eeq
 
\subsection{The Dirac Field}
The original spin-one-half field 
is the Dirac field which is a four-component
combination of the right-handed field $ \psi $
and the left-handed field $ \bar \chi $.
In the massless case, the action density
for the Dirac field is just the sum 
of (\ref{L1/2}) for the left-handed field $ \bar \chi $
and 
of (\ref{1/2}) for the right-handed field $ \psi $
\beq
{ \cal L}_K = 
\frac{ i }{ 2 } \partial_n \chi \s^n \bar \chi
- \frac{ i }{ 2 } \chi \s^n \partial_n \bar \chi 
+ \frac{ i }{ 2 } \partial_n \bar \psi \bar \s^n \psi
- \frac{ i }{ 2 } \bar \psi \bar \s^n \partial_n \psi ,
\label {dirac}
\eeq
and little is gained by the combination
\beq
\Psi_l = \left( \ba {c}
              \bar \chi ^\da \\
              \psi _b
              \ea \right) ,
\label {Dirac}
\eeq
except that the two field equations
(\ref {1/2 eq} ) and (\ref {L1/2 eq} )
can be written as
\beq
\c^n \partial_n \Psi ( x ) = 0 .
\label {d eq}
\eeq
The $ \c $ matrices are required to satisfy
the anti-commutation relations
\beq
\left\{ \c^n , \c^m \right\} = 2 g^{ n m } .
\label {Gammas}
\eeq
A useful set of $ \c $ matrices is 
\beq
\c^0 = -i \left( \ba {cc} 0 & 1 \\
                          1 & 0 
                          \ea \right)
\and
\c^i = -i \left( \ba {cc} 0 & \s^i \\
                     - \s^i & 0 
                          \ea \right) .
\label {gammas}
\eeq
\par
The real point of the Dirac field is that it
provides a simple way of giving mass to a field
that carries a conserved charge and
whose particles and anti-particles are different.
In terms of the two-component fields $ \psi $ and $ \chi $,
the action density for the massive case is
after an integration by parts
\beq
{ \cal L}_D = 
- i \chi \s^n \partial_n \bar \chi 
- i \bar \psi \bar \s^n \partial_n \psi
- m \chi \psi - m \bar \psi \bar \chi ,
\label {MDirac}
\eeq
which in a simpler notation is 
\beq
{ \cal L}_D = 
i \bar \chi^\dgr \left( \partial_0 - \vec \s \cdot \nabla \right) \bar \chi
+ i \psi^\dgr \left( \partial_0 + \vec \s \cdot \nabla \right) \psi
- m \bar \chi^\dgr \psi - m \psi^\dgr \bar \chi . 
\label {sMDirac}
\eeq
In Dirac's notation this action density is
\beq
{ \cal L}_D =
- \o \Psi \left( \c^n \partial_n + m \right) \Psi
\label {Dirac's ad}
\eeq
in which $ \o \Psi \equiv i \Psi ^\dgr \c^0 $.
The Dirac field of mass $ m $ satisfies
\beq
\left( \c^n \partial_n + m \right) \Psi = 0 .
\label {Dirac's eq}
\eeq
The corresponding two-component equations are
\bea
-i \s ^n_{ a \db } \, \partial_n \bar \chi ^\db - m \psi _a & = & 0 \\
-i \bar \s ^{n \da b } \, \partial_n \psi_b - m \bar \chi ^\da & = & 0 .
\label {2comp dirac eq}
\eea
\par
The expansion of the Dirac field in terms 
of operators $ b ( p , s ) $ that annihilate
the particles of the field $ \Psi $ and 
operators $ b _c ^ \dgr ( p , s ) $ that
create their anti-particles is
\beq
\Psi_l ( x ) = \sum_s \int \! \frac{ \dcp }{ ( 2 \pi ) ^ {3/2} }
\left[ U_l ( p , s ) e^ { ipx } b ( p , s )
+ V_l ( p , s ) e^ { - ipx } b_c ^ \dgr ( p , s ) \right]
\label {Dirac field}
\eeq
in which the sum over spin states $ s $ is from
$ - 1/2 $  to $ 1/2 $ and the index $ l $ runs from 1 to 4. 
The spinors $ U ( p , s ) $ and $ V ( p , s ) $ are normalized to 
\( \d_{ s s' } \)
and satisfy the equations
\beq
\left( i p^n \c_n + m \right) U( p , s ) = 0
\and
\left( i p^n \c_n - m \right) V( p , s ) = 0 
\label {D+-}
\eeq
as well as
\beq
\sum_{ s = - {1 \over 2} }^ {1 \over 2}
U( p , s ) U ^\dgr ( p , s )
= \frac{ i }{ 2 p^0 } \left( -i p^n \c_n + m \right) \c^0 
\label {UU}
\eeq
and
\beq
\sum_{ s = - {1 \over 2} }^ {1 \over 2}
V( p , s ) V ^\dgr ( p , s )
= \frac{ i }{ 2 p^0 } \left( -i p^n \c_n - m \right) \c^0 .
\label {VV}
\eeq

\section{Simple Chiral Multiplets}
\subsection{A Right-Handed Spinor and a Spinless Boson}
\par
The susy action density for a right-handed
spinor field $ \psi $ not interacting with a complex
spin-zero field $ A $ is
\begin{equation}
{ \cal L}_K = \frac{ i }{ 2 } \partial_n \bar \psi \bar \s^n \psi
- \frac{ i }{ 2 } \bar \psi \bar \s^n \partial_n \psi
- \partial_n \bar A \partial^n A + \bar F F .
\label {L_K}
\eeq 
The field equations are
\bea
i \, \bar \s^n \partial_n \psi & = & 0 \\
\partial_n \partial^n A & = & 0 \\
F & = & 0 
\label {F=0}.
\eea
By ignoring total derivatives and using the constraint
(\ref{F=0}), we may write the kinetic Lagrange density (\ref{L_K}) as
\beq
{ \cal L}_K = - i \bar \psi \bar \s^n \partial_n \psi
- \partial_n \bar A \partial^n A .
\label {LK}
\eeq
\subsection{Susy Invariance of the Free Field Equations}
The susy transformations on the
scalar field $ A $, the spinor field $ \psi_a $,
and the auxiliary field $ F $ are
\begin{eqnarray}
\d A &=& \sqr2 \x \psi \label {d A} \\ 
\d \psi &=& i \sqr2 \s^m \bar \x \partial_m A + \sqr2 \x F
\label {d psi}\\
\d F &=& i \sqr2 \bar \x \bar \s^m \partial_m \psi .
\label {d F}
\eea
The conjugate equations are
\begin{eqnarray}
\d \bar A &=& \sqr2 \bar \x \bar \psi \\
\d \bar \psi &=& - i \sqr2 \x \s^m \partial_m \bar A 
+ \sqr2 \bar \x \bar F \\
\d \bar F &=& - i \sqr2 \partial_m \bar \psi \bar \s^m \x .
\label {d bar F}
\eea
The factors of $ \sqrt{2} $ in these equations 
are due to a convention that probably ought to be
changed.
\par
Under the susy transformation (\ref{d A}--\ref{d bar F}),
the change in the Lagrange density 
\beq
{ \cal L}_K = \frac{ i }{ 2 } \partial_n \bar \psi \bar \s^n \psi
- \frac{ i }{ 2 } \bar \psi \bar \s^n \partial_n \psi
- \partial_n \bar A \partial^n A + \bar F F 
\eeq
is
\begin{eqnarray}
\d { \cal L}_K &=& \frac{ i }{ 2 } \partial_n \d \bar \psi \bar \s^n \psi
+ \frac{ i }{ 2 } \partial_n \bar \psi \bar \s^n \d \psi 
- \frac{ i }{ 2 } \d \bar \psi \bar \s^n \partial_n \psi
- \frac{ i }{ 2 } \bar \psi \bar \s^n \partial_n \d \psi \nn\\
& & \mbox{} - \partial_n \d \bar A \partial^n A
- \partial_n \bar A \partial^n \d A 
+ \d \bar F F + \bar F \d F 
\eea
or
\begin{eqnarray}
\d { \cal L}_K &=&  
\frac{ i }{ 2 } \partial_n \left( - i \sqr2 \x \s^m \partial_m \bar A
+ \sqr2 \bar \x \bar F \right)
\bar \s^n \psi \nn\\
& & \mbox{} + \frac{ i }{ 2 } \partial_n \bar \psi \bar \s^n 
\left( i \sqr2 \s^m \bar \x \partial_m A + \sqr2 \x F \right) \nn\\
& & \mbox{} - \frac{ i }{ 2 } \left( - i \sqr2 \x \s^m \partial_m \bar A
+ \sqr2 \bar \x \bar F \right)
\bar \s^n \partial_n \psi \nn\\
& & \mbox{} - \frac{ i }{ 2 } \bar \psi \bar \s^n \partial_n 
\left( i \sqr2 \s^m \bar \x \partial_m A + \sqr2 \x F \right) \nn\\
& & \mbox{} - \partial_n \left( \sqr2 \bar \x \bar \psi \right)
\partial^n A
- \partial_n \bar A 
\partial^n \left( \sqr2 \x \psi \right) \nn\\
& & \mbox{} + \left( - i \sqr2 \partial_m \bar \psi \bar \s^m \x \right) F
+ \bar F \left( i \sqr2 \bar \x \bar \s^m \partial_m \psi \right) .
\eea 
After a partial cancellation of terms involving $ \x F $,
the part of $ \d { \cal L}_K $ that depends upon $ \x $ is
\begin{eqnarray}
\d { \cal L}_{{\cal K}\x}  &=&
\rhalf \x \s^m \bar \s^n \psi \partial_n \partial_m \bar A 
- \rhalf \x \s^m 
\bar \s^n \partial_n \psi \partial_m \bar A 
- \sqr2 \x \partial^n \psi \partial_n \bar A \nn\\
& & \mbox{} - \frac{ i }{ \sqrt{2} } \bar \psi \bar \s^n 
\x \partial_n F 
- \frac{ i }{ \sqrt{2} } \partial_n \bar \psi \bar \s^n \x F .
\label {funnysigmas1}
\eea
We now use the identity (\ref{Pauli})
and the commutativity of partial derivatives
to rewrite the first term of the preceding expression 
\begin{eqnarray}
\rhalf \x \s^m \bar \s^n \psi \partial_n \partial_m \bar A
&=& \rhalf \x \left( - \s^n \bar \s^m - 2 \eta^{mn} I \right)
\psi \partial_n \partial_m \bar A \nn\\
&=& \mbox{} 
- \rhalf \x \s^m \bar \s^n \psi \partial_n \partial_m \bar A
- \sqr2 \x \psi \partial_n \partial^n \bar A .
\eea
Thus the change in the action density $ \d { \cal L}_{{\cal K}\x} $ is
\begin{eqnarray}
\d { \cal L}_{{\cal K}\x}  &=&
- \rhalf \x \s^m \bar \s^n \psi \partial_n \partial_m \bar A
- \sqr2 \x \psi \partial_n \partial^n \bar A \nn\\
& & \mbox{} - \rhalf \x \s^m
\bar \s^n \partial_n \psi \partial_m \bar A
- \sqr2 \x \partial^n \psi \partial_n \bar A \nn\\
& & \mbox{} - \frac{ i }{ \sqrt{2} } \bar \psi \bar \s^n
\x \partial_n F
- \frac{ i }{ \sqrt{2} } \partial_n \bar \psi \bar \s^n \x F .
\eea
So the change $ \d { \cal L}_{{\cal K}\x} $
is the total divergence
\begin{equation}
\d { \cal L}_{{\cal K}\x}  = \partial_n K^n_{{\cal K}\x}
\eeq
of the current 
\begin{equation}
K^n_{{\cal K}\x} = - \rhalf \x \s^m \bar \s^n \psi\partial_m \bar A
- \sqr2 \x \psi \partial^n \bar A 
- \frac{ i }{ \sqrt{2} } \bar \psi \bar \s^n \x F.
\label {funnysigmas2}
\eeq
The change in the free action density $ \d { \cal L}_K $
is the divergence
\begin{equation}
\d { \cal L}_K  = \partial_n K^n_K
\eeq
of the current
\begin{equation}
K^n_K = K^n_{K\x} + K^n_{K\bar\x} 
\eeq
in which $ K^n_{K\bar\x} = \left( K^n_{K\x} \right)^\dgr $\null.
Thus the equations of motion of the free field theory
are invariant under the susy transformation
(\ref{d A}--\ref{d bar F}).
\par
The Noether current associated with the
susy transformation (\ref{d A}--\ref{d bar F}) 
of the free action density is
\begin{eqnarray}
J^n_K &=& \frac{ i }{ 2 } \d \bar \psi \bar \s^n \psi
- \frac{ i }{ 2 } \bar \psi \bar \s^n \d \psi
- \d \bar A \partial^n A - \partial^n \bar A \d A \nn\\
&=& \mbox{} \frac{ i }{ 2 } 
\left( - i \sqr2 \x \s^m \partial_m \bar A
+ \sqr2 \bar \x \bar F \right) \bar \s^n \psi
- \frac{ i }{ 2 } \bar \psi \bar \s^n 
\left( i \sqr2 \s^m \bar \x \partial_m A + \sqr2 \x F \right) \nn\\
& & \mbox{} - \sqr2 \bar \x \bar \psi \partial^n A
- \partial^n \bar A \sqr2 \x \psi .
\eea
The part depending on $ \x $ is
\begin{equation}
J^n_{K\x} = 
\rhalf \x \s^m \bar \s^n \psi \partial_m \bar A
- \frac{ i }{ \sqr2 } \bar \psi \bar \s^n \x F
- \sqr2 \x \psi \partial^n \bar A .
\eeq
\par
The Noether current $ J^n_K $  satisfies
\begin{equation}
\d {\cal L}_K = \partial_n J^n_K ,
\eeq
and so the difference $ J^n_K - K^n_K $ of the two currents 
\begin{equation}
S^n_K = J^n_K - K^n_K
\eeq
is conserved
\begin{equation}
\partial_n S^n_K = 0 .
\eeq
The current $ S^n_K $ is the supercurrent
of the free Lagrange density $ {\cal L}_K $.
The part $ S^n_{K\x} $ that depends upon $ \x $ is simply
\begin{equation}
S^n_{K\x} = \sqr2 \x \s^m \bar \s^n \psi \partial_m \bar A .
\eeq
Thus the quantity $ \x Q_K $ is
\begin{equation}
\x Q_K = \int \! d^3x \, S^0_{K\x} 
= \int \! d^3x \, \sqr2 \x \s^m \bar \s^0 \psi \partial_m \bar A .
\eeq
And so the supercharges $ Q_{Ka} $ of the free theory are  
\bea
Q_{ K a } & = & \sqrt{2} \int \! d^3x \, \s^m_{a\db} \bar \s^{ 0 \db c } 
\psi_c \partial_m \bar A \nn\\
& = & - \sqrt{2} \int \! d^3x \, \s^m_{a\db}
\psi_b \partial_m \bar A .
\label {Q_K}
\eea
\subsection{A Left-Handed Spinor and a Spinless Boson}
The susy action density for a left-handed spinor $ \bar \chi $
not interacting with a complex spin-zero field $ A $ is 
\beq
{ \cal L}_K = \frac{ i }{ 2 } \partial_n \chi \s^n \bar \chi
- \frac{ i }{ 2 } \chi \s^n \partial_n \bar \chi 
- \partial_n \bar A \partial^n A + \bar F F .
\label {LL}
\eeq
The field equations are
\bea
i \, \s^n \partial_n \bar \chi & = & 0 \\
\partial_n \partial^n A & = & 0 \\
F & = & 0
\label {LLeq}.
\eea
By ignoring total derivatives and using the constraint
(\ref{LLeq}), we may write the kinetic Lagrange density (\ref{LL}) as
\beq
{ \cal L}_K = - i \chi \s^n \partial_n \bar \chi
- \partial_n \bar A \partial^n A .
\label {LLK}
\eeq
\par
The susy transformations 
for the fields $ A $, $ \chi^a $, and $ F $ are
\bea
\d A & = & \sqrt{2} \chi \x \label {dLL}\\
\d \chi & = & - i \sqrt{2} \bx \bs^m \p_m A + \sqrt{2} \x F \\
\d F & = & -i \sqrt{2} \p_m \chi \s^m \bx 
\eea
and for their conjugates are
\bea
\d \bar A & = & \sqrt{2} \bx \bar \chi \\
\d \bar \chi & = & i \sqrt{2} \bs^m \x \p_m \bar A + \sqrt{2} \bx \bar F \\
\d \bar F & = & i \sqrt{2} \x \s^m \p_m \bar \chi .
\label {dLLbar}
\eea
These transformations may seem different from those
of equations (\ref {d A}--{d bar F}).  But if we use
the identities (\ref{bxbs}--\ref {trans}) to write them 
for the fields $ A $, $ \chi_a $, and $ F $,
then we find that they are in effect the same:
\bea
\d A &=& \sqr2 \x \chi \label {d LA} \\
\d \chi &=& i \sqr2 \s^m \bar \x \partial_m A + \sqr2 \x F
\label {d Lchi}\\
\d F &=& i \sqr2 \bx \bar \s^m \partial_m \chi .
\label {d LF} \\
\d \bar A &=& \sqr2 \bar \x \bar \chi \label {d bar LA}\\
\d \bar \chi &=& - i \sqr2 \x \s^m \partial_m \bar A
+ \sqr2 \bar \x \bar F \label {d bar Lchi}\\
\d \bar F &=& - i \sqr2 \partial_m \bar \chi \bar \s^m \x .
\label {d bar LF}
\eea

\par
The action density (\ref {LL}) changes only by a total divergence
under the susy transformations (\ref {dLL}--\ref {dLLbar}).
Indeed the change in the action density (\ref {LL}) is
\bea
\d { \cal L}_K & = & \ihalf \p_n \d \chi \s^n \bar \chi
+ \ihalf \p_n \chi \s^n \d \bar \chi 
- \ihalf \d \chi  \s^n \partial_n \bar \chi 
- \ihalf \chi  \s^n \partial_n \d \bar \chi \nn\\
& & \mbox{} \quad
- \p_n \d \bar A \partial^n A - \p_n \bar A \partial^n \d A
+ \d \bar F F + \bar F \d F .
\label {dLl}
\eea
Explicitly this change is
\bea
\d { \cal L}_K & = & \ihalf \p_n 
\left( - i \sqrt{2} \bx \bs^m \p_m A + \sqrt{2} \x F \right)
\s^n \bar \chi 
\nn\\
&& \mbox{}
+ \ihalf \p_n \chi \s^n 
\left( i \sqrt{2} \bs^m \x \p_m \bar A + \sqrt{2} \bx \bar F \right) 
\nn\\ 
&& \mbox{} 
- \ihalf \left( - i \sqrt{2} \bx \bs^m \p_m A + \sqrt{2} \x F \right)
\s^n \partial_n \bar \chi 
\nn\\
&& \mbox{}
- \ihalf \chi  \s^n \partial_n 
\left( i \sqrt{2} \bs^m \x \p_m \bar A + \sqrt{2} \bx \bar F \right)
\nn\\ 
&& \mbox{}
- \p_n \left( \sqrt{2} \bx \bar \chi \right) \p^n A 
- \p_n \bar A \partial^n \left(  \sqrt{2} \chi \x \right)
\nn\\
&& \mbox{}
+ \left(  i \sqrt{2} \x \s^m \p_m \bar \chi \right) F
+ \bar F \left( -i \sqrt{2} \p_m \chi \s^m \bx \right) .
\label {DLl}
\eea
The part that depends upon $ \x $ is
\bea
\d { \cal L}_{K\x} & = & 
\mbox{} \irhalf \p_n \x F \s^n \bar \chi
- \rhalf \p_n \chi \s^n \bs^m \x \p_m \bar A
\nn\\
&& \mbox{}
- \irhalf \x F \s^n \partial_n \bar \chi
+ \rhalf \chi  \s^n \partial_n \bs^m \x \p_m \bar A
\nn\\
&& \mbox{}
- \sqrt{2} \p_n \bar A \partial^n \chi \x
+ i \sqrt{2} \x \s^m \p_m \bar \chi F 
\label {DLlx}
\eea
or
\bea
\d { \cal L}_{K\x} & = &
\mbox{} - \rhalf \p_n \chi \s^n \bs^m \x \p_m \bar A
+ \rhalf \chi  \s^n \bs^m \x \p_n \p_m \bar A
\nn\\
&& \mbox{}
- \sqrt{2} \p_n \bar A \partial^n \chi \x
+ \irhalf \p_n \left( \x F \s^n \bar \chi \right) .
\label {DLlxs}
\eea
By using formula (\ref {Pauli}),
we may write this change as the divergence
\beq
\d { \cal L}_{K\x} = \p_n K^n_{K \x}
\eeq
of the current
\beq
K^n_{ K \x} = 
- \rhalf \, \chi \s^n \bs^m \x \, \p_m \bar A
- \sqrt{2} \, \bar A \, \partial^n \chi \x
+ \irhalf \, \x \s^n \bar \chi \, F ,
\label {KL}
\eeq
which shows that the space-time integral
of the action density (\ref {LL})
is invariant under the susy transformations (\ref {dLL}--\ref {dLLbar}).
\par
The Noether current of the action density (\ref {LL})
under the susy transformations (\ref {dLL}--\ref {dLLbar}) is
\bea
J^n_K & = & \mbox{} 
\ihalf \d \chi \s^n \bar \chi
- \ihalf \chi \s^n \d \bar \chi
- \d \bar A \p^n A
- \p^n \bar A \d A \nn\\
& = & \mbox{}
\ihalf \left( - i \sqrt{2} \bx \bs^m \p_m A + \sqrt{2} \x F \right)
\s^n \bar \chi
- \ihalf \chi \s^n 
\left( i \sqrt{2} \bs^m \x \p_m \bar A + \sqrt{2} \bx \bar F \right) \nn\\
&& \mbox{}
- \left( \sqrt{2} \bx \bar \chi \right) \p^n A 
- \p^n \bar A \left( \sqrt{2} \x \chi \right) .
\label {JKL}
\eea
The part of this current that depends upon $ \x $ is
\beq
J^n_{K \x} =
\irhalf F \x \s^n \bar \chi
+ \rhalf \chi \s^n \bs^m \x \p_m \bar A
- \sqrt{2} \p^n \bar A \x \chi . 
\label {JKLx}
\eeq
The susy current $ S^n_\x $ is then
\beq
S^n_\x = J^n_{K \x} - K^n_{ K \x}
= \sqrt{2} \, \chi \s^n \bs^m \x \, \p_m \bar A .
\label {SLx}
\eeq
The super-charges of the free theory are therefore
\beq
Q^a_K = \sqrt{2} \int \! \dcx \, 
\left( \chi \s^0 \bs^m \right)^a \, \p_m \bar A
= \sqrt{2} \int \! \dcx \,
\chi^c \s^0_{c \db} \bs^{m \db a} \, \p_m \bar A .
\label {QL}
\eeq
In a simpler notation, the spinor of susy charges 
of a left-handed spinor $ \bar \chi $
not interacting with a complex scalar field $ A $ is
\beq
Q_K = \sqrt{2} \int \! \dcx \,
\chi \left( \p_0 + \vec \s \cdot \nabla \right) \bar A .
\label {QLs}
\eeq

\section{General Chiral Multiplets}
\subsection{A Pair of Right-Handed Fields}
For a single right-handed spin-one-half field $ \psi $ and a single
complex scalar field $ A $, the most general chiral multiplet
is described by the Lagrange density 
\bea
{ \cal L} & = &
\frac{ i }{ 2 } \partial_n \bar \psi \bar \s^n \psi
- \frac{ i }{ 2 } \bar \psi \bar \s^n \partial_n \psi
- \partial_n \bar A \partial^n A + \bar F F \nn\\
& & \mbox{} + F W' + \bar F \bar W'
- \half W'' \psi \psi - \half \bar W'' \bar \psi \bar \psi 
\label {L}
\eea
in which $ W( A ) $ is an analytic function
of the complex scalar field $ A $ and the primes
indicate differentiation with respect to the field
$ A $ or $ \bar A $
\beq
W' = \frac { \partial W(A) } { \partial A }, \quad
\bar W' = \frac { \partial \bar W ( \bar A ) } { \partial \bar A },
\quad {\hbox {\it etc.}} 
\eeq
The constraints on the auxiliary field $ F $ are
\beq
F = - \bar W' \quad {\hbox \mathrm{and}} \quad
\bar F = - W' .
\eeq
By implementing these constraints
and ignoring total derivatives, we may write
the Lagrange density (\ref{L}) as
\beq
{ \cal L} = - i \bar \psi \bar \s^n \partial_n \psi
- \partial_n \bar A \partial^n A 
- | W' |^2 - \half W'' \psi \psi - \half \bar W'' \bar \psi \bar \psi ,
\label {Lc}
\eeq
in which the term $ | W' |^2 $ is the self-interaction 
of the field $ A $ and the last two terms are Yukawa interactions.
The field equations are
\bea
-i \bar \s^n \partial_n \psi & = & \bar W'' \bar \psi \\
\partial_n \partial^n A & = & \bar W'' W' 
+ \half \bar W''' \bar \psi \bar \psi. 
\eea
\par
If the theory is to be renormalizable,
then the function $ W(A) $ should be
a polynomial in the field $ A $ that is at most cubic
\beq
W( A ) = c A + \half m A^2 + \third g A^3.
\eeq
in which we have omitted a possible constant term
because only derivatives of $ W $ appear in the
lagrangian. 
By a field redefinition $ A' = A + d $, 
where $ d $ is a constant,
one may also remove the linear term
$ c A $, as is usually done.
For this case,
\beq
W( A ) = \half m A^2 + \third g A^3,
\eeq
the Lagrange density (\ref{Lc}) is
\beq
{ \cal L} = - i \bar \psi \bar \s^n \partial_n \psi
- \partial_n \bar A \partial^n A
- | mA + g A^2 |^2
- \half \left( m + 2g A \right) \psi \psi
- \half \left( \bar m + 2 \bar g \bar A  \right) \bar \psi \bar \psi ,
\eeq
and the equations of motion are
\bea
-i \bar \s^n \partial_n \psi & = & 
\left( \bar m + 2 \bar g \bar A \right) \bar \psi \\
\partial_n \partial^n A & = & 
\left( \bar m + 2 \bar g \bar A \right)
\left( mA + g A^2 \right) + \bar g \bar \psi \bar \psi.
\eea
\subsection{Many Pairs of Right-Handed Fields}
The most general right-handed
chiral multiplet consists of $ n $
copies of the spinor, complex scalar, and
complex auxiliary fields $ \psi_i $, $ A_i $, and $ F_i $
of the simplest chiral multiplet.
The most general chiral Lagrange density is
\bea
{ \cal L} & = &
\frac{ i }{ 2 } \partial_n \bar \psi_i \bar \s^n \psi_i
- \frac{ i }{ 2 } \bar \psi_i \bar \s^n \partial_n \psi_i
- \partial_n \bar A_i \partial^n A_i + \bar F_i F_i \nn\\
& & \mbox{} + F_i W_i + \bar F_i \bar W_i
- \half W_{ij} \psi_i \psi_j - \half \bar W_{ij} \bar \psi_i \bar \psi_j
\label {LCG}
\eea
in which sums are understood over repeated indices
and in which $ W(A) $ is an analytic function
of the $ n $ complex scalar fields $ A_i $ with
\beq
W_i = \frac{\partial W } { \partial A_i }
\and
W_{ij} = \frac{\partial^2 W } { \partial A_i \partial A_j }.
\eeq
\par
As we shall see, the action density (\ref{LCG})
changes only by a total derivative when its
scalar fields $ A_i $, spinor fields $ \psi_i $,
and auxiliary fields $ F_i $ are subjected
to the supersymmetry transformations 
\begin{eqnarray}
\d A_i &=& \sqr2 \x \psi_i \label {d A_i} \\
\d \psi_i &=& i \sqr2 \s^m \bar \x \partial_m A_i + \sqr2 \x F_i
\label {d psi_i}\\
\d F_i &=& i \sqr2 \bar \x \bar \s^m \partial_m \psi_i 
\label {d F_i}
\eea
and to the conjugate susy transformations
\begin{eqnarray}
\d \bar A_i &=& \sqr2 \bar \x \bar \psi_i \\
\d \bar \psi_i &=& - i \sqr2 \x \s^m \partial_m \bar A_i
+ \sqr2 \bar \x \bar F_i \\
\d \bar F_i &=& - i \sqr2 \partial_m \bar \psi_i \bar \s^m \x .
\label {d bar F_i}
\eea
\par
We may express the auxiliary fields $ F_i $
in terms of the scalar fields $ A_i $ by 
resolving the constraints
\beq
0 = \frac { \partial {\cal L } } { \partial \bar F_i }
= F_i + \bar W_i
\and
0 = \frac { \partial {\cal L } } { \partial F_i }
= \bar F_i + W_i . 
\eeq
Eliminating the auxiliary fields and dropping
total derivatives, we may write the general chiral
action density as
\beq
{ \cal L} = - \bar \psi_i \bar \s^n \partial_n \psi_i
- \partial_n \bar A_i \partial^n A_i 
- | W_i |^2 
- \half W_{ij} \psi_i \psi_j 
- \half \bar W_{ij} \bar \psi_i \bar \psi_j.
\label {LCGSF}
\eeq
\par
If the theory is to be renormalizable, then the function
$ W(A) $ should be a polynomial in the $ n $ scalar fields
$ A_i $ that is no higher than cubic
\beq
W( A ) = c_i A_i + \half m_{ij} A_i A_j + \third g_{ijk} A_i A_j A_k.
\eeq
Since only derivatives of $ W $ appear in the lagrangian,
we have omitted a constant term.
The tensors $ m_{ij} $ and $ g_{ijk} $ must be totally
symmetric in their indices; any anti-symmetric piece 
would not contribute to $ W $.
By redefining the scalar fields, one may
remove the linear terms
\beq
W( A ) = \half m_{ij} A_i A_j + \third g_{ijk} A_i A_j A_k.
\label {W}
\eeq
\subsection{Invariance of the Right-Handed Chiral Action}
It will be useful to separate the Lagrange density (\ref{LCG})
into the kinetic part $ \cal L_K $ and the interaction part
$ \cal L_I $
\beq
\cal L = \cal L_K + \cal L_I 
\label {LCGa}
\eeq
where
\beq
{\cal L_K} = \frac{ i }{ 2 } \partial_n \bar \psi_i \bar \s^n \psi_i
- \frac{ i }{ 2 } \bar \psi_i \bar \s^n \partial_n \psi_i
- \partial_n \bar A_i \partial^n A_i + \bar F_i F_i 
\label {LCGK}
\eeq
and
\beq
{\cal L_I} = F_i W_i + \bar F_i \bar W_i
- \half W_{ij} \psi_i \psi_j - \half \bar W_{ij} \bar \psi_i \bar \psi_j.
\label {LCGI}
\eeq 
The kinetic Lagrange density (\ref{LCGK}) is the sum of $n$
copies of the Lagrange density (\ref{L_K}).  
And the susy transformation (\ref{d A_i}--\ref{d bar F_i}) 
is $n$ copies of the susy transformation (\ref{d A}--\ref{d bar F}).
Thus just as (\ref{L_K}) is invariant under (\ref{d A}--\ref{d bar F}),
so too the kinetic Lagrange density (\ref{LCGK}) changes only
by a total derivative under the susy transformation 
(\ref{d A_i}--\ref{d bar F_i}).
\par
We shall now examine the change in the rest of the lagrangian.
Under the susy transformation (\ref{d A_i}--\ref{d bar F_i}),
the part of the change in the interaction Lagrange 
density $ { \cal L_I} $ 
that involves $ \x $ is
\begin{eqnarray*}
\d { \cal L_{{\cal I}\x}} &=& F_i W_{ij} \d A_j + \d \bar F_i \bar W_i 
- \half W_{ijk} \d A_i \psi_j \psi_k - W_{ij} \psi_i \d \psi_j
- \bar W_{ij} \d \bar \psi_i \bar \psi_j \nn\\
&=& F_i W_{ij} \sqr2 \x \psi_j 
- i \sqr2 \partial_m \bar \psi_i \bar \s^m \x \bar W_i
- \rhalf W_{ijk} \x \psi_i \psi_j \psi_k \nn\\
& & \mbox{} - W_{ij} \psi_j \sqr2 \x F_i
+ i \bar W_{ij} \sqr2 \x \s^m \left( \partial_m \bar A_i \right) \bar \psi_j .
\eeas
in which we drop terms proportional to $ \bx $.
The first term cancels the fourth term.
\par
The term $ W_{ijk} \x \psi_i \psi_j \psi_k $
vanishes because $ W_{ijk} $ is symmetric.
Let us write this term more explicitly as 
\( {\cal W}_{ijk} \x^a \psi_{a i} \psi^b_j \psi_{b k} \).  
Since the spinor indices \( a \) and \( b \)
can assume only the two values 1 and 2,
it follows that in every term at least two of the
spinor indices must be the same.
The symmetry in \( i, j, k \) of \( {\cal W}_{ijk} \)
then implies that such terms, {\emph e.g.,}
\( {\cal W}_{ijk} \psi_{ 1 i } \psi_{ 1 j } \),
must vanish.
In particular
\begin{equation}
{\cal W}_{ijk} \psi_{ 1 i } \psi_{ 1 j } =
\frac{ 1 } { 2 } \left( {\cal W}_{ijk} + {\cal W}_{jik} \right)
\psi_{ 1 i } \psi_{ 1 j } =
\frac{ 1 } { 2 } \left( {\cal W}_{ijk} \psi_{ 1 i } \psi_{ 1 j }
+ {\cal W}_{jik} \psi_{ 1 i } \psi_{ 1 j } \right)
\label {W'''1}
\eeq
so that
\begin{equation}
{\cal W}_{ijk} \psi_{ 1 i } \psi_{ 1 j } =
\frac{ 1 } { 2 } \left( {\cal W}_{ijk} \psi_{ 1 i } \psi_{ 1 j }
+ {\cal W}_{ijk} \psi_{ 1 j } \psi_{ 1 i } \right) =
\frac{ 1 } { 2 } {\cal W}_{ijk} \left(
\psi_{ 1 i } \psi_{ 1 j } + \psi_{ 1 j } \psi_{ 1 i } \right) = 0 .
\label {W'''2}
\eeq 
\par
Thus using (\ref{trans}), we see that
the part of the change $ \d { \cal L}_{{\cal I}\x} $
in the interaction Lagrange 
density that depends upon $ \x $ is
\begin{equation}
\d { \cal L}_{{\cal I}\x} = 
-i \sqrt{2} \bar \psi_i \bar \s^m \x \bar W_{ij} \partial_m \bar A_j
- i \sqrt{2} \partial_m \bar \psi_i \bar \s^m \x \bar W_i
\eeq
which is the total divergence
\begin{equation}
\d { \cal L}_{{\cal I}\x} = \partial_m K^m_{{\cal I}\x} 
\eeq
of the current
\begin{equation}
K^m_{{\cal I}\x} = -i \sqrt{2} \bar \psi_i \bar \s^m \x \bar W_i .
\eeq
The change $ \d { \cal L_I} $ is the divergence
\begin{equation}
\d { \cal L_I} = \partial_n K^n_{\cal I} = 
\partial_n \left( K^n_{{\cal I}\x} 
+ \left( K^n_{{\cal I}\x} \right)^\dgr \right) .
\eeq
Since the change in the full action density (\ref {LCGa})
is a total divergence
\begin{equation}
\d { \cal L } = \partial_n \left( K^n_K + K^n_{\cal I} \right) ,
\eeq
the action and the equations
of motion are invariant under the
susy transformation (\ref{d A}--\ref{d bar F}).
\par
The species index $ i $ essentially serves to label the copies;
we shall suppress it in what follows, except when it is worth noticing.
\subsection{Supercharges of the Right-Handed Chiral Theory}
\par
Because there are no derivatives of the fields
in the interaction Lagrange
density $ { \cal L_I} $,
the Noether current of the interacting theory
is the same as that of the free theory.
Thus the part of the supercurrent $ S^n $
that depends upon $ \x $ is
\begin{eqnarray}
S^n_\x &=& J^n_{K\x} - K^n_{K\x} - K^n_{\cal I} \nn\\
&=& S^n_{K\x} - K^n_{\cal I} \nn\\
&=& \sqr2 \x \s^m \bar \s^n \psi \partial_m \bar A
+ i \sqr2 \bar \psi \bar \s^n \x \bar W' .
\eea
The divergence of this susy current 
is the difference between two equal expressions
for the change in the action density ---
the Noether form
and the more straightforward form
obtained by substituting the changes 
in the fields.
The susy current $ S^n_\x $ therefore is conserved:
\beq
\p_n S^n_\x = 0 .
\label {divSR}
\eeq
\par
The supercharge $ \x Q $ of this theory is 
the spatial integral of the time component
of this conserved current
\begin{equation}
\x Q = \int \! d^3x \, S^0_\x
= \sqr2 \int \! d^3x \, \left(
\x \s^m \bar \s^0 \psi \partial_m \bar A
+ i \bar \psi \bar \s^0 \x \bar W' \right) .
\eeq
By the identity (\ref{trans}) we have
$ \bar \psi \bar \s^0 \x = - \x \s^0 \bar \psi $,
and so the supercharges $ Q_a $ are
\beq
Q_a  = \sqrt{2} \int \! d^3x \, \left(
\s^m_{a\db} \bar \s^{ 0 \db c }
\psi_c \partial_m \bar A
- i \s^0_{a\db} \bar \psi^\db \bar W' \right) . \label {QR}
\eeq
In a mixed notation the supercharges are
\bea
Q_a & = & - \sqrt{2} \int \! d^3x \, \left(
\s^m_{a\db} \psi_b \partial_m \bar A
- i \bar \psi^\da \bar W' \right)
\label {Qup} \\
& = & - \sqrt{2} \int \! d^3x \, \left(
\s^m_{a\db} \psi_b \partial_m \bar A
- i \e^{\da \db} \bar \psi_\db \bar W' \right) 
\label {Qdown}
\eea
or more simply
\beq
Q = \sqrt{2} \int \! d^3x \, \left[ \, \left(
\p_0 \bar A - \vec \s \cdot \nabla \bar A \right) \psi
- \bar W' \, \s^2 \bar \psi \, \right] .
\label {QRss}
\eeq
The adjoint supercharges $ \bar Q_\da $ are
\beq
\bar Q_\da = \left( Q_a \right)^\dgr 
= \sqrt{2} \int \! d^3x \, \left(
\bar \psi_\dc \bs^{ 0 \dc b } \s^m_{ b \da } \, \p_m A
+ i \psi^b \s^0_{ b \da } \, W' \, \right) .
\label {QRbar}
\eeq
In a mixed notation they are
\beq
\bar Q_\da = - \sqrt{2} \int \! d^3x \, \left(
\bar \psi_\db \s^m_{b \da} \partial_m A
+ i \e^{\da \db} \psi_b W' \right) 
\label {QRbarm}
\eeq
or more simply
\beq
\bar Q_\da = \sqrt{2} \int \! d^3x \, \left[ \, 
\bar \psi \left(
\p_0 - \vec \s \cdot \nabla \right) A + W' \, \s^2 \psi
\, \right] .
\label {QRbarss}
\eeq

\subsection{A Pair of Left-Handed Fields}
The action density for a left-handed spin-one-half
field $ \bar \chi $ interacting with
a complex scalar field $ A $ is
\bea
{\cal L}_L & = & \mbox{}
\frac{ i }{ 2 } \partial_n \chi \s^n \bar \chi
- \frac{ i }{ 2 } \chi \s^n \partial_n \bar \chi
- \partial_n \bar A \partial^n A + \bar F F 
\label {LL0} \\
& & \mbox{}
+ F W' + \bar F \bar W' - \half W'' \chi \chi 
- \half \bar W'' \bar \chi \bar \chi ,
\label {LLI}
\eea
in which $ W( A ) $ is a cubic polynomial
in the field $ A $.  
The equations of motion are
\bea
\mbox{} - i \s^n \p_n \bar \chi & = & \mbox{} W'' \chi \\
\mbox{} - \p_n \p^n A & = & \mbox{} F W'' -  \half W''' \chi \chi \\
F & = & \mbox{} - \bar W' .
\label {eoml}
\eea
If we implement the constraint (\ref {eoml})
and ignore total derivatives,
then we may write the action density as
\beq
{\cal L}_L = \mbox{} - i \chi \s^n \partial_n \bar \chi
- \partial_n \bar A \partial^n A
- | W' |^2 - \half W'' \chi \chi
- \half \bar W'' \bar \chi \bar \chi 
\label {LLIc}
\eeq
and the equations of motion as
\bea
\mbox{} - i \s^n \p_n \bar \chi & = & \mbox{} W'' \chi \\
\mbox{} \p_n \p^n A & = & \mbox{} \bar W' W'' +  \half W''' \chi \chi .
\label {eomlc}
\eea

\subsection{Invariance of the Left-Handed Chiral Action}
Under the susy transformations (\ref {dLL}--\ref {dLLbar}),
the change of the interaction part (\ref {LLI}) of the 
action density is
\beq
\d {\cal L}_{LI} = \d F W' + F W'' \d A + \d \bar F \bar W'
+ \bar F \bar W'' \d \bar A - W'' \d \chi \chi 
- \bar W'' \bar \chi \d \bar \chi
\eeq
in which we have ignored the terms proportional to the
product of three $\chi$'s, which vanish as shown earlier
(\ref {W'''1}--\ref {W'''2})\null.
This change is
\bea
\d {\cal L}_{LI} & = & \mbox{}
\left( -i \sqrt{2} \p_m \chi \s^m \bx \right) W'
+ F W'' \left( \sqrt{2} \chi \x \right) 
+ \left( i \sqrt{2} \x \s^m \p_m \bar \chi \right) \bar W'
\nn\\
& & \mbox{} 
+ \bar F \bar W'' \left( \sqrt{2} \bx \bar \chi \right) 
- W'' \left( - i \sqrt{2} \bx \bs^m \p_m A + \sqrt{2} \x F \right) \chi
\nn\\
& & \mbox{} 
- \bar W'' \bar \chi
\left( i \sqrt{2} \bs^m \x \p_m \bar A + \sqrt{2} \bx \bar F \right) .
\label {dLIl}
\eea
Of this change, the part that depends upon $ \x $ is
\bea
\d {\cal L}_{LI\x}  & = & \mbox{}
F W'' \left( \sqrt{2} \chi \x \right)
+ \left( i \sqrt{2} \x \s^m \p_m \bar \chi \right) \bar W'
- W'' \sqrt{2} \x F \chi \nn\\
& & \mbox{}
- \bar W'' \bar \chi
\left( i \sqrt{2} \bs^m \x \p_m \bar A \right) .
\label {dLIlx}
\eea
By using the identity (\ref {trans}),
we may write this change as the divergence
\beq
\d {\cal L}_{LI\x} = \p_n K^n_{LI\x}
\label {dLIx}
\eeq
of the current
\beq
K^n_{LI\x} = i \sqrt{2} \x \s^n \bar \chi \, \bar W' .
\label {KLIx}
\eeq
This result and the earlier one (\ref {KL})
establish the invariance of the action
of this theory under susy transformations. 
\subsection{Supercharges of the Left-Handed Chiral Theory}
\par
The interaction (\ref {LLI}) contains
no derivatives of $ \chi $ or of $ A $
and so generates no extra Noether current. 
The difference between the Noether current (\ref {JKLx})
and sum of the currents $ K^n_{ K \x} $ (\ref {KL})
and $ K^n_{LI\x} $ (\ref {KLIx}) is therefore
the susy current of the interacting theory
(\ref {LL0}--\ref {LLI}):
\beq
S^n_\x = J^n_{K \x} - K^n_{ K \x} - K^n_{LI\x}
= \sqrt{2} \, \chi \s^n \bs^m \x \, \p_m \bar A 
- i \sqrt{2} \x \s^n \bar \chi \, \bar W' ,
\label {SLFx}
\eeq
and it is conserved
\beq
\p_n S^n_\x = 0 .
\label {divS}
\eeq
\par
The supercharges $ Q_a $
of the left-handed chiral theory are given by
\beq
\x Q = \int \! \dcx \, S^0_\x 
= \sqrt{2} \int \! \dcx \left(
\chi \s^0 \bs^m \x \, \p_m \bar A - i \x \s^0 \bar \chi \, \bar W'
\right) .
\label {xQL}
\eeq
The identity (\ref {sstrans})
implies that $ \chi \s^0 \bs^m \x = \x \s^m \bs^0 \chi $,
and so the supercharges $ Q_a $ are
\beq
Q_a = \sqrt{2} \int \! \dcx \left(
\s^m_{a \db} \bs^{ 0 \db c } \chi_c \, \, \p_m \bar A 
- i \s^0_{ a \db} \bar \chi^\db \, \bar W' \right) .
\label {QLfull}
\eeq
Since $ \s^0 = \bs^0 = - I $, we may write
the $ Q_a $ in a mixed notation as 
\bea
Q_a & = & \mbox{}
- \sqrt{2} \int \! \dcx \left(
\s^m_{ab} \chi_b \, \, \p_m \bar A
- i \bar \chi^\da \, \bar W' \right)
\label {QLs1} \\
& = & \mbox{}
- \sqrt{2} \int \! \dcx \left(
\s^m_{ab} \chi_b \, \, \p_m \bar A
- i \e^{ \da \db } \bar \chi_\db \, \bar W' \right)
\label {QLs2}
\eea
or in a simpler notation as
\beq
Q = \sqrt{2} \int \! \dcx \left[ \, \left(
\p_0 \bar A - \vec \s \cdot \nabla \bar A \right) \chi
- \bar W' \, \s^2 \bar \chi \, \right] .
\label {QLss}
\eeq
The adjoint supercharges are
\beq
\bar Q_\da = \sqrt{2} \int \! \dcx \left(
\bar \chi_\dc \bs^{ 0 \dc b } \s^m_{ b \da } \, \, \p_m A
+ i \chi^b \s^0_{ b \da } \, W' \right)
\label {QLbar}
\eeq
or more simply
\beq
\bar Q = \sqrt{2} \int \! \dcx \left[ \, 
\bar \chi \left( \p_0 A - \vec \s \cdot \nabla A \right)
+ W' \, \s^2 \chi \, \right] .
\label {QLbars}
\eeq

\section{The Supercharges}
\par
Supercharges satisfy the 
remarkable \{anti-\}commutation relations
\begin{equation}
\left\{ Q_a , \bar Q_\db \right\}
= 2 \s^m_{a\db} P_m 
\label {algebra1}
\eeq
\begin{equation}
\left\{ Q_a , Q_b \right\} = \left\{ \bar Q_\da , \bar Q_\db \right\}
= 0 
\label {Q Q 0}
\eeq
\begin{equation}
\left[ Q_a , P_n \right] = \left[ \bar Q_\da , P_n \right] =
\left[ P_m , P_n \right] = 0 .
\label {algebra2} 
\eeq
In particular
\begin{eqnarray}
\left\{ Q_a , \bar Q_\da \right\} &=& 
Q_1 \left( Q_1 \right)^\dgr + \left( Q_1 \right)^\dgr Q_1
+ Q_2 \left( Q_2 \right)^\dgr + \left( Q_2 \right)^\dgr Q_2 \nn\\
&=& 2 \Tr \left( \s^m \right) P_m
= -4 P_0 = 4 P^0 = 4 H .
\eea
The factors of 2 and 4 that litter these equations
are due to the extra factors of $ \sqr2 $ in the
conventional definition of the susy transformation 
(\ref{d A}--\ref{d bar F}) of the chiral multiplet.
\par
By using the canonical commutation relations,
one may show that the right-handed supercharges 
\bea
Q_a  & = & \sqrt{2} \int \! d^3x \, \left(
\s^m_{a\db} \bar \s^{ 0 \db c }
\psi_c \partial_m \bar A
- i \s^0_{a\db} \bar \psi^\db \bar W' \right) \\
\bar Q_\da & = & \sqrt{2} \int \! d^3x \, \left(
\bar \psi_\dc \bs^{ 0 \dc b } \s^m_{ b \da } \, \p_m A
+ i \psi^b \s^0_{ b \da } \, W' \, \right) ,
\eea
whose various forms were listed
in eqs.(\ref {QR}--\ref {QRbarss}), actually do satisfy 
the anti-commutation relations (\ref {algebra1}, \ref {algebra2}).
The the canonical equal-time commutation 
and anti-commutation relations obeyed by the fields
of the right-handed chiral multiplet are:
\begin{eqnarray}
\left[ A ( t, \vec x ) , \partial_0 \bar A ( t, \vec y ) \right]
&=& i \d ( \vec x - \vec y ) \label {A A dot} \\
\left[ \bar A ( t , \vec x ) , \partial_0 A ( t , \vec y ) \right]
&=& \mbox{} i \d ( \vec x - \vec y ) \\
\left\{ \psi_a ( t, \vec x ) , \bar \psi_\db ( t, \vec y ) \right\}
&=& \d_{ a \db } \d ( \vec x - \vec y ) ,
\label {psi psi bar}
\eea
\begin{equation} \left[ A ( t, \vec x ) , A ( t, \vec y ) \right] = 
\left[ A ( t, \vec x ) , \bar A ( t, \vec y ) \right] = 
\left[ \bar A ( t, \vec x ) , \bar A ( t, \vec y ) \right] = 0 ,
\label {A A}
\eeq
\begin{equation}
\left[ \partial_0 A ( t, \vec x ) , \partial_0 A ( t, \vec y ) \right] 
= 
\left[ \partial_0 A ( t, \vec x ) , \partial_0 \bar A ( t, \vec y ) \right] 
= \left[ \partial_0 \bar A ( t, \vec x ) , 
\partial_0 \bar A ( t, \vec y ) \right] = 0 ,
\eeq
and 
\begin{equation}
\left\{ \psi_a ( t, \vec x ) , \psi_b ( t, \vec y ) \right\}
= \left\{ \bar \psi_\da ( t, \vec x ) , \bar \psi_\db ( t, \vec y ) \right\}
= 0 .
\label {psi psi}
\eeq
\par
The anti-commutator (\ref{algebra1}) is
\beq
\left\{ Q_a , \bar Q_\db \right\}
= 2 \int \! d^3x d^3y \, 
\left\{ \left(
\s^m_{a\dc} \psi_c \partial_m \bar A
- i \e^{a c} \bar \psi_\dc \bar W' \right) ,
\left(
\bar \psi_{\dd} \s^n_{d \db} \partial_n A
+ i \e^{\db \dd} \psi_d W' \right)
\right\}
\label {ac}
\eeq
the sum of four terms.
The simplest term arises from the 
anti-commutator of the $ \bar W' $ term
with the $ W' $ term
\beq
\bar WW_{a\db} \equiv 2 \int \! d^3x d^3y \,
\left\{
- i \e^{a c} \bar \psi_\dc(x) \bar W'(x) ,
i \e^{\db \dd} \psi_d(y) W'(y)
\right\} .
\eeq
By using the anti-commutation relation (\ref{psi psi bar}),
we may write this term as
\bea
\bar WW_{a\db} & = & 2 \int \! d^3x d^3y \,
\e^{a c} \e^{\db \dd} \bar W'(x) W'(y) 
\d_{cd} \d( \vec x - \vec y ) \nn\\
& = & 2 \int \! d^3x \,
\e^{a c} \e^{\db \dc} |W'(x)|^2 \nn\\
& = & 2 \d_{a \db} \int \! d^3x \,
|W'(x)|^2 .
\label {W bar W}
\eea
The next simplest term arises from the anti-commutator
of the first $ \bar W' $ term with the $ \partial A $ term
\bea
\bar W A_{a\db} & \equiv & 2 \int \! d^3x d^3y \,
\left\{
- i \e^{\da \dc} \bar \psi_\dc(x) \bar W' ,
\bar \psi_{\dd}(y) \s^n_{d \db} \partial_n A(y)
\right\} \nn\\
& = & - 2 i \int \! d^3x d^3y \,
\left(
\e^{\da \dc} \bar \psi_\dc(x) \bar W' 
\bar \psi_{\dd}(y) \s^n_{d \db} \partial_n A(y) \right. \nn\\
& & \mbox{} \left. 
+ \bar \psi_{\dd}(y) \s^n_{d \db} \partial_n A(y)
\e^{\da \dc} \bar \psi_\dc(x) \bar W'
\right) .
\eea
Due to the vanishing of the anti-commutator (\ref{psi psi}),
we may write these terms as
\beq
\bar W A_{a\db} = - 2 i \int \! d^3x d^3y \,
\e^{\da \dc} \bar \psi_\dc(x) 
\bar \psi_{\dd}(y) \s^n_{d \db} 
\left[ \bar W' , \partial_n A(y) \right] .
\eeq
By referring to the commutators (\ref{A A dot}) and (\ref{A A}),
we see that only the $ n=0 $ part survives
\bea
\bar W A_{a\db} & = & - 2 i \int \! d^3x d^3y \,
\e^{\da \dc} \bar \psi_\dc(x)
\bar \psi_{\dd}(y) \s^0_{d \db}
\left[ \bar W' , \partial_0 A(y) \right] \nn\\
& = & 2 \int \! d^3x d^3y \,
\e^{\da \dc} \bar \psi_\dc(x)
\bar \psi_{\dd}(y) \s^0_{d \db}
\bar W'' \d( \vec x - \vec y ) \nn\\
& = & 2 \int \! d^3x \,
\e^{\da \dc} \bar \psi_\dc(x)
\bar \psi_{\dd}(x) \s^0_{d \db} \bar W'' \nn\\
& = & 2 \int \! d^3x \,
\bar \psi_\db(x) \bar \psi^\da(x) \bar W'' .
\label {W bar A}
\eea
By completing an analogous argument,
the reader may show that the adjoint term $ \bar A W $ is
\bea
\bar A W_{a\db} & \equiv & 2 i \int \! d^3x d^3y \,
\left\{
\s^m_{a\dc} \psi_c \partial_m \bar A ,
i \e^{b d} \psi_d W' 
\right\} \nn\\
& = & 2 \int \! d^3x \,
\psi^a(x) \psi_b(x) W'' .
\label {A bar W}
\eea
The last term is the full anti-commutator
of the free theory
\bea
\bar AA_{a\db} & \equiv & 2 \int \! d^3x d^3y \,
\left\{ 
\s^m_{a\dc} \psi_c(x) \partial_m \bar A(x) ,
\bar \psi_{\dd}(y) \s^n_{d \db} \partial_n A(y)
\right\} \nn\\
& = & 2 \int \! d^3x d^3y \,
\left(
\bar \psi_{\dd}(y) \s^n_{d \db} \partial_n A(y)
\s^m_{a\dc} \psi_c(x) \partial_m \bar A(x) \right. \nn\\
& & \qquad \qquad \mbox{} 
+ \left. \s^m_{a\dc} \psi_c(x) \partial_m \bar A(x)
\bar \psi_{\dd}(y) \s^n_{d \db} \partial_n A(y) \right) \nn\\
& = & 2 \int \! d^3x d^3y \,
\left( \bar \psi_{\dd}(y) \s^n_{d \db}
\s^m_{a\dc} \psi_c(x)
\left[
\partial_n A(y) , \partial_m \bar A(x)
\right] \right. \nn\\
& & \qquad \qquad \mbox{}
+ \left. \s^m_{a\dc} \s^n_{d \db} 
\d_{\dc d} \d( \vec x - \vec y )
\partial_m \bar A(x) \partial_n A(y) \right) . 
\label {jump}
\eea
By using the commutation relations
(\ref{A A dot}) and (\ref{A A}),
we see that the first term is non-zero
if both $ n = i \ne 0 $ and $ m = 0 $
or if both $ n = 0 $ and $ m = j \ne 0 $.
In the former case, we integrate by parts
over $ y $; in the latter case,
we integrate by parts over $ x $
\bea
\bar AA_{a\db} & = & 2 \int \! d^3x d^3y \,
\left( - \partial_i \bar \psi_{\dd}(y) \s^i_{d \db}
\s^0_{a\dc} \psi_c(x) 
\left[ A(y) , \partial_0 \bar A(x) \right] \right. \nn\\
& & \qquad \qquad \mbox{}
- \left. \bar \psi_{\dd}(y) \s^0_{d \db}
\s^j_{a\dc} \partial_j \psi_c(x) 
\left[ \partial_0 A(y) , \bar A(x) \right] \right. \nn\\
& & \qquad \qquad \mbox{}
+ \left. \s^m_{a\dc} \s^n_{c \db}
\d( \vec x - \vec y )
\partial_m \bar A(x) \partial_n A(y) \right) 
\label {land} \\
& = & 2 \int \! d^3x d^3y \,
\left( - \partial_i \bar \psi_{\dd}(y) \s^i_{d \db}
\s^0_{a\dc} \psi_c(x)
i \d( \vec x - \vec y ) \right. \nn\\
& & \qquad \qquad \mbox{}
- \left. \bar \psi_{\dd}(y) \s^0_{d \db}
\s^j_{a\dc} \partial_j \psi_c(x)
(-i) \d( \vec x - \vec y ) \right. \nn\\
& & \qquad \qquad \mbox{}
+ \left. \s^m_{a\dc} \s^n_{c \db}
\d( \vec x - \vec y )
\partial_m \bar A(x) \partial_n A(y) \right) \nn\\
& = & 2 \int \! d^3x \,
\left( -i \partial_i \bar \psi_{\dd}(x) \s^i_{d \db}
\s^0_{a\dc} \psi_c(x) 
+ i \bar \psi_{\dd}(x) \s^0_{d \db}
\s^j_{a\dc} \partial_j \psi_c(x) \right. \nn\\
& & \qquad \qquad \mbox{}
+ \left. \s^m_{a\dc} \s^n_{c \db}
\partial_m \bar A(x) \partial_n A(x) \right) \nn\\
& = & 2 \int \! d^3x \,
\left( i \partial_i \bar \psi_{\dd}(x) \s^i_{d \db}
\psi_a(x) 
- i \bar \psi_{\db}(x) 
\s^i_{a\dc} \partial_i \psi_c(x) \right. \nn\\
& & \qquad \qquad \mbox{}
+ \left. \s^m_{a\dc} \s^n_{c \db}
\partial_m \bar A(x) \partial_n A(x) \right) .
\label {A bar A}
\eea
\par
Collecting our four terms
(\ref{W bar W}), (\ref{W bar A}), (\ref{A bar W}),
and (\ref{A bar A}), we find for 
the anti-commutator (\ref{ac})
of $ Q_a $ with $ \bar Q_\db $ the expression
\bea
\left\{ Q_a , \bar Q_\db \right\} & = &
2 \int \! d^3x \,
\left( i \partial_i \bar \psi_{\dd}(x) \s^i_{d \db}
\psi_a(x)
- i \bar \psi_{\db}(x)
\s^i_{a\dc} \partial_i \psi_c(x) \right. \nn\\
& & \qquad \qquad \mbox{}
+ \s^m_{a\dc} \s^n_{c \db}
\partial_m \bar A(x) \partial_n A(x) 
+ \d_{a \db} |W'(x)|^2  \nn\\
& & \qquad \qquad \mbox{}
+ \left. \psi^a(x) \psi_b(x) W''
+ \bar \psi_\db(x) \bar \psi^\da(x) \bar W''
\right) .
\eea
By setting $ a = b $, summing from $ 1 $ to 2, 
and using the trace identity (\ref{traces}),
we obtain
\bea
\left\{ Q_a , \bar Q_\da \right\} & = &
4 \int \! d^3x \,
\left( \ihalf \partial_i \bar \psi (x) \s^i \psi(x)
- \ihalf \bar \psi(x) \s^i \partial_i \psi(x) 
+ \partial_n \bar A(x) \partial_n A(x) \right. \nn\\
& & \qquad \qquad \mbox{}
+ \left. \half \psi \psi W'' + \half \bar \psi \bar \psi \bar W'' 
+ |W'(x)|^2 \right) .
\eea
Thus the famous relation
\beq
\left\{ Q_a , \bar Q_\da \right\} = 4 H .
\label {4H}
\eeq
\par
We may select the momentum operator $ P_3 $
by taking the trace with $ \s^3 $:
\bea
\left\{ Q_a , \bar Q_\db \right\} \s^3_{b\da} & = &
2 \int \! d^3x \,
\left( i \partial_i \bar \psi_{\dd}(x) \s^i_{d \db}
\s^3_{b\da} \psi_a(x)
- i \bar \psi_{\db}(x) \s^3_{b\da}
\s^i_{a\dc} \partial_i \psi_c(x) \right. \nn\\
& & \qquad \mbox{}
+ \s^m_{a\dc} \s^n_{c \db} \s^3_{b\da}
\partial_m \bar A(x) \partial_n A(x)
+ \d_{a \db} \s^3_{b\da} |W'(x)|^2  \nn\\
& & \qquad \mbox{}
+ \left. \psi^a(x) \s^3_{b\da} \psi_b(x) W''
+ \bar \psi_\db(x) \s^3_{b\da} \bar \psi^\da(x) \bar W''
\right) .
\eea
The terms involving the potential $ W $ all vanish.
The kinetic term for the scalar field $ A $ contains
the trace $ \Tr \left( \s^m \s^n \s^3 \right) 
= -2 \d_{m0} \d_{n3} - 2 \d_{m3} \d_{n0} $\null.
Thus integrating by parts, we have
\bea
\left\{ Q_a , \bar Q_\db \right\} \s^3_{b\da} & = &
2 \int \! d^3x \,
\left( i \partial_i \bar \psi \left\{ \s^i , \s^3 \right\} \psi
-2 \partial_0 \bar A \partial_3 A
-2 \partial_3 \bar A \partial_0 A \right) \nn\\
& = & \mbox{}
4 \int \! d^3x \,
\left( i \partial_3 \bar \psi \psi
- \partial_0 \bar A \partial_3 A 
- \partial_3 \bar A \partial_0 A \right) ,
\eea
which is the relation
\beq
\left\{ Q_a , \bar Q_\db \right\} \s^3_{b\da}
= 4 P_3 .
\eeq

\par
To verify that the supercharges
anti-commute, we use the commutation
relations (\ref{A A}) and (\ref{psi psi})
to reduce $ \{ Q_a , Q_b \} $ to
\bea
 \{ Q_a , Q_b \} & = &
2 \int \! d^3x d^3y \, \left(
\left\{ 
\s^m_{a\dc} \psi_c(x) \partial_m \bar A(x) ,
- i \e^{\db \dd} \bar \psi_{\dd}(y) \bar W' \right\}
\right. \nn\\
& & \qquad \qquad \mbox{}
+ \left. \left\{
- i \e^{\da \dc} \bar \psi_{\dc}(x) \bar W' ,
\s^n_{b\dd} \psi_d(y) \partial_n \bar A(y) 
\right\} \right) .
\eea
Now using the anti-commutation relation (\ref{psi psi bar}),
we get
\bea
 \{ Q_a , Q_b \} & = &
2 \int \! d^3x d^3y \, \left(
\partial_m \bar A(x) \bar W' 
\s^m_{a\dc} (-i) \e^{\db \dd} 
\d_{cd} \d( \vec x - \vec y )
\right. \nn\\
& & \qquad \qquad \mbox{}
+ \left.
\bar W' \partial_n \bar A(y) 
(-i) \e^{\da \dc} \s^n_{b\dd} 
\d_{cd} \d( \vec x - \vec y ) \right) \nn\\
& = & - 2 i \int \! d^3x \,
\partial_n \bar A(x) \bar W'
\left( \s^n_{a\dc} \e^{\db \dc} 
+ \e^{\da \dc} \s^n_{b\dc} \right) .
\eea
For $ n = 0 $, 
the $ \s $'s and $ \e $'s come to
$ \e^{\db \da} + \e^{\da \db} = 0 $,
and so only the spatial terms survive
\bea
\{ Q_a , Q_b \} & = &
- 2 i \int \! d^3x \,
\partial_i \bar A(x) \bar W'
\left( \s^i_{a\dc} \e^{\db \dc}
+ \e^{\da \dc} \s^i_{b\dc} \right) .
\eea
But this quantity is a space integral
of a total spatial divergence,
and so it vanishes
\beq
\{ Q_a , Q_b \} = 
- 2 i \int \! d^3x \,
\partial_i \left( \bar W
\left( \s^i_{a\dc} \e^{\db \dc}
+ \e^{\da \dc} \s^i_{b\dc} \right) \right)
= 0 .
\eeq 
It is worth checking, however,
whether the surface terms that we
here neglect as well as those that
we neglected in deriving eq.(\ref{land})
from eq.(\ref{jump}) actually do vanish.
In particular, when susy is spontaneously broken,
the minimum of $ |W'| $ is positive,
and there is at least one massless fermion.
\section{The Supercharges As Generators}
\subsection{The Supercharges Lack Auxiliary Fields}
The field equations 
\beq
0 = \frac{ \partial {\cal L} } { \partial \bar F }
= F + \bar W'
\quad {\hbox \mathrm{and}} \quad
0 = \frac{ \partial {\cal L} } { \partial F }
= \bar F + W'
\eeq
express the auxiliary fields $ F $ and $ \bar F $
in terms of the fields $ A $ and $ \bar A $ as
\beq
F = - \bar W' 
\quad {\hbox \mathrm{and}} \quad
\bar F = - W' .
\eeq
The supercharges $ Q_a $ and $ \bar Q_\db $
involve the fields $ A $ and $ \psi $
explicitly
\beq
Q_a = - \sqrt{2} \int \! d^3x \, \left(
\s^m_{a\db} \psi_b \partial_m \bar A
- i \e^{\da \db} \bar \psi_\db \bar W' \right) 
\eeq
\beq
\bar Q_\da = - \sqrt{2} \int \! d^3x \, \left(
\bar \psi_\db \s^m_{b \da} \partial_m A
+ i \e^{\da \db} \psi_b W' \right) .
\eeq
But in these formulas,
the auxiliary fields $ F $ and $ \bar F $
appear only as $ - \bar W' $ and $ - W' $\null.
\subsection{The Supercharges Generate Susy Transformations}
If as the generator $ G( \x ) $ of a susy transformation,
we use 
\beq
G( \x ) = \x Q + \bar Q \bx,
\label {G}
\eeq
then we find that the change in the
field $ A(x) $ is
\bea
\d_\x A( x ) \equiv \left[ i G( \x ), A( x ) \right] & = &
i \sqrt{2} \int \! d^3y \, 
\x^a \s^m_{a\db} \psi_b( y ) 
\left[ A( x ) , \partial_m \bar A( y ) \right] \nn\\
& = & \mbox{} - \sqrt{2} \int \! d^3y \,
\x^a \s^0_{a\db} \psi_b( y )
\d( \vec x - \vec y ) \nn\\
& = & \mbox{} \sqrt{2} \x^a \psi_a( x ) = \sqrt{2} \x \psi ,
\label {dA}
\eea
which is (\ref{d A})\null.
But if we apply $ G( \x ) $ to the field $ \psi(x) $,
then we get
\bea
\d_\x \psi_a(x) & \equiv & \left[ i G( \x ), \psi_a(x) \right] \nn\\
& = & \mbox{} \sqrt{2} \! \int \! \! d^3y 
\left[ \psi_a(x), \x^b \e^{\db \dc} \bar \psi_\dc(y) \bar W'
+ i \bar \psi_\dc(y) \s^m_{c \db} \bx^\db \partial_m A(y) \right] 
\nn\\
& = & \mbox{} \sqrt{2} \left(
- \x^b \e^{\db \da} \bar W'
+ i \s^m_{a \db} \bx^\db \partial_m A(x) \right)
\nn\\
& = & \mbox{}
\sqrt{2} \left( \x^b \e_{b a} \bar W'
+ i \s^m_{a \db} \bx^\db \partial_m A(x) \right)
\nn\\
& = & \mbox{} \sqrt{2} i \s^m_{a \db} \bx^\db \partial_m A(x)
- \sqrt{2} \x_a \bar W'
\label {dpsi}
\eea
which agrees with (\ref{d psi})
if we use $ - \bar W' = F $\null.
\subsection{Iterated Susy Transformations}
We may now ask what happens when we 
perform a second susy transformation
\bea
\d_\eta \d_\x A(x) & \equiv & \left[ \, i G( \eta ) ,
\left[ i G( \x ) , A(x) \right] \, \right] 
= \left[ i G( \eta ) , \sqrt{2} \x^a \psi_a( x ) \right] \nn\\ 
& = & \mbox{} \sqrt{2} \x^a \left[ i G( \eta ) , \psi_a( x ) \right]
= 2 i \x^a \s^m_{a \db} \bar \eta^\db \partial_m A(x)
- 2 \x^a \eta_a \bar W' .
\eea
So the difference $ \d_\eta \d_\x A - \d_\x \d_\eta A $ is 
\bea
\left[ \d_\eta , \d_\x \right] A(x)
& \equiv & \left[ \, i G( \eta ) , \left[ i G( \x ) , A(x) \right] 
\, \right] - 
\left[ \, i G( \x ) , \left[ i G( \eta ) , A(x) \right] 
\, \right] \nn\\
& = & \mbox{} 2 i \x \s^m \bar \eta \partial_m A(x) - 2 \x \eta \bar W'
- 2 i \eta \s^m \bx \partial_m A(x) + 2 \eta \x \bar W' \nn\\
& = & \mbox{} 2 i \left( \x \s^m \bar \eta  
- \eta \s^m \bx \right) \partial_m A(x) .
\label {ddA}
\eea
\par
The algebra closes on the field $ A $, as it must. 
For by the Jacobi identity, the difference 
\beq
\left[ \d_\eta , \d_\x \right] A(x) = 
\left[ \, i G( \eta ) , \left[ i G( \x ) , A(x) \right]
\, \right] -
\left[ \, i G( \x ) , \left[ i G( \eta ) , A(x) \right]
\, \right]
\eeq
is
\beq
\left[ \d_\eta , \d_\x \right] A(x)
= \left[ \, A(x) , \left[ i G( \x ) , i G( \eta ) \right] \, \right] . 
\eeq
Now by using the definition (\ref{G}) of the generators 
and successively eqs.(\ref{Q Q 0}) and
(\ref{algebra1}) of the susy algebra, 
we may write the commutator of the generators as
\bea 
\left[  i G( \x ) , i G( \eta ) \right] & = &
\left[ i \x Q + i \bar Q \bx , i \eta Q + i \bar Q \bar \eta \right]
\nn\\
& = & \mbox{} 
\left[ \eta Q , \bar Q \bx \right] 
- \left[ \x Q , \bar Q \bar \eta \right] 
\nn\\
& = & \mbox{}
\eta^a \left\{ Q_a , \bar Q_\db \right\} \bx^{\db}
- \x^a \left\{ Q_a , \bar Q_\db \right\} \bar \eta^{\db}
\nn\\
& = & \mbox{}
\eta^a 2 \s^n_{a\db} P_n \bx^{\db}
- \x^a 2 \s^n_{a\db} P_n \bar \eta^{\db} \nn\\
& = & \mbox{} 2 \eta \s^n P_n \bx
- 2 \x \s^n P_n \bar \eta .
\eea
By virtue of the commutation relations
\beq
\partial_n A(x) = i \left[ A(x) , P_n \right],
\eeq
the difference 
\beq
\left[ \d_\eta , \d_\x \right] A(x)
= \left[ A(x) , 2 \eta \s^n P_n \bx - 2 \x \s^n P_n \bar \eta \right] 
\eeq
is
\beq
\left[ \d_\eta , \d_\x \right] A(x)
= 2 i \left( \x \s^n \bar \eta - \eta \s^n \bx \right)
\partial_n A(x) ,
\eeq
which is eq.(\ref{ddA}).
\par
To see how the algebra closes on the field $ \psi $,
we examine the double susy transformation 
\bea
\d_\eta \d_\x \psi_a(x) 
& \equiv & \left[ \, i G( \eta ) , \left[ \, i G( \x ) ,
\psi_a(x) \, \right] \, \right] \nn\\
& = & \mbox{} \left[ \, i G( \eta ) ,
\sqrt{2} i \s^m_{a \db} \bx^\db \partial_m A(x)
- \sqrt{2} \x_a \bar W' \, \right] \nn\\
& = & \mbox{} \sqrt{2} i \s^m_{a \db} \bx^\db
\partial_m \left[ \, i G( \eta ) , A(x) \, \right]
- \sqrt{2} \x_a \left[ \, i G( \eta ) \, \bar W' \, \right] \nn\\
& = & \mbox{} \sqrt{2} i \s^m_{a \db} \bx^\db
\partial_m \sqrt{2} \eta^c \psi_c( x )
- \sqrt{2} \x_a \bar W'' \left[ \, i G( \eta ) , \bar A \, \right] \nn\\
& = & \mbox{} 2 i \s^m_{a \db} \bx^\db
\eta^c \partial_m \psi_c( x ) 
- \sqrt{2} \x_a \bar W'' \sqrt{2} \bar \psi_{\dc} \bar \eta^{\dc} \nn\\
& = & \mbox{} 2 i \s^m_{a \db} \bx^\db
\eta^c \partial_m \psi_c( x ) - 2 \x_a \bar \eta_{\dc} \bar \psi^{\dc} 
\bar W'' .
\eea
The Heisenberg equation of motion
for the field $ \psi $ is
\beq
-i \bar \s^n \partial_n \psi - \bar W'' \bar \psi = 0 .
\label {psi eq}
\eeq
So we may write the second variation $ \d_\eta \d_\x \psi_a(x) $ as
\beq
\d_\eta \d_\x \psi_a(x) = 2 i \s^m_{a \db} \bx^\db
\eta^c \partial_m \psi_c( x )
+ 2 i \x_a \bar \eta_{\dc} \bs^{n \dc d} \partial_n \psi_d(x) . 
\label {d d psi}
\eeq
We may manipulate the first term into the successive forms
\bea
2 i \s^m_{a \db} \bx^\db \eta^c \partial_m \psi_c
& = & \mbox{} - 2 i \eta \s^{m} \bx \partial_m \psi_a
+ \sum_{c \ne a} \left(
2i \s^m_{ a \db } \bx^\db \eta^c \partial_m \psi_c
+ 2i \eta^c \s^m_{ c \db } \bx^\db \partial_m \psi_a 
\right) \nn\\
& = & \mbox{} - 2 i \eta \s^{m} \bx \partial_m \psi_a 
+ 2 i \sum_{c \ne a} \left(
\s^m_{ c \db } \partial_m \psi_a 
- \s^m_{ a \db } \partial_m \psi_c 
\right) \eta^c \bx^\db .
\label {first term}
\eea
The terms in the restricted sum with $ c = a $ actually cancel
\bea
\lefteqn{ \!\!\!\!\!\!\! \!\!\!\!\!\!\! 
2 i \sum_{c \ne a} \left(
\s^m_{ c \db } \partial_m \psi_a
- \s^m_{ a \db } \partial_m \psi_c
\right) \eta^c \bx^\db \nn } \\
& = & \mbox{} 2 i \left( \s^m_{ c \db } \partial_m \psi_a
- \s^m_{ a \db } \partial_m \psi_c
\right) \eta^c \bx^\db \nn\\
& = & \mbox{} 2 i \e_{ \db \dd } \left(
\e_{ c e } \bs^{ m \dd e } \partial_m \psi_a
- \e_{ a e } \bs^{ m \dd e } \partial_m \psi_c 
\right) \eta^c \bx^\db .
\label {sum}
\eea
Both for $ a = 1 $ and for $ a = 2 $,
this sum is
\bea
2 i \sum_{c \ne a} \left(
\s^m_{ c \db } \partial_m \psi_a
- \s^m_{ a \db } \partial_m \psi_c
\right) \eta^c \bx^\db 
& = & \mbox{} 2 i \e_{ \db \dd } 
\bs^{ m \dd c } \partial_m \psi_c \e_{ f a } \eta^f \bx^{\db} \nn\\
& = & \mbox{} 2 i \bs^{ m \dd c } \partial_m \psi_c
\eta_a \bx_{\db} .
\label {sum2}
\eea
By inserting Kronecker deltas,
and next using first (\ref{comp}) and then (\ref{trans})
and last (\ref{afcomp}),
we may write this last term as
\bea
2 i \bs^{ m \dd c } \partial_m \psi_c \eta_a \bx_{\db}
& = & 2 i \bs^{ m \dd c } \partial_m \psi_c \eta_g \bx_{\dh}
\d_a^g \d_{\db}^{\dh} \nn\\
& = & \mbox{} i \s^n_{ a \dd } \bs^{ m \dd c } \partial_m \psi_c
\bx_{\dh} \bs^{ \dh g }_n \eta_g \nn\\
& = & \mbox{} - i \left( \s^n \bs^m \partial_m \psi \right)_a
\eta \s_n \bx \nn\\
& = & \mbox{} 
2 i \eta_a \bx \bs^m \partial_m \psi(x) .
\label {sum3}
\eea
Gathering the results (\ref{d d psi}--\ref{sum3}),
we may write the second variation of the field $ \psi_a(x) $ 
in the symmetrical form
\beq
\d_\eta \d_\x \psi_a(x) = - 2 i \eta \s^{m} \bx \partial_m \psi_a(x)
+ 2 i \eta_a \bx \bs^m \partial_m \psi(x)
+ 2 i \x_a \bar \eta \bs^m \partial_m \psi(x) .
\eeq
Thus the double difference 
$ \d_\eta \d_\x \psi - \d_\x \d_\eta \psi $ is
\beq
\left[ \d_\eta , \d_\x \right] \psi_a(x) =
2 i \left( \x \s^m \bar \eta - \eta \s^m \bx \right) \partial_m \psi_a(x) ,
\eeq
which shows that the algebra also closes
on the field $ \psi $, as it must.
\section{Susy without Grassmann Variables}
Exponentials of the generator
\( G( \x ) = \x Q + \bar Q \bx \)
are not unitary operators
because they involve Grassmann variables.
Can one avoid these anti-commuting variables?
\par
Let us consider using generators
\( G( z ) \) that are complex linear
forms in the supercharges \( Q \) and \( \bar Q \)
\beq
G( z ) = z Q + \bar Q \bar z
\label {G(z)}
\eeq
where \( z^a \) is a complex spinor.
\par
Now the change in the field $ A(x) $ is
\bea
dA( x ) \equiv \left[ \, i G( z ) , A( x ) \, \right] & = &
\mbox{} - i \sqrt{2} \int \! d^3y \,
z^a \s^m_{a\db} \psi_b( y )
\left[ \, \partial_m \bar A( y ) , A( x ) \, \right] \nn\\
& = & \mbox{} - \sqrt{2} \int \! d^3y \,
z^a \s^0_{a\db} \psi_b( y )
\d( \vec x - \vec y ) \nn\\
& = & \mbox{} \sqrt{2} z^a \psi_a( x ) = \sqrt{2} z^a \psi_a ,
\label {dzA}
\eea
which is the same as (\ref {dA})
except that the Grassmann spinor \( \x \) 
has been replaced by the complex spinor \( z \).
The conjugate change is
\beq
d\bar A = \sqrt{2} \bar \psi_\da \bar z^\da .
\label {dzbarA}
\eeq
\par
This procedure will not work, however,
for the Fermi field \( \psi \).
Instead we must write \( d_z \psi \) as an anti-commutator.
There are several ways of doing this.
If we want a single rule for the
change in the product of two spinor fields
irrespective of whether they transform like
$ \psi $ or like $ \bar \chi $, then
we can not have $ d \psi $ be the adjoint
of $ d \bar \psi $.  We choose to have
$ d \psi $ be the adjoint
of $ d \bar \psi $.  We shall have four
different rules for the change in the
product of two spinor fields.  We define
\bea
d\psi_a(x) & \equiv & - \left\{ i G( z ) , \psi_a(x) \right\} \nn\\
& = & \mbox{} \sqrt{2} \! \int \! \! d^3y
\left\{ z^b \e^{\db \dc} \bar \psi_\dc(y) \bar W'
+ i \bar \psi_\dc(y) \s^m_{c \db} \bz^\db \partial_m A(y) ,
\psi_a(x) \right\}
\nn\\
& = & \mbox{} \sqrt{2} \left(
 z^b \e^{\db \da} \bar W'
+ i \s^m_{a \db} \bz^\db \partial_m A(x) \right)
\nn\\
& = & \mbox{}
\sqrt{2} \left( - z^b \e_{b a} \bar W'
+ i \s^m_{a \db} \bz^\db \partial_m A(x) \right)
\nn\\
& = & \mbox{} \sqrt{2} i \s^m_{a \db} \bz^\db \partial_m A(x)
+ \sqrt{2} z_a \bar W' .
\label {dzpsi}
\eea
The change in the conjugate $\bar \psi $ 
is the conjugate of the change in $ \psi $:
\bea
d\bar \psi_\da & \equiv & \left\{ i G( z ) , \bar \psi_\da \right\}  
= \left( - \left\{ i G( z ) , \psi_a(x) \right\} \right)^\dgr 
= ( d_z \psi_a )^\dgr \nn\\
& = & \mbox{} - \sqrt{2} i z^b \s^m_{b \da} \p_m \bar A 
+ \sqrt{2} \bar z_\da W' .
\label{dzbarpsi}
\eea
Although these formulas differ from expression (\ref {dpsi})
for $ \d\psi $ and its conjugate for $ \d\bar\psi $
by the signs of their second terms, and of course
by the replacement of a Grassmann spinor $ \x $ by 
a complex one $ z $,
we shall see that these sign differences
are appropriate and that supersymmetry can be implemented
by unitary transformations acting on the
states and physical operators of the theory.
\par
The key point is that the physical operators of the theory
contain even powers of the Fermi fields.
Thus the change in the generic product
$ \psi \phi $ is
\bea
d(\psi \phi) & = & 
\left [ \, i G ( z ) , \psi \phi \, \right ] \nn\\ 
& = & \mbox{} i G ( z ) \, \psi \phi - \psi \phi \, i G ( z ) \nn\\
& = & \mbox{} i G ( z ) \, \psi \phi + \psi \, i G (z ) \, \phi
- \psi \, i G (z ) \, \phi - \psi \phi \, i G (z ) \nn\\
& = & \mbox{} 
\left \{ i G (z ) , \psi \right \} \phi 
- \psi \left \{ i G (z ) , \phi \right \}
\nn\\
& = & \mbox{} - d_z \psi \, \phi + \psi \, d_z \phi ,
\label {dpsiphi}
\eea
in which the spinor indices, which may be different, are suppressed. 
It is easy to see that the other three rules are:
\bea
d( \bar \psi \bar \phi ) = 
\left [ \, i G(z) , \bar \psi \bar \phi \, \right ] 
& = & d \bar \psi \, \bar \phi
- \bar \psi \, d \bar \phi 
\label {barpsibarphi} \\
d( \bar \psi \phi ) = 
\left [ \, i G(z) , \bar \psi \phi \, \right ] 
& = & d \bar \psi \, \phi 
+ \bar \psi \, d \phi 
\label {barpsiphi} \\
d( \psi \bar \phi ) = 
\left [ \, i G(z) , \psi \bar \phi \, \right ] 
& = & - d \psi \, \bar \phi
- \psi \, d  \bar \phi .
\label {psibarphi}
\eea
\par
Let us consider now the effect of these
transformations on the action density (\ref {Lc})
or equivalently on the general chiral action density (\ref {LCG})
with the index $i$ suppressed and with the auxiliary
field $ F $ replaced by $ - \bar W' $,
\beq
{ \cal L} = \ihalf \p_n \bar \psi \bar \s^n \psi
- \ihalf \bar \psi \bar \s^n \partial_n \psi
- \partial_n \bar A \partial^n A
- | W' |^2 - \half W'' \psi \psi - \half \bar W'' \bar \psi \bar \psi .
\label {Lcn}
\eeq
The change in $ { \cal L} $ due to the 
changes $ d_z A $ and $ d_z \psi $ and their conjugates is
\bea
\lefteqn{
d{ \cal L } = \left [ \, i G(z) , { \cal L } \, \right] =
\ihalf \p_n d_z \bar \psi \bar \s^n \psi
+  \ihalf \p_n \bar \psi \bar \s^n d_z \psi
- \ihalf d_z \bar \psi \bar \s^n \partial_n \psi
- \ihalf \bar \psi \bar \s^n \partial_n d_z \psi } \hspace {0.9in} \nn \\
& & \mbox{} - \partial_n d_z \bar A \partial^n A
- \partial_n \bar A \partial^n d_z A 
- \bar W' W'' d_z A - W' \bar W'' d_z \bar A \nn\\
& & \mbox{} - \half W''' d_z A \psi \psi
- \half \bar W''' d_z \bar A \bar \psi \bar \psi 
+ \half W'' d_z \psi \psi - \half W'' \psi d_z \psi \nn\\
& & \mbox{} - \half \bar W'' d_z \bar \psi  \bar \psi
+ \half \bar W'' \bar \psi d_z \bar \psi .
\label {dLcn}
\eea 
The part of $ d{ \cal L} $ that depends upon $ z $ is
\bea
d_z { \cal L} & = &
\ihalf \p_n ( - \sqrt{2} i z \s^m \p_m \bar A ) \bar \s^n \psi
+  \ihalf \p_n \bar \psi \bar \s^n ( \sqrt{2} z \bar W' ) \nn\\
& & \mbox{}
- \ihalf ( - \sqrt{2} i z \s^m \p_m \bar A ) \bar \s^n \partial_n \psi
- \ihalf \bar \psi \bar \s^n \partial_n ( \sqrt{2} z \bar W' ) \nn\\
& & \mbox{} - \partial_n \bar A \partial^n ( \sqrt{2} z \psi )
- \bar W' W'' ( \sqrt{2} z \psi ) \nn\\
& & \mbox{} - \half W''' ( \sqrt{2} z \psi ) \psi \psi
+ \half W'' ( \sqrt{2} z \bar W' ) \psi 
- \half W'' \psi ( \sqrt{2} z \bar W' ) \nn\\
& & \mbox{} - \half \bar W'' ( - \sqrt{2} i z \s^m \p_m \bar A ) \bar \psi
+ \half \bar W'' \bar \psi ( - \sqrt{2} i z \s^m \p_m \bar A ) .
\label {dLcnd}
\eea
As in eqs.(\ref {W'''1}) and (\ref {W'''2}),
the term proportional to $ W''' $ vanishes.
Also the two terms proportional 
to $ \bar W' \, W'' $ cancel.
Finally the last two terms may be written as
\beq
\ihalf \bar \psi^\da \s^m_{b\da} z^b \p_m \bar W' = 
\ihalf \bar \psi_\dc \, \e^{\da\dc} \s^m_{b\da} \e^{bd} z_d \, \p_m \bar W'
= \ihalf \bar \psi \bs^m z \, \p_m \bar W'
\eeq
and as
\beq
- \ihalf \bar \psi_\da z^b \s^m_{b\dc} \, \e^{\da\dc} \p_m \bar W' =
- \ihalf \bar \psi_\da \e^{\da\dc} \e^{bd} \s^m_{b\dc} z_d \, \p_m \bar W'
= \ihalf \bar \psi \bs^m z \, \p_m \bar W' .
\eeq
So the change $ d_z { \cal L} $ in the action density is 
\bea
d_z { \cal L} & = &
{1 \over \sqrt{2} } z \s^m \bs^n \psi \p_n \p_m \bar A
- {1 \over \sqrt{2} } z \s^m \bs^n  \p_n \psi \p_m \bar A \nn\\
& & \mbox{} - \sqrt{2} z \p^n \psi \p_n \bar A
+ \ihalf \p_n \left( \bar \psi \bs^n z \bar W' \right) .
\label {daL}
\eea
Just as in eqs.(\ref {funnysigmas1}) and (\ref {funnysigmas2}),
we may write this change $ d_z { \cal L} $ as the total
divergence
\beq
d_z { \cal L} = \p_n K_z^n
\eeq
of the current
\beq
K_z^n = - { 1 \over \sqrt{2} } z \s^m \bs^n \p_m \bar A
- \sqrt{2} z \psi \p^n \bar A 
+ \ihalf \bar \psi \bs^n z \bar W' ,
\label {Kz}
\eeq
which shows that the action is invariant 
under the unitary transformation
\beq
U(z) = e^{-iG(z)}
\label {U}
\eeq
at least for infinitesimal values of the
complex spinor $ z $. 
\par
Thus unitary operators without Grassmann variables
can implement supersymmetry transformations
upon the action and other
operators that involve only even powers of Fermi fields.
But the effect (\ref {dzpsi})
of a susy transformation
upon a single Fermi field does not seem
to admit such a representation.
\par
Also if we define physical states $ | \psi \rangle $
as those that under a rotation of angle
$ 2 \pi $ about any axis $ \hat \theta $
suffer at most a phase change
\beq
e^{ -i 2 \pi \hat \theta \cdot \vec J } | \psi \rangle
= e^{ i \phi } | \psi \rangle ,
\label {def of physical state}
\eeq
then the unitary operator (\ref {U}),
being the exponential of an operator $ G(z) $
that is odd or fermionic,
transforms all physical states that 
are not annihilated by $ G(z) $
into states that are not physical.

\section{The Free Supercharges $ Q_K $}
\subsection{Expansions}
We may expand the supercharges $ Q_{Ka} $
of the free theory in terms of
creation and annihilation operators 
by using the expansions of the 
field operators $ \psi_a $
and $ A $\null.
Since the free field $ \psi_a $
satisfies the free-field equation
\beq
-i \bar \s^n \partial_n \psi = 0 ,
\eeq
we may expand it as
\beq
\psi_a( x ) = 
\int \! 
\frac{ d^3p }{ \sqrt{ ( 2 \pi )^3 } } \, 
\left[
e^{ ip \cdot x } u(p)_a b(p) + e^{ - ip \cdot x } v(p)_a c^\dgr(p) 
\right] 
\label {psi exp}
\eeq
in which the spinors $ u(p) $ and $ v(p) $
are eigenvectors of $ \vec \s \cdot \vec p $
\beq
\vec \s \cdot \vec p \, u(p) =  | \vec p | \, u(p)
\quad {\hbox \mathrm{and}} \quad
\vec \s \cdot \vec p \, v(p) = | \vec p | \, v(p),
\label {spinor ev eqs}
\eeq
which are normalized $ u(p)^\dgr u(p) = v(p)^\dgr v(p) = 1$.
The expansion of the scalar
field $ A $ is
\beq
A(x) = \int \! 
\frac{ d^3k }{ \sqrt{ ( 2 \pi )^3  2 | \vec k | } } \,
\left[
e^{ ik \cdot x } a(k) + e^{ - ik \cdot x } a_c^\dgr(k) 
\right],
\label {A exp}
\eeq
in which the operator
$ a_c^\dgr $ creates the particle
that is the anti-particle of the
particle created by $ a^\dgr $\null.
By substituting the expansion
(\ref{psi exp}) and the adjoint
of (\ref{A exp}) into the formula (\ref{Q_K})
for the supercharges of the free theory,
we find after some elementary manipulations
\beq
Q_{{\cal K}a} =
2 \, i \int \! d^3p \,
\sqrt{ | \vec p | } \, \left[
\, u(p)_a \, b(p) \, a^\dgr(p) - v(p)_a \, c^\dgr(p) \, a_c(p) \, \right].
\label {Q_K exp}
\eeq
Since every term
in the supercharge $ Q_{{\cal K}a} $
and in its adjoint $ \bar Q_{{\cal K}\da} $
contains an annihilation operator,
it is clear that both $ Q_{{\cal K}a} $
and $ \bar Q_{{\cal K}\da} = Q_{{\cal K}a}^\dgr $
annihilate the no-particle state
$ | 0 \rangle $, which is the vacuum
of the free-field theory
\beq
Q_{{\cal K}a} | 0 \rangle = 0
\quad {\hbox \mathrm{and}} \quad
Q_{{\cal K}a}^\dgr | 0 \rangle = 0,
\label {Q_K vac = 0}
\eeq
which incidentally shows
that susy is unbroken
in the free theory.
\subsection{Functional Methods}
It may be useful to see how 
the equations (\ref{Q_K vac = 0})
look in terms of the 
wave function of the vacuum
of the free theory. 
Let the state $ | \cA \rangle $
be a simultaneous eigenstate
of the field operators $ A $
and $ \bar A $ at time $ t = 0 $
\beq
A( 0, \vec x ) | \cA \rangle = 
\cA( \vec x )  | \cA \rangle
\quad {\hbox \mathrm{and}} \quad
\bar A( 0, \vec x ) | \cA \rangle =
\bar \cA( \vec x )  | \cA \rangle .
\eeq
To define the analogous state for the
Fermi field $ \psi $,
we follow Weinberg~\cite{SW},
{\it mutatis mutandis\/}, and write
\beq
| \vp \rangle = \exp\left( \int \! d^3x \, \psi^\dgr_a(0, \vec x)
\vp_a( \vec x ) \right) 
\left( \prod_{\vec x b} \psi_b( 0, \vec x ) \right) | 0 \rangle 
\label {vp state}
\eeq
in which the sum over $ a $ is from 1 to 2.
The state $ | \vp \rangle $ is an eigenstate of the operator
$ \psi_a( 0, \vec x ) $
\beq
\psi_a( 0, \vec x ) | \vp \rangle
= \vp_a( \vec x ) | \vp \rangle.
\label {psi vp}
\eeq
We may also construct the state
\beq
| \bar \vp \rangle = \exp\left( \int \! d^3x \, \bar \vp_a( \vec x )
\psi_a(0, \vec x) \right)
\left( \prod_{\vec x b} \psi_b^\dgr ( 0, \vec x ) \right) | 0 \rangle
\label {vp bar state}
\eeq
which is an eigenstate of $ \psi_a^\dgr ( 0, \vec x ) $
\beq
\psi_a^\dgr ( 0, \vec x ) | \bar \vp \rangle
= \bar \vp_a( \vec x ) | \bar \vp \rangle.
\label {psi bar vp}
\eeq
We shall write the adjoint states as
\beq
\langle \bar \vp | =
\langle 0 | \left( \prod_{\vec x b} \psi_b^\dgr ( 0, \vec x ) \right)
\exp\left( \int \! d^3x \, \bar \vp_\da( \vec x ) \psi_a(0, \vec x)
\right) 
\label {vp adj state}
\eeq
which satisfies 
\beq
\langle \bar \vp | \, \bar \psi_\da( 0 , \vec x ) =
\left( \psi_a( 0, \vec x ) | \vp \rangle \right)^\dgr
= \langle \bar \vp | \, \bar \vp_\da ( \vec x )
\label {bar psi}
\eeq
and as
\beq
\langle \vp | =
\langle 0 | \left( \prod_{\vec x b} \psi_b ( 0, \vec x ) \right)
\exp\left( \int \! d^3x \, \psi_a^\dgr (0, \vec x) \vp_a( \vec x ) 
\right) 
\label {vp bar adj state}
\eeq
which satisfies 
\beq
\langle \vp | \, \psi_a( 0 , \vec x ) =
\left( \psi_a^\dgr ( 0, \vec x ) | \bar \vp \rangle \right)^\dgr
= \langle \vp | \, \vp_a ( \vec x ) .
\label {bar psi bar}
\eeq

But to obtain the effect from the right
of $ \psi $ on the state $ \langle \bar \vp | $
or of $ \psi^\dgr $ on the state $ \langle \vp | $,
one differentiates
\beq
\langle \bar \vp | \psi_a( 0 , \vec x ) =
\frac{ \d }{ \d \bar \vp_\da } \, \langle \bar \vp |
\quad {\hbox \mathrm{and}} \quad
\langle \vp | \psi_a^\dgr( 0 , \vec x ) =
- \frac{ \d }{ \d \vp_a } \langle \vp |
\label {other evs}
\eeq
Both the states $ | \vp \rangle $
and the states $ | \bar \vp \rangle $
form complete sets of states.
\par
Let us now compute the inner product
of the no-particle state $ | 0 \rangle $
with the eigenstates of the Fermi fields
$ \psi $ and $ \psi^\dgr $\@. 
To this end it will be useful first
to calculate the integrals that occur
in the exponentials (\ref{vp state}) and (\ref{vp bar state})\@.
We may expand the field $ \psi $ at time $ t = 0 $
in terms of the annihilation operators $ b(\vec p) $
and the creation operators $ c^\dgr( - \vec p) $
\beq
\psi_a( 0, \vec x ) =
\int \!
\frac{ d^3p }{ \sqrt{ ( 2 \pi )^3 } } \,
e^{ i \vec p \cdot  \vec x } \left[ 
u(\vec p)_a b(\vec p) + v( - \vec p)_a c^\dgr( - \vec p)
\right] .
\label {psi t=0 exp}
\eeq
We may also expand the Grassmann field $ \vp( \vec x ) $
as
\beq
\vp_a( \vec x ) = 
\int \!
\frac{ d^3k }{ \sqrt{ ( 2 \pi )^3 } } \,
e^{ i \vec k \cdot  \vec x }
\sum_{s=1}^2 w( \vec k, s )_a \, \vp( k , s ) 
\label {G field}
\eeq
in which the variables $ \vp( k , s ) $
are grassmannian and the vectors $ w( \vec k, s ) $ 
are chosen to obey the orthonormality relations
\bea
\sum_{a=1}^2 w^*( \vec p, 1)_a u(\vec p)_a = 1
& {\hbox \mathrm{and}} &
\sum_{a=1}^2 w^*( \vec p, 1)_a v( - \vec p)_a = 0 \nn\\
\sum_{a=1}^2 w^*( \vec p, 2)_a u(\vec p)_a = 0
& {\hbox \mathrm{and}} &
\sum_{a=1}^2 w^*( \vec p, 2)_a v( - \vec p)_a = 1 .
\label {w u v}
\eea
Then the integrals are
\beq
\int \! d^3x \, \psi^\dgr_a(0, \vec x) \vp_a( \vec x ) =
\int \! d^3p\, \left[ 
b^\dgr( \vec p ) \vp( \vec p , 1 )
+ c( - \vec p ) \vp(\vec p , 2 )
\right]
\label {int}
\eeq
and its adjoint
\beq
\int \! d^3x \, \bar \vp_a( \vec x ) \psi_a(0, \vec x) =
\int \! d^3p\, \left[
\bar \vp( \vec p , 1 ) b( \vec p ) 
+ \bar \vp(\vec p , 2 ) c^\dgr( - \vec p ) 
\right] ,
\label {int adj}
\eeq
in which sums over $ a $ from 1 to 2 are understood.
\par
We may now evaluate the inner product
\beq
\langle 0 | \vp \rangle =
\langle 0 | \exp\left( \int \! d^3x \, \psi^\dgr_a(0, \vec x)
\vp_a( \vec x ) \right)
\left( \prod_{\vec x b} \psi_b( 0, \vec x ) \right) | 0 \rangle.
\eeq
In the infinite product,
only the negative-frequency parts 
\beq
\psi_b^{(-)}( 0, \vec x ) =
\int \!
\frac{ d^3p }{ \sqrt{ ( 2 \pi )^3 } } \,
e^{ i \vec p \cdot  \vec x } v( - \vec p)_a c^\dgr( - \vec p)
\eeq
survive
\beq
\langle 0 | \vp \rangle =
\langle 0 | \exp\left( \int \! d^3x \, \psi^\dgr_a(0, \vec x)
\vp_a( \vec x ) \right)
\left( \prod_{\vec x b} \psi_b^{(-)}( 0, \vec x ) \right) | 0 \rangle.
\eeq
Substituting our formula (\ref{int})
for the integral, we get
\beq
\langle 0 | \vp \rangle =
\langle 0 | \exp\left( \int \! d^3p\, \left[
b^\dgr( \vec p ) \vp( \vec p , 1 )
+ c( - \vec p ) \vp(\vec p , 2 )
\right]
\right)
\left( \prod_{\vec x b} \psi_b^{(-)}( 0, \vec x ) \right) | 0 \rangle.
\eeq
If at each momentum $ \vec p $ we now expand 
the exponential, remembering that 
$ \vp $ is Grassmann, then we find
\beq
\langle 0 | \vp \rangle =
\langle 0 | \left( \prod_{\vec p}
\left[ 1 + d^3p \left( b^\dgr( \vec p ) \vp( \vec p , 1 )
+ c( - \vec p ) \vp(\vec p , 2 ) \right)
\right] \right)
\left( \prod_{\vec x a} \psi_a^{(-)}( 0, \vec x ) \right) | 0 \rangle.
\eeq
None of the $ b^\dgr( \vec p ) $ terms survive,
and so apart from an overall factor
independent of $ \vp $, we have
\beq
\langle 0 | \vp \rangle =
\prod_{\vec p} \vp(\vec p , 2 ) .
\label {0 vp}
\eeq 
By a similar argument we find
\bea
\langle 0 | \bar \vp \rangle & = &
\langle 0 | \exp\left( \int \! d^3x \, \bar \vp_a( \vec x )
\psi_a(0, \vec x) \right)
\left( \prod_{\vec x b} \psi_b^\dgr ( 0, \vec x ) \right) | 0 \rangle
\nn\\
& = & \langle 0 | \exp\left( \int \! d^3p\, \left[
\bar \vp( \vec p , 1 ) b( \vec p )
+ \bar \vp(\vec p , 2 ) c^\dgr( - \vec p ) \right] \right) 
\left( \prod_{\vec x b} \psi_b^{(+)\dgr} ( 0, \vec x ) \right) | 0 \rangle
\nn\\
& = & \prod_{\vec p} \bar \vp(\vec p , 1 ) .
\label {0 vp bar}
\eea
\par
The bosonic wave-function of the vacuum
is well known to be
\beq
\langle A, \bar A | 0 \rangle =
\exp \left( - \int \! d^3x \, 
\bar A( \vec x ) \sqrt{ - \nabla^2 } A( \vec x ) \right) 
\eeq
or equivalently
\beq
\langle A, \bar A | 0 \rangle =
\exp \left( - \int \! d^3x \,
A( \vec x ) \sqrt{ - \nabla^2 } \bar A( \vec x ) \right) .
\eeq
Thus the full wave-functions of the vacuum are 
\beq
\langle \vp , A, \bar A | 0 \rangle =
\left( \prod_{\vec p} \vp(\vec p , 1 ) \right)
\exp \left( - \int \! d^3x \,
A( \vec x ) \sqrt{ - \nabla^2 } \bar A( \vec x ) \right)
\label {vp A vac}
\eeq
and
\beq
\langle \bar \vp , A, \bar A | 0 \rangle =
\left( \prod_{\vec p} \bar \vp(\vec p , 2 ) \right)
\exp \left( - \int \! d^3x \,
\bar A( \vec x ) \sqrt{ - \nabla^2 } A( \vec x ) \right).
\label {bar vp A vac}
\eeq
\par
We may now verify in this functional formalism
that the free supercharges $ Q_{{\cal K}a} $ and $ \bar Q_{{\cal K}\da} $
annihilate the no-particle state $ | 0 \rangle $.
By referring to eq.(\ref{Q_K}), we have
\beq
\langle \vp , A, \bar A | Q_{{\cal K}a} | 0 \rangle =
\langle \vp , A, \bar A | ( - \sqrt{2} ) \int \! d^3x \, \s^m_{a\db}
\psi_b ( 0, \vec x ) \partial_m \bar A ( 0, \vec x ) | 0 \rangle .
\eeq
The time derivative $ \partial_0 \bar A ( \vec x ) $ is the
canonical momentum operator $ \pi( \vec x ) $
which may be represented as a functional derivative
\beq
\langle \vp , A, \bar A | \partial_0 \bar A ( 0, \vec x )
= \langle \vp , A, \bar A | \pi( 0, \vec x ) 
= \frac { \d } { i \d A ( \vec x ) } \langle \vp , A, \bar A | .
\label {pi}
\eeq
Thus we have
\beq
\langle \vp , A, \bar A | Q_{{\cal K}a} | 0 \rangle =
- \sqrt{2} \langle \vp , A, \bar A |
\! \! \int \! d^3x \s^i_{a\db} \psi_b (0 , \vec x )
\partial_i \bar A (0 , \vec x ) - \psi_a (0 , \vec x )
\pi (0 , \vec x ) | 0 \rangle .
\eeq
Now using the eigenvalue relation (\ref{vp bar adj state})
and the functional relation (\ref{pi}), we find
\beq
\langle \vp , A, \bar A | Q_{{\cal K}a} | 0 \rangle =
- \sqrt{2} \!\! \int \!\! d^3x \!\! \left[
\s^i_{a\db} \vp_b ( \vec x ) \partial_i \bar A ( \vec x )
- \vp_a ( \vec x ) \frac { \d } { i \d A ( \vec x ) } \right]
\!\! \langle \vp , A, \bar A | Q_{{\cal K}a} | 0 \rangle .
\eeq
By using our formula (\ref{vp A vac}) for the
wave-function of the vacuum, we have
\bea
\lefteqn{ \!\!\!\!\!\!\!\!\!\!\!\!\!\!\!\!\!\!\!\!\!\!\!\!\!\!\!\!\!\!\!
\!\!\!\!
\langle \vp , A, \bar A | Q_{{\cal K}a} | 0 \rangle =
- \sqrt{2} \int \! d^3x \, \left[
\s^i_{a\db} \vp_b ( \vec x ) \partial_i \bar A ( \vec x )
- \vp_a ( \vec x ) \frac { \d } { i \d A ( \vec x ) } \right] \nn}\\
\qquad \qquad \qquad \qquad & & \mbox{} \times 
\left( \prod_{\vec p} \vp(\vec p , 1 ) \right)  
\exp \left( - \int \! d^3y \,
A( \vec y ) \sqrt{ - \nabla^2 } \bar A( \vec y ) \right) \nn\\
\!\! & = & - \sqrt{2} \int \! d^3x \, \left[
\s^i_{a\db} \vp_b ( \vec x ) \partial_i \bar A ( \vec x )
- i \vp_a ( \vec x ) \sqrt{ - \nabla^2 } \bar A ( \vec x ) \right] \nn\\
& & \mbox{} \times
\left( \prod_{\vec p} \vp(\vec p , 1 ) \right) 
\exp \left( - \int \! d^3y \,
A( \vec y ) \sqrt{ - \nabla^2 } \bar A( \vec y ) \right) \nn
\eea
or after integrating by parts
\beq
\langle \vp , A, \bar A | Q_{{\cal K}a} | 0 \rangle =
i \sqrt{2} \int \!\! d^3x  
\bar A ( \vec x ) \left[ \left( -i \vec \s \cdot \nabla 
+ \sqrt{ - \nabla^2 } \right) \vp ( \vec x ) \right]_a 
\! \langle \vp , A, \bar A | Q_{{\cal K}a} | 0 \rangle .
\eeq
\par
The effect of the differential operator 
$ -i \vec \s \cdot \nabla + \sqrt{ - \nabla^2 } $ 
on the Grassmann field $ \vp $ is
\beq
\left( -i \vec \s \cdot \nabla
+ \sqrt{ - \nabla^2 } \right) \vp ( \vec x )
= \int \!
\frac{ d^3k }{ \sqrt{ ( 2 \pi )^3 } } \,
\left( \vec \s \cdot \vec k + | \vec k | \right) 
e^{ i \vec k \cdot  \vec x }
\sum_{s=1}^2 w( \vec k, s ) \, \vp( k , s ) .
\eeq
By their definitions (\ref{w u v}), the spinors
$ w( \vec k, 1 ) $ and $ w( \vec k, 2 )$ respond
to $ \vec \s \cdot \vec k + | \vec k | $ the
same ways as do the spinors $ u( \vec k ) $ and $ v( - \vec k ) $
respectively.  Thus by the spinor eigenvalue
equations (\ref{spinor ev eqs}), 
the operator $ -i \vec \s \cdot \nabla + \sqrt{ - \nabla^2 } $
projects out the part of $ \vp $ that is 
proportional to $ w( \vec k , 1 ) $
\beq
\left( -i \vec \s \cdot \nabla
+ \sqrt{ - \nabla^2 } \right) \vp ( \vec x )
= \int \!
\frac{ d^3k }{ \sqrt{ ( 2 \pi )^3 } } \,
2 | \vec k | e^{ i \vec k \cdot  \vec x }
w( \vec k, 1 ) \vp( \vec k , 1 ) .
\label {proj on vp}
\eeq
Hence on the vacuum the free supercharge $ Q_K $ 
produces a doubling of the Grassmann variable $ \vp( \vec k , 1 ) $
\bea
\langle \vp , A, \bar A | Q_{{\cal K}a} | 0 \rangle & = &
i \sqrt{2} \int \! d^3x \,
\bar A ( \vec x ) \int \!
\frac{ d^3k }{ \sqrt{ ( 2 \pi )^3 } } \,
2 | \vec k | e^{ i \vec k \cdot  \vec x }
w( \vec k, 1 )_a \vp( \vec k , 1 )
\nn\\
& & \times 
\left( \prod_{\vec p} \vp(\vec p , 1 ) \right)
e^{ - \int \! d^3y \,
A( \vec y ) \sqrt{ - \nabla^2 } \bar A( \vec y ) }
\;\; = \;\; 0 
\eea
which vanishes.
\par
The free adjoint supercharges $ Q^\dgr_{{\cal K}} $
may be shown to annihilate the free vacuum $ | 0 \rangle $
by similar functional manipulations.
Indeed because $ \langle \bar \vp, A, \bar A | 0 \rangle $
is proportional by (\ref{bar vp A vac})
to the product of $ \bar \vp( \vec k , 2 ) $
for all $ \vec k $, the matrix element 
\bea
\langle \bar \vp, A, \bar A | Q^\dgr_{{\cal K}\da} | 0 \rangle & = & 
- \sqrt{2} \int \! d^3x \, 
\langle \bar \vp, A, \bar A |
\bar \psi_\db \s^m_{b\da} \partial_m A | 0 \rangle \nn\\
& = & - \sqrt{2} \int \! d^3x \,
\left[ \bar \vp \left( \vec \s \cdot \nabla A - 
\frac { \d } { i \d \bar A } \right) \right]_\da
\langle \bar \vp, A, \bar A | 0 \rangle \nn\\
& = &  i \sqrt{2} \int \! d^3x \,
\left[ \bar \vp \left( i \vec \s \cdot \nabla A
+ \sqrt{ - \nabla^2 } A \right) \right]_\da
\langle \bar \vp, A, \bar A | 0 \rangle \nn\\
& = &  i \sqrt{2} \int \! d^3x \,
A \left( - i \partial_i \bar \vp \s^i + \sqrt{ - \nabla^2 } \bar \vp 
\right)_\da 
\langle \bar \vp, A, \bar A | 0 \rangle \nn\\
& = &  i \sqrt{2} \int \! 
\frac { d^3x d^3k }{ \sqrt{ ( 2 \pi )^3 } }
A e^{-i \vec k \cdot \vec x }
2 | \vec k | w^\dgr_\da ( \vec k , 2 ) \bar \vp( \vec k , 2 )
\langle \bar \vp, A, \bar A | 0 \rangle \nn\\
& = & 0
\eea
is proportional to a sum of squares of $ \bar \vp( \vec k , 2 ) $ 
and therefore vanishes.

\section{Abelian Supersymmetric Gauge Theories}
\subsection{The Vector Multiplet}
The action density of the 
supersymmetric $ U(1) $ gauge theory is
\beq
{\cal L} = - \fourth v_{mn} v^{mn} -i \bar \l \bs^m \partial_m \l
+ \half D^2 
\label {Labelian}
\eeq
in which $ v_n $ is the gauge field
and 
\beq
v_{mn}  = \partial_m v_n - \partial_n v_m 
\label {vmn}
\eeq
is the Maxwell field-strength tensor,
which is often written as $ F_{mn} $.
Because the gaugino or photino field $ \l $
is, like $ v_n $, in the adjoint representation,
its covariant derivative 
is the same as its ordinary derivative.
The theory thus describes two non-interacting free fields
after the constraint $ D = 0 $ 
on the auxiliary field $ D $ is implemented.
\par
The fields $ v_m $, $ \l $, and $ D $
form part of a vector supermultiplet.
One may make the other fields of the supermultiplet
vanish by performing a supergauge transformation
to the Wess-Zumino gauge.
Under a general supersymmetry transformation,
these extra fields do not remain zero.
But if one augments an arbitrary supersymmetry transformation
by a related supergauge transformation,
one may keep the extra fields zero and restore
the Wess-Zumino gauge.
The action density (\ref{Labelian}) changes
by at most a total derivative under such an augmented
supersymmetry transformation
\bea
\d v_m & = & - i \bar \l \bs^m \x + i \bx \bs^m \l \\
\d \l & = & \s^{mn} \x v_{mn} + i \x D \\
\d D & = & - \x \s^m \partial_m \bar \l - \partial_m \l \s^m \bx 
\label {absst}
\eea
and is invariant under the ordinary gauge transformation
\bea
\d v_m & = & \partial_m \omega \\
\d \l & = & 0 \\
\d D & = & 0 
\label {abgt}
\eea
in which the function $ \omega(x) $ is a scalar.

\subsection{The Fayet-Iliopoulos $D$ Term}
Because the auxiliary field $ D $ changes only
by a space-time derivative under the
supersymmetry transformation (\ref {absst})
and not at all under the gauge transformation (\ref {abgt}),
the space-time integral of $ D $ is invariant and may be
added to the action.  This extra piece $ \xi D $ in the action,
conventionally multiplied by the constant $ \xi $,
is known as the Fayet-Iliopoulos $D$ term.
\par
With the Fayet-Iliopoulos $D$ term, the action density
\beq
{\cal L} = - \fourth v_{mn} v^{mn} -i \bar \l \bs^m \partial_m \l
+ \half D^2 + \xi D 
\label {LabelianFID}
\eeq
entails for the auxiliary field $ D $ the constraint
\beq
D = - \xi ,
\label {Dxi}
\eeq
and may be written as 
\beq
{\cal L} = - \fourth v_{mn} v^{mn} -i \bar \l \bs^m \partial_m \l
- \half \xi^2 .
\label {LFDos}
\eeq
The energy density thus acquires the positive term
\beq
V = \half \xi^2 ,
\label {VFD}
\eeq
and supersymmetry is spontaneously broken.

\subsection{The General Abelian Gauge Theory}
The action density for a general supersymmetric abelian
gauge theory is
\bea
{\cal L} & = & - \fourth v_{mn} v^{mn} \label {avv} \\
&   & \mbox{}
-i \bar \l \bs^m {\partial}_m \l \label {all} \\
&   & \mbox{}
+ \half D ^2 \label {aDD} \\
&   & \mbox{}
+ \xi D \label {axD} \\
&   & \mbox{}
- \overline{ \cD_m A^i } \; \cD^{m} A^i \label {aAA} \\
&   & \mbox{}
- i \bar \psi^i \bs^m \cD_m \psi^i \label {apsipsi} \\
&   & \mbox{}
+ \bar F_i F_i \label {aFF} \\
&   & \mbox{}
- i \sqrt{2} g_i \left( \bar A_i \psi_i \l
- \bar \l \bar \psi_i A_i \right) \label {aapsil} \\
&   & \mbox{}
- g_i D \bar A_i A_i \label {aDAA} \\
&   & \mbox{}
+ m_{ij} \left( A_i F_j - \half \psi_i \psi_j \right)
+ \bar m_{ij} \left( \bar A_i \bar F_j
- \half \bar \psi_i \bar \psi_j \right) \label {aAF psipsi} \\
&   & \mbox{}
+ g_{ijk} \left( F_i A_j A_k - \psi_i \psi_j A_k \right) 
+ \bar g_{ijk} \left(
\bar F_i \bar A_j \bar A_k
- \bar \psi_i \bar \psi_j \bar A_k \right) ,
\label {aFAA psipsiA}
\eea
in which $ v_m $ is the gauge field,
$ \l $ is the gaugino field, and the chiral fields
$ \psi_i $ and $ A_i $ carry charge $ g_i $.
The indices $ i $, $ j $, and $ k $ which label
chiral multiplets with abelian charges $ g_i $, $ g_j $, and $ g_k $
are summed over.
The symmetric tensors $ m_{ij} $ and $ g_{ijk} $
must be invariant under the action of the gauge group.
The covariant derivatives\footnote{The
notation used here for the covariant derivatives
follows that of Weinberg~\cite{SWI,SWII}
in which the covariant derivative
of a field that annihilates a particle
of charge $ q $ is $ ( \p_m - i q A_m ) \psi $\@.}
are
\bea
\cD_m A_i & = & \partial_m A_i - ig_i v_m A_i \\
\cD_m \psi_i  & = & \partial_m \psi_i - ig_i v_m \psi_i \\
\cD_m \bar \psi_i  & = & \partial_m \bar \psi_i + ig_i v_m \bar \psi_i \\
v_{mn} & = & \partial_m v_n^a - \partial_n v_m^a \,.
\eea

\subsection{Super $ QED $}
An important example of an abelian gauge theory
is super $ QED $ with a Fayet-Iliopoulos
$ D $ term.  The action density is a special case
of the general abelian action density (\ref {avv}--\ref {aFAA psipsiA})
in which the index $ i $ runs from 1 to 2
\bea
{\cal L}_{sqed} & = & \mbox{} - \fourth v_{mn} v^{mn} 
-i \bar \l \bs^m {\partial}_m \l 
+ \half D ^2 
+ \xi D \nn\\
&   & \mbox{}
- \overline{ \cD_m A^i } \; \cD^{m} A^i 
- i \bar \psi^i \bs^m \cD_m \psi^i 
+ \bar F_i F_i \nn\\
&   & \mbox{}
- i \sqrt{2} g_i \left( \bar A_i \psi_i \l
- \bar \l \bar \psi_i A_i \right) 
- g_i D \bar A_i A_i \nn\\
&   & \mbox{}
+ m_{ij} \left( A_i F_j - \half \psi_i \psi_j \right)
+ \bar m_{ij} \left( \bar A_i \bar F_j
- \half \bar \psi_i \bar \psi_j \right) \label {Lsqed} 
\eea
with $ \psi^1 $ being the right-handed electron
and $ \psi^2 $ being the right-handed positron.
Thus $ g_1 = - q < 0 $ and
$ \cD_m \psi^1 = ( \partial_m + i q v_m ) \psi^1 $;
$ g_2 = q > 0 $ and
$ \cD_m \psi^2 = ( \partial_m - i q v_m ) \psi^2 $;
and the only non-zero values of the symmetric tensor
$ m_{ij} $ are $ m_{12} = m_{21} = m $\@.
\par
We may also write this action density in terms of 
the right- and left-handed electron fields $ e_R = \psi^1 $ 
and $ \bar e_L = \bar \psi^2 $\@.
By using the identity (\ref{trans}),
integrating by parts, and dropping surface terms,
we see that the action density
\beq
-i \bar \psi^2 \bs^m \cD_m \psi^2 
= -i \bar \psi^2 \bs^m ( \partial_m - i q v_m ) \psi^2
\eeq
is equivalent to
\beq
-i \psi^2 \s^m \cD_m \bar \psi^2 
= -i \psi^2 \s^m ( \partial_m + i q v_m ) \bar \psi^2 . 
\eeq
So if we denote the right-handed selectron field  
as $ \t e_R = A^1 $ and the left-handed selectron field
as $ \t e_L = \bar A^2 $, then we may write 
the action density (\ref {Lsqed}) in the form
\bea
{\cal L}_{sqed} & = & \mbox{} - \fourth v_{mn} v^{mn} 
-i \bar \l \bs^m {\partial}_m \l 
+ \half D ^2 
+ \xi D \nn \\
&   & \mbox{}
- \overline{ \cD_m \t e_R } \; \cD^{m} \t e_R 
- \overline{ \cD_m \t e_L } \; \cD^{m} \t e_L 
- i \bar e_R \bs^m \cD_m e_R
-i e_L \s^m \cD_m \bar e_L \nn\\
&   & \mbox{}
+ \overline{ \g e_R } \g e_R 
+ \overline{ \g e_L } \g e_L 
+ i \sqrt{2} q \left( \overline{ \t e_R } e_R \l
- \bar \l \bar e_R \t e_R \right) \nn\\
&   & \mbox{}
- i \sqrt{2} q \left( \t e_L e_L \l
- \bar \l \bar e_L \overline{ \t e_L } \right) 
+ e D ( \overline{ \t e_R } \t e_R - \overline{ \t e_L } \t e_L ) \nn \\
&   & \mbox{}
+ m \left( \t e_R \overline{ \g e_L } 
+ \overline{ \t e_L } \g e_R  
- e_R e_L \right) + \bar m \left(  
\overline{ \t e_R } \g e_L  
+ \t e_L \overline{ \g e_R }  
- \bar e_R \bar e_L \right)
\label {Esqed} 
\eea
in which we have used for the auxiliary fields of the
right- and left-handed electron the notation
$ \g e_R = F_1 $ and $ \g e_L = \bar F_2 $\@. 
Because all the matter fields 
in this expression 
have negative charge $ -q $,
all the covariant derivatives in it are 
$ \cD_m = {\partial}_m + i q v_m $\@.
It is clear that we may choose the phase of the
fields $ e_R $, $ \g e_R $, and $ \t e_R $
so as to render the mass parameter $ m $
real and non-negative whatever the arbitrary phases of
the corresponding left-handed-electron fields
might be.  So from now on we shall take $ m \ge 0 $.
\par
The constraints on the auxiliary fields are:
\bea
D & = &  - \xi - q \left( | \t e_R |^2 - | \t e_L |^2 \right) \nn\\
\g e_R & = & - m \t e_L \nn\\
\g e_L & = & - m \t e_R ,
\label {De1e2are}
\eea
and when these constraints are enforced the action density
takes the form
\bea
{\cal L}_{sqed} & = & \mbox{} - \fourth v_{mn} v^{mn} 
-i \bar \l \bs^m {\partial}_m \l \nn \\
&   & \mbox{}
- \overline{ \cD_m \t e_R } \; \cD^{m} \t e_R 
- \overline{ \cD_m \t e_L } \; \cD^{m} \t e_L 
- i \bar e_R \bs^m \cD_m e_R
-i e_L \s^m \cD_m \bar e_L \nn\\
&   & \mbox{}
+ i \sqrt{2} q \left( \overline{ \t e_R } e_R \l
- \bar \l \bar e_R \t e_R \right) 
- i \sqrt{2} q \left( \t e_L e_L \l
- \bar \l \bar e_L \overline{ \t e_L } \right) \nn\\
&   & \mbox{}
- m ( e_R e_L - \bar e_R \bar e_L ) \nn\\
&   & \mbox{}
- \half \left[ \xi + q \left( | \t e_R |^2 - | \t e_L |^2 \right) 
\right]^2 - m^2 \left( | \t e_R |^2 + | \t e_L |^2 \right) .
\label {Csqed} 
\eea
Thus the scalar potential is
\beq
V = \half \left[ \xi + q \left( | \t e_R |^2 - | \t e_L |^2 \right) 
\right]^2 + m^2 \left( | \t e_R |^2 + | \t e_L |^2 \right) .
\label {Vsqed}
\eeq
If $ \xi = 0 $, but $ m \ne 0 $, 
then the mean values $ \langle \t e_R \rangle $ 
and $ \langle \t e_L \rangle $
of the selectron fields in the vacuum vanish, 
and neither supersymmetry nor gauge symmetry is broken.
If $ \xi \ne 0 $, but $ m = 0 $, 
then gauge symmetry, but not supersymmetry, is broken
in a three-parameter valley of vacua which satisfy 
\beq
\langle \Omega | \left( | \t e_L |^2 - | \t e_R |^2 \right) | \Omega \rangle
= \frac { \xi } { q } \,.
\label {Valsqed}
\eeq
When both $ \xi $ and $ m $ are non-zero
with $ | m |^2 > | \xi q | $,
then the mean values $ \langle \t e_R \rangle $
and $ \langle \t e_L \rangle $ both vanish in the vacuum,
and gauge symmetry is exact, but
the mean value of the potential in the vacuum is
\beq
\langle V \rangle = \half \xi^2 ,
\eeq
and supersymmetry is broken. 
On the other hand,
when both $ \xi $ and $ m $ are non-zero
with $ m^2 < | \xi q | $,
then for $ \xi > 0 $
there is a one-parameter ring
of vacua $ \Omega $ in which 
the mean values of the scalar fields satisfy
\bea
\langle \Omega | \t e_R | \Omega \rangle & = & 0 \\
| \langle \Omega | \t e_L | \Omega \rangle | & = &
\sqrt{ \frac { \xi } { q } - \frac { m^2 }{ q^2 } } \,.
\label {selvev}
\eea
The $ U(1) $ gauge symmetry is broken because
this mean value gives mass to the gauge boson $ v_m $,
which absorbs the derivative
of the phase of the left-handed selectron field
as its longitudinal component.
But the overall phase of the left-handed selectron field
is still arbitrary; we now choose it so that
its mean value in the vacuum is non-negative:
\beq
    h = \langle \Omega | \t e_L | \Omega \rangle \ge 0.
\label {rselvev}
\eeq
For $ \xi < 0 $ one may 
use these formulae 
provided one interchanges 1 with 2 and replaces
$ \xi $ by $ |\xi|$.
In these vacua the mean value of the potential is
\beq
\langle V \rangle = \frac { m^2 |\xi| }{ q }
- \frac { m^4 }{ 2q^2 } > 0 ,
\eeq
and so both gauge symmetry 
and supersymmetry are broken.
\subsection{The Goldstino}
When supersymmetry is spontaneously broken,
some of the fermions acquire masses along
with some of the gauge bosons,
but at least one of the fermions remains massless.
This massless fermion is called the goldstino.
We may illustrate this effect by computing 
the tree-level masses of the various particles
of super $QED$ for $ \xi > 0 $ and $ m > 0 $\@. 
\par
In the case $ m^2 > \xi q > 0 $, 
the mean values of the selectron fields 
in the vacuum vanish. 
Thus the gauge symmetry is unbroken,
and the gauge boson $ v_m $ remains massless.
Also there is no mixing between the electrons $ e_L $ and $ e_R $
and the photino $ \l $, and so the photino
remains massless:  it is the goldstino.
The right- and left-handed electron fields
form a Dirac electron of mass 
\beq
m_e = m .
\eeq
The scalar potential $ V $ contains the mass terms
\beq
\left( m^2 + \xi q \right) | \t e_R |^2
+ \left( m^2 - \xi q \right)  | \t e_L |^2 .
\label {c1masses}
\eeq
Because in (\ref {Csqed}) the kinetic action
of the complex selectron fields contains no
prefactor of $ 1/2 $, we may identify 
the masses of the selectrons as
\beq
m^2_{ \t e_R } =  m^2 + \xi q 
\eeq
and
\beq
m^2_{ \t e_L } =  m^2 - \xi q .
\eeq

\par
In the case $ \xi q > m^2 > 0 $, 
both gauge symmetry and supersymmetry are broken.
The simplest mass to identify is that of the gauge boson $ v_m $.  
In the action density (\ref {Csqed}), the kinetic term
of the left-handed selectron,
$ - \overline{ \cD_m \t e_L } \; \cD^{m} \t e_L $,
contains the mass term of the gauge boson  
\beq
 - q^2 h^2 v_m v^m 
= - \frac {1}{2} m_v^2 v_m v^m 
\label {m_v}
\eeq
which arises from the mean value 
\beq
h = \langle \Omega | \t e_L | \Omega \rangle 
= \sqrt{ \frac { \xi } { q } - \frac { m^2 }{ q^2 } } \ge 0 
\eeq
of the left-handed selectron
which is given by (\ref {selvev})
and which we have chosen to be real and non-negative
(\ref {rselvev})\@.
Thus the mass squared of the gauge boson $ v_m $ is
\beq
m_v^2 = 2 q^2 h^2 = 2 q \xi - 2 m^2 ,
\label {vqedmass}
\eeq
the $U(1)$ gauge symmetry is broken, and charge is not conserved.
\par
To find the masses of the selectrons,
we write the left-handed selectron as
\beq
\t e_L = h + r 
\label {unigauge}
\eeq
in which the field $ r $ is hermitian.
Then by expressing formula (\ref {Vsqed}) for the potential $ V $
in terms of $ r $, $ h $, and $ \t e_R $,
we see that 
the terms linear in $ r $ cancel by (\ref {selvev}),
and that $ V $ contains the mass terms 
\beq
2 m^2 | \t e_R |^2 + 2 q^2 h^2 r^2 
= m_{\t e_R}^2 | \t e_R |^2 + m_r^2 r^2 .
\label {selmass}
\eeq
Because in (\ref {Csqed}) the kinetic action
of the complex selectron fields contains no 
prefactor of $ 1/2 $, there is none in this equation.
Thus the mass squared of the right-handed selectron is
\beq
m_{\t e_R}^2 = 2 m^2 ,
\label {eRmass}
\eeq
and that of the radial part $r$ of the left-handed selectron is
\beq
m_r^2 =  2 q^2 h^2 = 2 q \xi - 2 m^2 = m_v^2.
\label {rmass}
\eeq
\par
To find the masses of the fermions of this model,
we write down their linearized equations of motion
by differentiating the action density:
\bea
i \bs^m \partial_m \l & = & i \sqrt{2} q h \, \bar e_L\nn\\
i \bs^m \partial_m e_R & = & - m \, \bar e_L \nn\\
i \bs^m \partial_m e_L & = & i \sqrt{2} q h \, \bl 
- m \, \bar e_R \nn\\
i \s^m \partial_m \bl & = & - i \sqrt{2} q h \, e_L \nn\\ 
i \s^m \partial_m \bar e_R & = & - m \, e_L \nn\\
i \s^m \partial_m \bar e_L & = &  - i \sqrt{2} q h \, \l
- m \, e_R .
\label {lineqssqed}
\eea
The Pauli identity (\ref {Pauli}) 
implies that the product of the two differential 
operators in the above sextet of equations is
the operator of Le Rond d'Alembert
\beq
i \s^m \partial_m i \bs^n \partial_n = \eta^{mn} \partial_m  \partial_n
= \Box .
\label {dalam}
\eeq
Thus the second-order linearized field equations are:
\bea
\Box \, \l & = & 2 q^2 h^2 \, \l 
- i \sqrt{2} q h m \, e_R = m_v^2 \, \l - i m_v m \, e_R \nn\\
\Box \, e_R & = &  i \sqrt{2} q h m \, \l 
+ m^2 \, e_R = i m_v m \, \l + m^2 \, e_R \nn\\
\Box \, e_L & = & \left( m^2 + 2 q^2 h^2 \right) \, e_L 
= \left( m^2 + m_v^2 \right) \, e_L .
\label {seclineqssqed}
\eea
It is clear that the squared mass of $ e_L $ is
\beq
m^2_{e_L} = m^2 + m_v^2 .
\label {mel}
\eeq
The eigenvalues of the matrix of squared masses
\beq
\pmatrix{  m_v^2 & - i m_v m \cr
            i m_v m &  m^2 }
\label {masmat}
\eeq
are $ 0 $ and 
\beq
m_{e'}^2 = m^2 + m_v^2 = m_{e_L}^2 .
\label {me'}
\eeq
The massive eigenvector is the field $ e'_R $
\beq
e'_R = \frac { m_v \, \l + i m \, e_R }
{ m_{e'} }
\label {e'}
\eeq
which with the field $ \bar e'_L = \bar e_L $ 
forms a Dirac spinor $ e' $ of mass $ m_{e'} $\@.
It is interesting to note that although
the coupling of the left-handed electron 
to the massive photon is $ - q $,
that of the right-handed electron is less 
\beq
-q_R = -q \frac { m } { m_{e'} }
= \frac { -q } { \sqrt{ 1 + \frac { m_v^2 } { m^2 } } } 
= -q \frac { m } { \sqrt{ 2 \xi q - m^2 } }
\eeq
and small if $ \xi q >> m^2 $\@.
\par
The massless eigenvector is the Goldstone spinor or goldstino
\beq
\psi = \frac { i m \, \l + m_v \, e_R } 
{ m_{e'} } .
\label {goldstino}
\eeq
\par
Let us add up the squared masses of the bosons
and subtract the squared masses of the fermions;
we find
\bea
\sum_i (-1)^{2j_i} (2j_i+1) m_i^2 & = & 2 m^2_{\t e_R} + m_r^2
- 4 m_{e'}^2 + 3 m_v^2 \cr
& = & \mbox{}
4 m^2 + m_v^2 - 4 ( m^2 + m_v^2 ) + 3 m_v^2 \cr
& = & \mbox{}
0 ,
\label {exItmass}
\eea
which is an example of the remarkable formula
of S.~Ferrara, L.~Girardello, and F.~Palumbo~\cite{italians}.   
\par
The previous case in which $ m^2 > \xi q > 0 $
provides a second example of the vanishing of the 
super-trace of the squared masses.
In this case the masses are:  $ m_v = m_\l = 0 $,
$ m_e = m $, $ m^2_{ \t e_R } = m^2 + \xi q $,
and $ m^2_{ \t e_L } = m^2 - \xi q $\@.
Thus the supertrace again vanishes:
\bea
\sum_i (-1)^{2j_i} (2j_i+1) m_i^2 & = &
2 m^2_{ \t e_R } + 2 m^2_{ \t e_L } - 4 m_e^2 \cr
& = &
2 ( m^2 + \xi q ) + 2 ( m^2 - \xi q )
- 4 m^2 \cr
& = & 0 .
\eea

\section{Non-Abelian Susy Gauge Theories}

\subsection{The Non-Abelian Vector Multiplet}
For an arbitrary non-abelian gauge group,
the action density of the supersymmetric pure gauge theory is
\beq
{\cal L} = - \fourth v^a_{mn} v_a^{mn} 
-i \bar \l^a \bs^m {\cal D}_m^{ab} \l^b 
+ \half D^a D^a 
\label {LpurenonA}
\eeq
in which the indices $ a, b,$ and $c$ here
represent Yang-Mills indices in the adjoint
representation of the gauge group.
The term ``pure'' means that the gauge fields
$ v^a_m $ and $ \l^a $ are not coupled 
to ``matter'' fields, and that there are no
matter fields in the theory.
The Yang-Mills field strength $ v^a_{mn} $ is
\beq
v^a_{mn} = \partial_m v_n^a - \partial_n v_m^a
+ g t^{abc} v_m^b v_n^c ,
\label {vamn}
\eeq
and the covariant derivative of the gaugino
field $ \l^a $ is
\beq
\cD_m^{ab} \l^b = \partial_m \d_{ab} \l^b + g t^{acb} v_m^c \l^b ,
\label {Dl}
\eeq
in which the real numbers $ t^{abc} $ are the structure
constants of the gauge group and therefore
also the generators of the adjoint representation.
\par
The action density (\ref{LpurenonA})
changes by at most a total derivative
under the augmented supersymmetry transformation
\bea
\d v_m^a & = & - i \bar \l^a \bs^m \x + i \bx \bs^m \l^a \\
\d \l^a & = & \s^{mn} \x v_{mn}^a + i \x D^a \\
\d D^a & = & - \x \s^m \cD_m^{ab} \bar \l^b - \cD_m ^{ab}\l^b \s^m \bx.
\eea
\par
The constraints imposed on the auxiliary fields $ D^a $ 
by the action density (\ref {LpurenonA}) are
\beq
D^a = 0 .
\eeq
\subsection{Adding Matter Multiplets}
\par
One may add to the action density (\ref{LpurenonA})
terms for a right-handed chiral multiplet
\bea
{\cal L}_R & = & \mbox{} - \overline{ \cD_m^{ij} A_j } \; \cD^{mik} A_k
- i \bar \psi_i \bs^m \cD_m^{ij} \psi_j
+ \bar F_i F_i \nn\\
&   & \mbox{}
- i \sqrt{2} g \left( \bar A_i T^a_{ij} \psi_j \l^a
- \bar \l^a \bar \psi_i T^a_{ij} A_j \right)
- g D^a \bar A_i T^a_{ij} A_j \label {rhcm}
\eea
in which the covariant derivatives are:
\bea
\cD_m^{ij} A_j & = & 
\delta_{ij} \partial_m A_j - ig v^a_m T^a_{ij} A_j 
\label {cDbR} \\
\cD_m^{ij} \psi_j  & = & 
\delta_{ij} \partial_m \psi_j - ig v^a_m T^a_{ij} \psi_j ,
\label {cDcR}
\eea
and the hermitian matrices $ T^a_{ij} $ are the 
generators of the gauge group in the representation
according to which the fields $ \psi_j $ and $ A_i $
transform.
\par
From these formulae for the right-handed multiplet
$ (A_i, \psi_i, F_i ) $,
we may infer the terms for a left-handed multiplet
$ ( B_i , \bar \chi_i , G_i ) $\@.
By integrating by parts and using the identity (\ref {trans}),
we see that the action density
\beq
-i \bar \psi_i \bar \sigma^m 
( \delta_{ij} \partial_m -i g v^a_m T^a_{ij} ) \psi_j
\eeq
is equivalent to 
\beq
-i \psi_j \sigma^m ( \delta_{ij} \partial_m + i g v^a_m T^a_{ij} )
\bar \psi_i .
\eeq
Thus if we wish to define the 
covariant derivatives of the fields $  B_i $
and $ \bar \chi_i $ of the left-handed multiplet as
\bea
\cD_m^{ij} B_j & = & 
\delta_{ij} \partial_m B_j - ig v^a_m T^a_{ij} B_j 
\label {cDbL} \\
\cD_m^{ij} \bar \chi_j & = & \d_{ij} \partial_m \bar \chi_j 
- ig v^a_m T^a_{ij} \bar \chi_j
\label {cDcL}
\eea
which is similar to the definitions (\ref {cDbR}--\ref {cDcR})
for the covariant derivatives of the fields of the 
right-handed multiplet, then
we must replace $ T^a_{ij} $ 
by $ - T^a_{ji} = - T^{a*}_{ij} $
throughout the terms that refer to the
left-handed multiplet $ ( B, \bar \chi, G ) $\@.
This is permissible since the generators $ T^a $
and $ -T^{a*} $ have the same structure constants.
The action density for the left-handed 
multiplet $ ( B, \bar \chi, G ) $ is thus:
\bea
{\cal L}_L & = & \mbox{} 
- \overline{ \cD_m^{ij} B_j } \; \cD^{mik} B_k
- i \chi_i \s^m \cD_m^{ij} \bar \chi_j
+ \bar G_i G_i \nn\\
&   & \mbox{}
- i \sqrt{2} g \left( \bar \l^a \bar B_i T^a_{ij} \bar \chi_j 
- \chi_i T^a_{ij} B_j \l^a \right)
+ g D^a \bar B_i T^a_{ij} B_j . \label {lhcm}
\eea

\subsection{The General Non-Abelian Gauge Theory}
After field shifts to omit the terms linear
in the complex scalar fields
and after some integrations by parts,
the most general renormalizable, supersymmetric,
gauge-invariant Lagrange density is
\bea
{\cal L} & = & - \fourth v^a_{mn} v_a^{mn} \label {vv} \\
&   & \mbox{} 
-i \bar \l^a \bs^m {\cal D}_m^{ab} \l^b \label {ll} \\
&   & \mbox{} 
+ \half D^a D^a \label {DD} \\
&   & \mbox{}
+ \xi_a D^a \label {xD} \\
&   & \mbox{}
- \overline{ \cD_m^{ij} A^j } \; \cD^{mik} A^k \label {AA} \\
&   & \mbox{}
- i \bar \psi^i \bs^m \cD_m^{ij} \psi^j \label {psipsi} \\
&   & \mbox{}
+ \bar F_i F_i \label {FF} \\
&   & \mbox{}
+ \mu_i F_i + \bar \mu_i \bar F_i \label {mF} \\
&   & \mbox{}
- i \sqrt{2} g \left( \bar A_i T^a_{ij} \psi_j \l^a
- \bar \l^a \bar \psi_i T^a_{ij} A_j \right) \label {apsil} \\
&   & \mbox{}
- g D^a \bar A_i T^a_{ij} A_j \label {DAA} \\
&   & \mbox{}
+ m_{ij} \left( A_i F_j - \half \psi_i \psi_j \right) 
+ \bar m_{ij} \left( \bar A_i \bar F_j 
- \half \bar \psi_i \bar \psi_j \right) \label {AF psipsi} \\
&   & \mbox{}
+ g_{ijk} \left( F_i A_j A_k - \psi_i \psi_j A_k \right) 
+ \bar g_{ijk} \left(
\bar F_i \bar A_j \bar A_k 
- \bar \psi_i \bar \psi_j \bar A_k \right) ,
\label {FAA psipsiA}
\eea 
in which the indices $ a, b,$ and $c$ here 
are Yang-Mills indices in the adjoint
representation of the gauge group
and the indices $ i, j,$ and $k$ here are
Yang-Mills indices in an arbitrary
representation of the gauge group.
In the Fayet-Iliopoulos term (\ref {xD}),
the sum must be restricted to those generators
of the gauge group that commute with all the
generators of the gauge group.
(Such generators are said to form invariant abelian
subalgebras.)
In the O'Raifeartaigh term (\ref {mF}),
the sum must be restricted to those chiral fields
that are singlets under all gauge transformations,
\emph{i.e.,} to fields that are completely neutral
and interact only with gravity or with other fields
not in the present model.
The symmetric tensors $ m_{ij} $ and $ g_{ijk} $
must be invariant under the action of the
gauge group.
\par
The Yang-Mills terms are in vector notation:
\bea
\cD_m A & = & \partial_m A - ig v^a_m T^a A \\
\cD_m \psi  & = & \partial_m \psi - ig v^a_m T^a \psi \\
\cD_m^{ab} \l^b & = & \partial_m \l^a + g t^{abc} v_m^b \l^c \\
v_{mn}^a & = & \partial_m v_n^a - \partial_n v_m^a 
+ g t^{abc} v_m^b v_n^c 
\eea
in which $ v_m^a $ is the gauge field,
$ \l^a $ is the gaugino field,
$ \psi $ is the chiral Fermi field,
$ A $ is the chiral scalar field,
$ T^a $ is a generator of the gauge group
in the arbitrary representation, the 
$ t^{abc} $ are the structure constants of the gauge group
(and therefore also the generators of the adjoint
representation), and $g$ is the coupling constant 
of the gauge group.
\par
The first term (\ref{vv}) is the gauge-field action.
The second term (\ref{ll}) is the gauged kinetic action
of the gauginos.
The third term (\ref{DD}) is the auxiliary fields
of the vector multiplet.
The fourth term (\ref{AA}) is the gauged kinetic action
of the scalar fields (squarks and sleptons and Higgs).
The fifth term (\ref{psipsi}) is the gauged kinetic action
of the chiral Fermi fields (quarks and leptons and higgsinos).
The sixth term (\ref{FF}) is the auxiliary fields
of the chiral multiplet.
The seventh term (\ref{apsil}) is a Yukawa interaction
of the chiral scalar, chiral Fermi, and gaugino fields.
The eighth term (\ref{DAA}) couples the auxiliary fields
of the vector multiplet with the chiral scalar fields.
The ninth term (\ref{AF psipsi}) is a hermitian
combination of mass terms that couple the chiral
scalar and auxiliary fields and the chiral Fermi fields. 
The last term (\ref{FAA psipsiA}) is a hermitian
combination of Yukawa interactions among the chiral 
scalar and auxiliary fields and the chiral Fermi 
and scalar fields.
\par
Under the augmented \susy transformation 
\bea
\d A^i & = & \sqrt{2} \x \psi^i \\
\d \psi^i & = & i \sqrt{2} \s^m \bx \cD_m^{ij} A^j + \sqrt{2} \x F^i \\
\d F^i & = & i \sqrt{2} \bx \bs^m \cD_m^{ij} \psi^j 
- 2i g T^a_{ij} A^j \bx \bar \l^a \\
\d v_m^a & = & - i \bar \l^a \bs^m \x + i \bx \bs^m \l^a \\
\d \l^a & = & \s^{mn} \x v_{mn}^a + i \x D^a \\
\d D^a & = & - \x \s^m \cD_m^{ab} \bar \l^b - \cD_m ^{ab}\l^b \s^m \bx ,
\eea
the Lagrange density (\ref{vv}--\ref{FAA psipsiA})
changes only by a total derivative.  
The Yang-Mills group indices $ a, b, c, i, j, k $
are placed where they fit, and no distinction 
is made between raised and lowered Yang-Mills indices.
\par
The constraints are
\beq
0 = \frac { \partial \cL }{ \partial D^a }
= \xi_a + D^a - \sum g \bar A_i T^a_{ij} A_j
\label {0 = Da + }
\eeq
in which the sum is over all the matter multiplets
that couple to \( D^a \) as well as over \( i \) and \( j \),
\beq
0 = \frac { \partial \cL }{ \partial F_i }
= \mu_i + \bar F_i + m_{ij} A_j + g_{ijk} A_j A_k ,
\label {0 = Fbar + }
\eeq
and
\beq
0 = \frac { \partial \cL }{ \partial \bar F_i }
=  \bar \mu_i + F_i + \bar m_{ij} \bar A_j 
+ \bar g_{ijk} \bar A_j \bar A_k .
\label {0 = F + }
\eeq
By implementing these constraints,
we may remove the auxiliary fields $ D^a $, $ F_i $, and $ \bar F_i $
from the action density (\ref{vv}--\ref{FAA psipsiA}).
The resulting expression is the most general
supersymmetric, gauge-invariant Lagrange density
into which we must fit the minimal supersymmetric
standard model apart from terms that explicitly break 
supersymmetry:
\bea
{\cal L} & = & - \fourth v^a_{mn} v_a^{mn}
-i \bar \l^a \bs^m {\cal D}_m^{ab} \l^b 
- \overline{ \cD_m^{ij} A^j } \; \cD^{mik} A^k
- i \bar \psi^i \bs^m \cD_m^{ij} \psi^j \label {GLKs}\\
&   & \mbox{} 
- \half \left( \xi_a - \sum g \bar A_i T^a_{ij} A_j \right)^2
- i \sqrt{2} g \left( \bar A_i T^a_{ij} \psi_j \l^a
- \bar \l^a \bar \psi_i T^a_{ij} A_j \right) \label {GLIs}\\
&   & \mbox{} 
- \half m_{ij} \psi_i \psi_j - \half \bar m_{ij} \bar \psi_i \bar \psi_j
- g_{ijk} \psi_i \psi_j A_k - \bar g_{ijk} \bar \psi_i \bar \psi_j \bar A_k
\label {GLpsipsi}\\
&   & \mbox{}
- | \mu_i + m_{ij} A_j + g_{ijk} A_j A_k |^2.
\label {GLFF} 
\eea
The first line (\ref{GLKs}) consists of the kinetic
action densities of the gauge, gaugino, and matter fields.
These terms will be present in any SSM.
The only arbitrariness in them is the choice
of the gauge group and of the representations
into which we fit the known fields.
In the MSSM, the gauge group
is $ SU(3)_c \otimes SU(2)_L \otimes U(1)_Y $,
and the particles are the usual three generations
of 15 suspects, plus the gauge fields, plus two Higgs
doublets, plus all the superpartners.
The second line (\ref{GLIs}) contains the
quartic and Yukawa interaction terms that 
are required by susy and by the first line (\ref{GLKs}),
as well as the Fayet-Iliopoulos term 
(proportional to $ \xi_a $), which is absent
when the gauge group is semi-simple
(\emph{i.e.,} when its Lie group
has no invariant abelian subalgebras).
The third line (\ref{GLpsipsi}) contains completely 
arbitrary fermion mass terms and Yukawa interactions.
The fourth line (\ref{GLFF}) consists of mass terms 
for the scalar matter fields and of
cubic and quartic self-interactions among these fields; 
these terms are dependent upon
the preceding line (\ref{GLpsipsi}) which is 
itself wholly arbitrary, and upon the O'Raifeartaigh term
(proportional to $ \mu_i $), 
which occurs only when there is a chiral field
that is completely neutral.  
Yet the gauge group of the standard model is not 
semi-simple; and if neutrinos are Dirac fermions,
then the right-handed neutrino is a 
chiral field that is completely neutral.
\par
One of the principal benefits of supersymmetry
is the cancellation of quartic and quadratic divergences.
The most obvious of the obviated divergences
are the first two terms of the zero-point energy 
\beq
E_0 = \sum_{i} ( - 1 )^{2j} ( 2j + 1 )
\int \! \dck \, \half \sqrt{ k^2 + m_i^2 }
\label {E_0}
\eeq
in which $m_i$ is the mass of particle $i$ of spin $j$.
If
\beq
\sum_{i} ( - 1 )^{2j} ( 2j + 1 ) = 0 ,
\label {inf4}
\eeq
then the quartic divergence cancels; if
\beq
\sum_{i} ( - 1 )^{2j} ( 2j + 1 ) m_i^2 = 0 ,
\label {inf2}
\eeq
then the quadratic divergence also cancels.
When \susy is a symmetry of the action
but not a symmetry of the vacuum,
the cancellations (\ref {inf4}) and (\ref {inf2})
in general persist at least at tree level as shown by
Ferrara, Girardello, and Palumbo~\cite{italians}\null.
\par
Other sum rules similar to (\ref {inf2})
but specific to a single chiral supermultiplet  
also hold when \susy is broken spontaneously.
These more specific sum rules 
may be incompatible with observed masses,
and theories that evade them
--- such as ones with a Fayet--Iliopoulos $D$-term,
\emph{e.g.} (\ref {LabelianFID}), ---
often have their own problems.
But by introducing mirror fermions,
so that every left-handed fermion
has a right-handed partner with
the same behavior under gauge transformations,
one may be able to construct
a theory that respects the sum rules (\ref {inf4})
and (\ref {inf2}) and still has 
an acceptable phenomenology.
\par
Most phenomenologists, however, have resorted to the 
use of terms that explicitly break \susy
but in ways that do not give rise to new,
field-dependent quadratic divergences. 
The possible terms that can be added
to break \susy softly are Majorana mass terms
for the gauginos
\beq
\d \cL_m = 
- \half m \, \l \l - \half \bar m \, \bl \bl ,
\label {gaugino mass term}
\eeq  
cubic analytic polynomials in the scalar fields
\beq
\d \cL_A = c_0 + c_i A_i + c_{ij} A_i A_j + c_{ijk} A_i A_j A_k 
+ \bar c_0 + \bar c_i \bar A_i + \bar c_{ij} \bar A_i \bar A_j
+ \bar c_{ijk} \bar A_i \bar A_j \bar A_k ,
\label {cubics}
\eeq
and quadratic forms in the scalar fields
\beq
\d \cL_{A^*A} = - m_{ij} \bar A_i A_j . 
\label {AbarA}
\eeq
Such terms can arise in the low-energy limit
of a theory in which local \susyp, \emph{ i.e.,} supergravity, 
is spontaneously broken at high energy.
\par
Some dimension-three terms 
do generate awkward quadratic divergences:
explicit mass terms for matter fermions
and cubic forms that mix scalar fields and their 
hermitian conjugates are not soft~\cite{GG}\null.
\par
It should be noted, however,
that the explicit breaking of \susy
entails a quadratic divergence 
in the zero-point energy.

\section{Superfield Notation}
\subsection{Superfields} 
Superfields are functions of 
space-time and Grassmann coordinates
$ x, \th, \bar \th $.
They provide an efficient way
of finding action densities
and of computing Feynman diagrams.
By using superfield notation,
one may symbolize compactly 
the structural parts of the action densities
that were described in the long
formulas of the preceding section.
But the superfield formalism is a very 
technical subject, and so I shall focus on its
use as notation. 
\par
Because Grassmann coordinates are anti-commuting numbers,
a general superfield \( F( x , \th , \bar \th ) \)
may be expressed as a polynomial in the Grassmann coordinates
with highest term \( \th \th \, \bar \th \bar \th \)
\bea
F( x , \th , \bar \th ) & = &
f(x) + \th \, \phi(x) + \bar \th \, \bar \chi(x) \nn\\
& & \mbox{} + \th \th \, m(x) + \bar \th \bar \th \, n(x)
+ \th \s^m \bar \th \, v_m(x) \nn\\
& & \mbox{} + \th \th \, \bar \th \bar \l (x)
+ \bar \th \bar \th \, \th \psi(x) 
+ \th \th \, \bar \th \bar \th \, d(x) .
\label {sF}
\eea 
This long, reducible expression involves nine
different fields.

\subsection{Chiral Superfields}
A simpler superfield is the 
right-handed chiral superfield \( \Phi \)
which is a function of $ x^m + i \th \s^m \bar \th $
and $ \th $
\bea
\Phi ( x^m + i \th \s^m \bar \th , \th ) & = &
A(x) + i \th \s^m \bar \th \, \p_m A(x) 
+ \fourth \th \th \, \bar \th \bar \th \, \Box A(x) 
\qquad \qquad \nn\\
& & \mbox{} + \sqrt{2} \, \th \psi(x)
- { i \over \sqrt{2} } \th \th \, \p_m  \psi(x) \s^m \bar \th 
+ \th \th \, F(x) .
\label {Phi}
\eea
We recognize the fields of the right-handed 
chiral multiplet (\ref {L_K}).
In fact if we let $ y^m = x^m + i \th \s^m \bar \th $,
then we may write the superfield \( \Phi \) in the form
\beq
\Phi ( y , \th ) = 
A( y ) + \sqrt{2} \, \th \psi( y ) 
+ \th \th \, F( y ),
\label {Phy}
\eeq
in which the argument of the auxiliary field $ F $
could as well be $ x $.
\par
The adjoint superfield $ \Phi^\dgr $
is a left-handed superfield 
which is a function of $ x^m - i \th \s^m \bar \th $
and $ \bar \th $
\bea
\Phi^\dgr ( x^m - i \th \s^m \bar \th , \bar \th ) & = &
A^\dgr(x) - i \th \s^m \bar \th \, \p_m A^\dgr(x)
+ \fourth \th \th \, \bar \th \bar \th \, \Box A^\dgr(x) 
\quad \qquad \nn\\
& & \mbox{} + \sqrt{2} \, \bar \th \bar \psi(x)
+ { i \over \sqrt{2} } \bar \th \bar \th \, \th \s^m \p_m \bar \psi(x) 
+ \bar \th \bar \th \, F^\dgr(x) .
\label {Phi+}
\eea
With $ z^m = x^m - i \th \s^m \bar \th $,
the adjoint superfield $ \Phi^\dgr $ may be
written as
\beq
\Phi^\dgr ( z , \bar \th ) =
A^\dgr(z) + \sqrt{2} \, \bar \th \bar \psi(z) 
+ \bar \th \bar \th \, F^\dgr(z).
\label {Phz}
\eeq
\par
One of the neat things about superfields
is that the supersymmetry transformations
(\ref {d A}--\ref {d bar F}), \emph{etc.}, 
can be thought of as translations in superspace
\bea
{x'} ^{ m} & = & x^m + i \th \s ^m \bx -i \x \s ^m \bar \th \nn\\
\th' & = & \th + \x \nn\\
\bar \th' & = & \bar \th + \bx .
\label {SS}
\eea
For instance after this translation,
the part of $ \Phi $ that is independent of
$ \th $ and of $ \bar \th $ is
\beq
A(x) + \d A(x) = A(x) + \sqrt{2} \, \x \psi (x) ;
\label {sdA}
\eeq
the part that depends linearly upon $ \th $ is
\beq
\sqrt{2} \, \th \left( \psi(x) + \d \psi(x) \right)
= \sqrt{2} \, \th \left( \psi(x) + i \sqrt{2} \, 
\s^m \bx \p_m A(x) + \sqrt{2} \, \x F (x) \right) ,
\label {sdpsi}
\eeq
in which the second term of $ \d \psi $ has
arisen both from the translation of the argument
$ x $ of $ A(x) $ and from the translation of $ \bth $;
and the part that depends quadratically upon $ \th $ is
\beq
\th \th \left( F (x) + \d F (x) \right) =
\th \th \left( F (x) + i \sqrt{2} \, \bx \bs^m \p_m \psi \right),
\label {sdF}
\eeq
in which the change $ \d F (x) $ 
has arisen both from the translation of the argument $ x $
of $ \psi(x) $ and from the translation of $ \bth $\null.
\par
Under a susy transformation, the
auxiliary field $ F ( x ) $ 
of a chiral superfield 
changes only by a total derivative,
and its space-time integral is a susy invariant.  
Such a term, stripped of Grassmann variables,
may therefore be used as an invariant part
of a \susyc action density. 
Many susy invariants are of this form.
\par
The product of two right-handed chiral superfields 
\bea
\Phi_1( y , \th ) \Phi_2( y , \th ) & = &   
\left[ A_1 (y) + \sqrt{2} \th \psi_1 (y) + \th \th F_1 (y) \right]
\nn\\
& & \times 
\left[ A_2 (y) + \sqrt{2} \th \psi_2 (y) + \th \th F_2 (y) \right]
\eea
is a right-handed chiral superfield:
\bea
\Phi_1( y , \th ) \Phi_2( y , \th ) 
& = & A_1 (y) A_2 (y) + \sqrt{2} \th \left[ 
A_1 (y) \psi_2 (y) + A_2 (y) \psi_1 (y) \right]
\nn\\
& & \mbox{} + \th \th \left[
A_1 (y) F_2 (y) + A_2 (y) F_1 (y) - \psi_1 (y) \psi_2 (y) \right]
\quad
\label {prod2chis}
\eea
in which the identity (\ref {thth})
was used to write $ 2 \th \psi_1 \th \psi_2 $ 
as $ - \th \th \psi_1 \psi_2 $\null. 
Similarly the product of two left-handed chiral superfields
\bea
\Phi_1^\dgr( z , \bth ) \Phi_2^\dgr( z , \bth ) & = &
\left[ A_1^\dgr (z) + \sqrt{2} \bth \psi_1^\dgr (z) 
+ \bth \bth F_1^\dgr (z) \right] \nn\\
& & \times
\left[ A_2^\dgr (z) + \sqrt{2} \bth \psi_2^\dgr (z) 
+ \bth \bth F_2 (z)^\dgr \right]
\eea
is a left-handed chiral superfield:
\bea
\Phi_1^\dgr( z , \bth ) \Phi_2^\dgr( z , \bth )
& = & A_1^\dgr (z) A_2^\dgr (z) + \sqrt{2} \bth \left[
A_1^\dgr (z) \psi_2^\dgr (z) + A_2^\dgr (z) \psi_1^\dgr (z) \right]
\nn\\
& & \mbox{} + \bth \bth \left[
A_1^\dgr (z) F_2^\dgr (z) + A_2^\dgr (z) F_1^\dgr (z) 
- \psi_1^\dgr (z) \psi_2^\dgr (z) \right] . \quad
\label {prod2chibars}
\eea

Thus under a \susy transformation,
the $ \th \th $ component of the product of two chiral superfields 
\beq
\Phi_i \Phi_j |_{ \th \th } =
A_i F_j + A_j F_i - \psi_i \psi_j 
= A_i F_j + A_j F_i 
- \thalf \left( \psi_i \psi_j + \psi_j \psi_i \right)
\label {PhiPhi}
\eeq
changes only by a total derivative,
and its space-time integral therefore is a susy invariant,
one encountered earlier (\ref {AF psipsi})\null.
Similarly the $ \th \th $ component
of the product of three chiral superfields is
a susy invariant:
\beq
\Phi_i \Phi_j \Phi_k |_{ \th \th } =
F_i A_j A_k + F_j A_k A_i + F_k A_i A_j
- \psi_i \psi_j A_k - \psi_j \psi_k A_i - \psi_k \psi_i A_j ,
\label {PhiPhiPhi}
\eeq
which occurs in (\ref {FAA psipsiA})\null.
Such $ \th \th $ components sometimes are called
``F terms'' because the $ \th \th $ component 
of the superfield $ \Phi $ is $ F $,
as shown by (\ref {Phi}). 
\par
Chiral superfields commute, and so
the expressions (\ref {PhiPhi}) and (\ref {PhiPhiPhi})
for $ \Phi_i \Phi_j $ and $ \Phi_i \Phi_j \Phi_k $
are symmetric in $ i, j $ and in $ i, j, k $, respectively.
This symmetry is the reason why 
the matrices $ m $ and $ g $ the superpotential (\ref {W})
are symmetric.
It follows that combinations like $ \e^{ij} \Phi_i \Phi_j $ vanish.  
But one may construct ``anti-symmetric" forms like
\beq
\e^{ij} \Phi_i \Psi_j = \Phi_1 \Psi_2 - \Phi_2 \Psi_1
\label {PhiPsi}
\eeq
that do not vanish.
At the risk of overemphasizing this point, 
which I find confusing, let us 
distinguish the two superfields $ \Phi $ and $ \Psi $
by indices and write the preceding expression as
\beq
\e^{ij} \Phi_i \Psi_j = \e^{ij} \Phi_{ 1 i } \Phi_{ 2 j } .
\eeq
Then the $ A F $ part of this ``anti-symmetric" form is
\beq
A_{ 1 1 } F_{ 2 2 } + F_{ 1 1 } A_{ 2 2 }
- A_{ 1 2 } F_{ 2 1 } - F_{ 1 2 } A_{ 2 1 } 
= A_{ 1 1 } F_{ 2 2 } + A_{ 2 2 } F_{ 1 1 } 
- A_{ 1 2 } F_{ 2 1 } - A_{ 2 1 } F_{ 1 2 }
\eeq
which is anti-symmetric under the interchange
of the first indices and of the second indices
but is symmetric under the interchange of the
double indices.
\par
The kinetic part of the chiral action density (\ref {LCGK})
apart from total derivatives
is the $ \th \th \, \bth \bth $ component
of the product $ \Phi_i ^\dgr \Phi_i $
\beq
\Phi_i ^\dgr \Phi_i |_{ \th \th \, \bth \bth } =
\frac{ i }{ 2 } \partial_n \bar \psi_i \bar \s^n \psi_i
- \frac{ i }{ 2 } \bar \psi_i \bar \s^n \partial_n \psi_i
- \partial_n \bar A_i \partial^n A_i + \bar F_i F_i .
\label {Phi+Phi}
\eeq
Such $ \th \th \, \bth \bth $ components
often are called ``D terms'' because the
$ \th \th \, \bth \bth $ component
of the vector superfield $ V $
in the Wess-Zumino gauge
is $ D $, as shown by (\ref {V})\null.
\par
\subsection{Vector Superfields}
In what is called the Wess-Zumino gauge,
the vector superfield is 
\beq
V = \th \s^m \bth \, v_m(x) 
+ i \th \th \, \bth \bl (x) - i \bth \bth \, \th \l (x)
+ \half \th \th \, \bth \bth \, D(x) ,
\label {V}
\eeq
in which both $ v_m $ and $ D $ are hermitian
fields so that the superfield $ V $ is also hermitian.
\par
Superfield notation is less transparent 
for gauge fields than for chiral fields.
One first defines the differential operators
\beq
D_a = { \p \over \p \th^a } + i \s _{ a \da } ^m 
\bth ^\da { \p \over \p x^m }
\and
\bar D_\da = - { \p \over \p \bth^\da }
- i \th^a \s_{ a \da } ^m  { \p \over \p x^m } ,
\label {DbD}
\eeq
in which the differentiation with respect to
$ \th^a $ is from the left, \emph{e.g.,}
\beq
{ \p \over \p \th^a } \, \th \th = 
{ \p \over \p \th^a } \, \th^b \th_b = \th_a .
\eeq
In terms of the real superfield $ V $,
one then defines the chiral superfields
\beq
W_a = - \fourth \bar D \bar D \, D_a V
\and
\bar W_\da = - \fourth D D \, \bar D_\da V .
\eeq
The $ \th \th $ component of $ W W $ is
\beq
W W |_{ \th \th } =
- \half v^{mn} v_{mn} - 2i \l \s^m \p_m \bl 
+ D^2 + { i \over 4 } v^{mn} v^{lk} \e_{mnlk} ,
\label {WW}
\eeq
in which $ D $ is the auxiliary field,
not the differential operator,
and $ \e_{mnlk} $ is the totally
anti-symmetric tensor in four indices
with $ \e_{0123} = -1 $.
The meager notational payoff for these spectacularly
elaborate definitions is that we may now write
the abelian action density (\ref {Labelian})
in the form
\beq
\fourth \left( W^a W_a |_{ \th \th } 
+ \bar W_\da \bar W^\da |_{ \bth \bth } \right) = 
- \fourth v^{mn} v_{mn} - i \l \s^m \p_m \bl
+ \half D^2 .
\eeq
 
\section{The Minimal Supersymmetric Standard Model}
The gauge group of the standard model 
is $ SU(3)_c \otimes SU(2)_L \otimes U(1)_Y $
which is spontaneously broken to $ SU(3)_c \otimes U(1)_{em} $\@.
There are three families or generations of quarks and leptons,
each with 15 or 16 particles depending on whether
there are right-handed neutrinos.
The first family consists of 6 left-handed quarks 
which form a triplet under $ SU(3)_c $ and a doublet
under $ SU(2)_L $ and carry hypercharge $ Y = 1/3 $,
quantum numbers which may be written as $ ( 3 , 2 , 1/3 )$\@.
The hypercharge $ Y $ is chosen so as to satisfy
the Gell-Mann--Nishijima relation
\beq
Q = I_3 + \half Y .
\label {GM-N}
\eeq
The other 6 first-family quarks form
a right-handed color triplet $ u_R $
that is a singlet under $ SU(2)_L $ and has $ Y = 4/3 $
and a second right-handed color triplet $ d_R $
that is also a singlet under $ SU(2)_L $ and has $ Y = - 2/3 $\@. 
The leptons are colorless and form a left-handed
doublet under $ SU(2)_L $ with $ Y = - 1 $,
a right-handed $ SU(2)_L $ singlet $ e_R $
with $ Y = - 2 $, and possibly a 
right-handed  $ SU(2)_L $ singlet $ \nu_R $
with $ Y = 0 $\@.
These quantum numbers may be restated as
\beq
\begin {array}{ccccc}
\pmatrix{ \vec u_L \cr \vec d_L \cr } &\quad& \vec u_R &\quad& \vec d_R \\ 
\quad &\quad& \quad &\quad& \quad \\
( 3 , 2 , \frac{1}{3} ) &\quad& ( 3 , 1 , \frac{4}{3} ) 
&\quad& ( 3 , 1 , - \frac{2}{3} ) 
\end {array}
\eeq
for the quarks and as
\beq
\begin {array}{ccccc}
\pmatrix{ \nu_L \cr e_L \cr } &\quad& \nu_R &\quad& e_R \\ 
\quad &\quad& \quad &\quad& \quad \\
( 1 , 2 , - 1 ) &\quad& ( 1 , 1 , 0 ) &\quad& ( 1 , 1 , - 2 ) 
\end {array}
\eeq
for the leptons.
\par
The standard model agrees with all experiments,
except possibly in the neutrino sector
where it is easy to add right-handed neutrino spinors 
to make mass terms.
So in building the minimal supersymmetric standard model
(MSSM), one tries to rock this peaceful boat
as little as possible.
It would be nice if we could put
the known particles into supersymmetric multiplets
without adding any super particles.
Alas, exactly the opposite is true:
nobody has identified a single pair 
of particles that are exchanged by \susyp. 
So \susy doubles the number of particles,
as well as the number of typos.
Thus to every particle, we attach a super particle.
For instance to the left-handed quark field $q_L$
we associate a ``left-handed'' squark field $\tilde{q}_L$, 
and to the right-handed quark field $d_R$,
we associate a ``right-handed'' field $\tilde{d}_R$\null.
Actually the situation is slightly worse:
we must add a second Higgs doublet and
its super partners.
\subsection{Why Two Higgs Superfields?}
There are three reasons why one must have two Higgs
doublets and the spartners.
The simplest reason is that the spartner
of the Higgs doublet $(h^+, h^0)$ of the standard model
is a doublet of fermions $(\t{h}^+,\t{h}^0)$.
These higginos have $Y=1$ and would generate a $U(1)$ anomaly
since the trace over all fermions would now be non zero,
\beq
\tr Y^3 = 2 .
\label {anomaly}
\eeq
One may solve this problem by adding
a second Higgs superfield
with spin-zero fields $(h^0, h^-)$ and higgsinos
$(\t{h}^0,\t{h}^-)$ with $Y=-1$.
Now we are back to
\beq
\tr Y^3 = 0 
\label {noanomaly}
\eeq
as in the standard model.
\par
The second reason why two Higgs fields are needed
is to give mass to the up quarks, $u$, $c$, and $t$.
The quark mass terms
of the standard model
are of the form
\beq
c_u \, \e_{ij} \, q_{Li} h_j \, u_R^\dgr
+ c_d \, h^\dgr_i q_{Li} \, d^\dgr_R + \hc,
\label {SMquarkmasses}
\eeq
in which $ h $ is the Higgs doublet $(\t{h}^+,\t{h}^0)$,
$ q_L $ is the doublet of left-handed quark fields
$ ( u_L, d_L ) $, and the family indices have been
suppressed. 
If we promote all the fields
of the first term to superfields,
using the conventional notation
$ H_2 = ( H_2^+ , H_2^0 ) $
and $ Q_L = ( U_L , D_L ) $,
then there is no problem with the 
mass term for the up quarks
\beq
\left. c_u \, \e_{ij} \, Q_{Li} H_{2j} \, U_R^\dgr \, \right|_{\bth\bth}
\label {MSSMupquarkmass}
\eeq
which is the product of three left-handed chiral superfields.
The problem is with the mass term for the down quarks.
A product of three superfields is invariant under \susy
only if all three superfields are of the same chirality. 
Now $ Q_{Li} $ and $ D_R^\dgr $ are left handed,
but $ H_{2i}^\dgr $ is right handed.
So one introduces a second left-handed doublet
of Higgs superfields $ H_1 = ( H_1^0 , H_1^- ) $
with $ Y = -1 $ and writes
the down-quark mass term as
\beq
\left. c_d \, \e_{ij} \, Q_{Li} H_{1j} \, D_R^\dgr \, \right|_{\bth\bth} .
\label {MSSMdownquarkmass}
\eeq
\par
The third reason why one needs two Higgs superfield doublets,
$H_1$ and $H_2$, is that the electroweak gauge bosons 
$W_i$ and $B$ of $SU(2)_L \otimes U(1)_Y$ are the spin-one
fields of the superfields $V_i$ and $V$.  Three of these
four massless gauge bosons become the massive vector bosons
$W^+$, $W^-$, and $Z$ by absorbing three massless spin-zero bosons;  
three of their massless spin-one-half superpartners $\t W_i$
and $\t B$ must therefore become massive 
by absorbing three massless spin-one-half fields.  
Were there only one Higgs superfield doublet,
there would only be two massless spin-one-half fields
as candidates for absorption.  With a second
Higgs superfield doublet, there is one 
massless spin-one-half field to spare. 
\par
After the spontaneous breakdown of $SU(2)_L \otimes U(1)_Y$
to $ U(1)_{EM} $, the physical fields are:
the massive charged $W^\pm$'s, the massive $Z$, and
the massless photon, all vector bosons;
three massive Dirac fields $\t W^\pm$, $ \t Z $,
and one massless neutral chiral spinor,
all spin-one-half fields;
and two charged and three neutral bosons,
all of spin zero.
Of course, these spin-one-half and spin-zero fields
may mix with other fields of the MSSM.

\subsection{The Electroweak Superpotential}
Since the rotation group, $ SU(3)_c $, and $ U(1)_{EM} $ 
are unbroken, the fields that might assume
non-zero mean values in the vacuum are
likely to be the spin-zero, colorless, and neutral. 
The left-handed purely electroweak chiral superfields of the MSSM 
are the two Higgs doublets $H_1$ and $H_2$,
the lepton doublet $L = ( N^0 , E^- )$,
and the adjoint $ E_R^\dgr $ of the superfield 
of the right-handed electron.
In view of the neutrino experiments
which are suggestive of neutrino masses,
it makes sense to include the adjoint $ N_R^\dgr $ 
of the superfield of the right-handed neutrino. 
Then avoiding the letter $W$ and
again suppressing family indices,
we may write the electroweak superpotential as
\bea
P & = & c_1 \, \e_{ij} \, H_{ 1 i } H_{ 2 j }
+ c_2 \, \e_{ij} \, L_{ i } H_{ 2 j }
+ c_3 \, \e_{ij} \, L_{ i } H_{ 1 j } E_R^\dgr \nn\\
& & \mbox{}
+ c_4 \, N_R^\dgr 
+ c_5 \, N_R^\dgr N_R^\dgr 
+ c_6 \, N_R^\dgr N_R^\dgr N_R^\dgr \nn\\
& & \mbox{}
+ c_7 \, \e_{ij} \, H_{ 1 i } H_{ 2 j } N_R^\dgr 
+ c_8 \, \e_{ij} \, L_i H_{ 2 j } N_R^\dgr .
\label {EPP}
\eea
Keeping in mind that $ \e_{12} = -1 $, 
one may expand this gauge-invariant and
susy-invariant superpotential 
in terms of the components of the several doublets 
\bea
P & = & c_1 \, \left( H_1^- H_2^+ - H_1^0 H_2^0 \right)
+ c_2 \, \left( E_L H_2^+ - N_L H_2^0 \right) \nn\\
& & \mbox{}
+ c_3 \, \left( E_L H_1^0 - N_L H_1^- \right) E_R^\dgr
+ c_4 \, N_R^\dgr 
+ c_5 \, N_R^\dgr N_R^\dgr 
\nn\\ & & \mbox{}
+ c_6 \, N_R^\dgr N_R^\dgr N_R^\dgr 
+ c_7 \, \left( H_1^- H_2^+ - H_1^0 H_2^0 \right) N_R^\dgr 
\nn\\ & & \mbox{}
+ c_8 \, \left( E_L H_2^+ - N_L H_2^0 \right) N_R^\dgr .
\label {EPPexp}
\eea
We may add to the action density the $\bth \bth$ component of $ P $
and its hermitian adjoint.
\par
One conventionally distinguishes
superpartners by tildes, writing the selectron as $\t e$.
Let us extend this notation
to auxiliary fields, distinguishing them with a grave accent,
$\g e$, so that the superfield of the left-handed electron
takes the form
\beq
E_L = \t e_L + \sqrt{2} \, \bth e_L + \bth \bth \, \g e_L .
\label {e_L}
\eeq 
Then the  bosonic part of the $\bth \bth$ component of $ P $ is 
\bea
\left. P \, \right|_{\bth\bth b} & = & 
c_1 \, \left( h_1^- \g h_2^+ + \g h_1^- h_2^+
- h_1^0 \g h_2^0 - \g h_1^0 h_2^0 \right) \nn\\ 
& & \mbox{} + c_2 \, 
\left( \t e_L \g h_2^+ + \g e_L h_2^+ 
- \t \nu_L \g h_2^0 - \g \nu_L h_2^0 \right)
\nn\\ & & \mbox{} + c_3 \, \left[
\left( \t e_L \g h_1^0 + \g e_L h_1^0 
- \t \nu_L \g h_1^- - \g \nu_L h_1^- \right)
\t e_R^\dgr 
+ \left( \t e_L h_1^0 - \t \nu_L h_1^- \right) \g e_R^\dgr \right] \nn\\
& & \mbox{} 
+ c_4 \, \g \nu_R^\dgr 
+ 2 c_5 \, \t \nu_R^\dgr \g \nu_R^\dgr 
+ 3 c_6 \, \t \nu_R^\dgr \t \nu_R^\dgr \g \nu_R^\dgr \nn\\
& & \mbox{} +  c_7 \, \left[
\left( h_1^- \g h_2^+ + \g h_1^- h_2^+ 
- h_1^0 \g h_2^0 - \g h_1^0 h_2^0 \right)
\t \nu_R^\dgr \right. \nn\\
& & \mbox{} \qquad \left.
+ \left( h_1^- h_2^+ - h_1^0 h_2^0 \right) \g \nu_R^\dgr \right] 
\nn\\ & & \mbox{}
+ c_8 \, \left[
\left( \t e_L \g h_2^+ + \g e_L h_2^+ 
- \t \nu_L \g h_2^0 - \g \nu_L h_2^0 \right)
\t \nu_R^\dgr \right. \nn\\
& & \mbox{} \qquad \left.
+ \left( \t e_L h_2^+ - \t \nu_L h_2^0 \right) \g \nu_R^\dgr \right] . 
\label {EWWexpcomp}
\eea
\par
Let us collect the terms of the action density
that involve the auxiliary field $\g \nu_R$
of the right-handed neutrino:
\bea
\cL_{\g \nu_R} & = & \g \nu_R^\dgr \g \nu_R
+ c_4 \, \g \nu_R^\dgr + \bar c_4 \, \g \nu_R
+ 2 c_5 \, \t \nu_R^\dgr \g \nu_R^\dgr
+ 2 \bar c_5 \, \t \nu_R \g \nu_R \nn\\ & & \mbox{}
+ 3 c_6 \, \t \nu_R^\dgr \t \nu_R^\dgr \g \nu_R^\dgr
+ 3 \bar c_6 \, \t \nu_R \t \nu_R \g \nu_R \nn\\ & & \mbox{}
+  c_7 \, \left( h_1^- h_2^+ - h_1^0 h_2^0 \right) \g \nu_R^\dgr
+  \bar c_7 \, 
\left( h_1^+ h_2^- - h_1^{0\dgr} h_2^{0\dgr} \right) \g \nu_R 
\nn\\ & & \mbox{}
+ c_8 \, \left( \t e_L h_2^+ - \t \nu_L h_2^0 \right) \g \nu_R^\dgr
+ \bar c_8 \, \left( \t e_L^\dgr h_2^- - \t \nu_L^\dgr h_2^{0\dgr}\right) 
\g \nu_R .
\label {nuRterms}
\eea
The constraint on $\g \nu_R$ is
\beq
\g \nu_R = - c_4 - 2 c_5 \t \nu_R^\dgr 
- 3 c_6 \left( \t \nu_R^\dgr \right)^2
- c_7 \left( h_1^- h_2^+ - h_1^0 h_2^0 \right) 
- c_8 \left( \t e_L h_2^+ - \t \nu_L h_2^0 \right) .
\label {gnuR}
\eeq
So the action density $ \cL_{\g \nu_R} $ becomes
\beq
\cL_{\g \nu_R} = - \left| 
c_4  + 2 c_5 \t \nu_R^\dgr + 3 c_6 \left( \t \nu_R^\dgr \right)^2
+ c_7 \left( h_1^- h_2^+ - h_1^0 h_2^0 \right) 
+ c_8 \left( \t e_L h_2^+ - \t \nu_L h_2^0 \right) \right|^2 .
\label {LnuR}
\eeq
\par
The part of the action density that involves
the auxiliary field $ \g h_1^0 $ is
\bea
\cL_{\g h_1^0} & = & 
\g h_1^{0\dgr} \g h_1^0  
- c_1 \, \g h_1^0 h_2^0 
- \bar c_1 \, \g h_1^{0\dgr} h_2^{0\dgr} 
+ c_3 \, \t e_L \t e_R^\dgr \, \g h_1^0 
+ \bar c_3 \, \t e_L^\dgr \t e_R \, \g h_1^{0\dgr} 
\nn\\ & & \mbox{}
- c_7 \, \g h_1^0 h_2^0 \, \t \nu_R^\dgr 
- \bar c_7 \, \g h_1^{0\dgr} h_2^{0\dgr} \, \t \nu_R .
\label {gh10terms}
\eea
So $ \g h_1^0 $ must be
\beq
\g h_1^0 = \bar c_1 \, h_2^{0\dgr}
- \bar c_3 \, \t e_L^\dgr \t e_R 
+ \bar c_7 \, h_2^{0\dgr} \, \t \nu_R ,
\label {gh10}
\eeq
and $ \cL_{\g h_1^0} $ becomes
\beq
\cL_{\g h_1^0} = - \left|
\bar c_1 \, h_2^{0\dgr}
- \bar c_3 \, \t e_L^\dgr \t e_R
+ \bar c_7 \, h_2^{0\dgr} \, \t \nu_R \right|^2 .
\label {Lh10}
\eeq
\par
The action density for the auxiliary field $ \g h_2^0 $ is
\bea
\cL_{\g h_2^0} & = &
\g h_2^{0\dgr} \g h_2^0
- c_1 \, h_1^0 \g h_2^0 
- \bar c_1 \, h_1^{0\dgr} \g h_2^{0\dgr} 
- c_2 \, \t \nu_L \g h_2^0 
- \bar c_2 \, \t \nu_L^\dgr \g h_2^{0\dgr} 
\nn\\ & & \mbox{}
- c_7 \, h_1^0 \g h_2^0 \, \t \nu_R^\dgr
- \bar c_7 \, h_1^{0\dgr} \g h_2^{0\dgr} \, \t \nu_R 
- c_8 \t \nu_L \t \nu_R^\dgr \g h_2^0
- \bar c_8 \t \nu_L^ \dgr \t \nu_R \g h_2^{0\dgr} .
\label {gh20terms}
\eea
So $ \g h_2^0 $ must be
\beq
\g h_2^0 = \bar c_1 \, h_1^{0\dgr} 
+ \bar c_2 \, \t \nu_L^\dgr 
+ \bar c_7 \, h_1^{0\dgr} \t \nu_R 
+ \bar c_8 \, \t \nu_L^ \dgr \t \nu_R ,
\label {gh20}
\eeq
and $ \cL_{\g h_2^0} $ must be
\beq
\cL_{\g h_2^0} = - \left|
\bar c_1 \, h_1^{0\dgr} 
+ \bar c_2 \, \t \nu_L^\dgr 
+ \bar c_7 \, h_1^{0\dgr} \t \nu_R 
+ \bar c_8 \, \t \nu_L^ \dgr \t \nu_R \right|^2 .
\label {Lh20}
\eeq
\par
The action density for the remaining
neutral chiral auxiliary field $ \g \nu_L $ is
\beq
\cL_{\g \nu_L} = \g \nu_L^\dgr \g \nu_L
- c_2 \g \nu_L h_2^0 
- \bar c_2 \, \g \nu_L^\dgr h_2^{0\dgr} 
- c_3 \g \nu_L h_1^- \t e_R^\dgr 
- \bar c_3 \, \g \nu_L ^\dgr h_1^+ \t e_R 
- c_8 \g \nu_L \t \nu_R^\dgr h_2^0 
- \bar c_8 \g \nu_L^ \dgr \t \nu_R h_2^{0 \dgr}.
\label {gnuLterms}
\eeq
The constraint on $ \g \nu_L $ is
\beq
\g \nu_L = \bar c_2 \, h_2^{0\dgr} 
+ \bar c_3 \, h_1^+ \t e_R + \bar c_8 \t \nu_R h_2^{0 \dgr} .
\label {gnuL}
\eeq
So the action density $ \cL_{\g \nu_L} $ is
\beq
\cL_{\g \nu_L} = - \left|
\bar c_2 \, h_2^{0\dgr}
+ \bar c_3 \, h_1^+ \t e_R 
+ \bar c_8 \t \nu_R h_2^{0 \dgr} \right|^2 .
\label {LgnuL}
\eeq
\par
Since charge is conserved,
we shall be interested mainly 
in the purely neutral part of the
above contributions to the scalar potential
and in other purely neutral terms that
arise from the $SU(2)_L$ auxiliary fields $ \vec D $
\beq
\cL_{\vec D} = - \half 
\left( \sum g_2 \bar A_i \thalf \vec \s_{ij} A_j \right)^2 
\label {LvD}
\eeq
and from the $U(1)_Y$ auxiliary field $D$
\beq
\cL_D = - \half 
\left( -\xi + \sum g_1 \thalf y_i |A_i|^2 \right)^2 \,.
\label {LD}
\eeq
The purely neutral part of $ \cL_{\vec D} $ is
\beq
\cL_{\vec D}^0 = - \frac{1}{8} g_2^2
\left( | h_1^0 |^2 - | h_2^0 |^2 + | \t \nu_L |^2 \right)^2 ,
\label {LvD0}
\eeq 
while that of $\cL_D$ is
\beq
\cL_D^0 =  - \frac{1}{2} \left[ - \xi + \frac{ g_1 }{ 2 } 
\left( - | h_1^0 |^2 + | h_2^0 |^2 - | \t \nu_L |^2 \right)
\right]^2 .
\label {LD0}
\eeq
\par
If we now gather all these purely neutral terms, 
then we may write the purely neutral scalar potential as
\bea
V_0 & = & \mbox{}
\left| 
c_4  + 2 c_5 \t \nu_R^\dgr + 3 c_6 \left( \t \nu_R^\dgr \right)^2
- c_7 \, h_1^0 h_2^0 - c_8 \, \t \nu_L h_2^0 \right|^2
+ \left| \bar c_1 + \bar c_7 \, \t \nu_R \right|^2 
\left| h_2^0 \right|^2
\nn\\ & & \mbox{} 
+ \left|
\bar c_1 \, h_1^{0\dgr}
+ \bar c_2 \, \t \nu_L^\dgr
+ \bar c_7 \, h_1^{0\dgr} \t \nu_R 
+ \bar c_8 \, \t \nu_L^ \dgr \t \nu_R \right|^2
+ \left| \bar c_2 + \bar c_8 \t \nu_R \right|^2
\left| h_2^0 \right|^2
\nn\\ & & \mbox{}
+ \frac{1}{2} \left[ \xi + \frac{ g_1 }{ 2 }
\left( | h_1^0 |^2 - | h_2^0 |^2 + | \t \nu_L |^2 \right)
\right]^2
\nn\\ & & \mbox{}
+ \frac{1}{8} g_2^2 
\left( | h_1^0 |^2 - | h_2^0 |^2 + | \t \nu_L |^2 \right)^2 .
\eea 
The last two terms in this expression for $ V_0 $
are positive and can not both be zero unless $ \xi = 0 $\@.
Thus if $ \xi \ne 0 $, then $ V_0 > 0 $,
and \susy is broken spontaneously.
It is also clear that if $ | \xi | $ is 
sufficiently large, then the minimum
of the potential will be drawn toward
\beq
| h_1^0 |^2 - | h_2^0 |^2 + | \t \nu_L |^2 = 
- 2 \frac { \xi }{ g_1 } ,
\label {attractor}
\eeq
so that gauge symmetry will also be broken spontaneously.

\section*{Acknowledgements}
This work was partially supported by
the U.S. Department of Energy.
I am grateful for conversations
with Michael Gold and Reinhard Stotzer.

\end{document}